\documentclass[submitting]{nst}

\usepackage{subfigure,dcolumn}
\usepackage[T2A,T1]{fontenc}
\usepackage[russian,english]{babel}

\usepackage{listings}
\usepackage{multirow}
\usepackage{hyperref}
\usepackage{float}
\usepackage{amssymb}
\usepackage{natbib}
\usepackage{float}
\usepackage{ulem}
\usepackage{soul}
\usepackage{jabbrv}
\begin{document}

\title{Effectiveness Study of Calibration and Correction Algorithms on the Prototype of the POLAR-2/LPD Detector}\thanks{Supported by the National Key R\&D Program of China 2022YFA1602204, the National Natural Science Foundation of China (Grant Nos. 12175241, 12221005, 12027803), and by the Fundamental Research Funds for the Central Universities}

\author{Di-Fan Yi}
\affiliation{School of Physical Science, University of Chinese Academy of Sciences, Beijing, 100049, China}
\affiliation{School of Physical Science and Technology, Guangxi University, Nanning 530004, China}
\author{Qian Liu}
\email[Corresponding author, ]{Qian Liu, liuqian@ucas.ac.cn}
\affiliation{School of Physical Science, University of Chinese Academy of Sciences, Beijing, 100049, China}
\author{Hong-Bang Liu}
\affiliation{School of Physical Science and Technology, Guangxi University, Nanning 530004, China}
\author{Fei Xie}
\affiliation{School of Physical Science and Technology, Guangxi University, Nanning 530004, China}
\author{Huan-Bo Feng}
\affiliation{School of Physical Science and Technology, Guangxi University, Nanning 530004, China}
\author{Zu-ke Feng}
\affiliation{School of Physical Science and Technology, Guangxi University, Nanning 530004, China}
\author{Jin Li}
\affiliation{Institute of High Energy Physics, Chinese Academy of Sciences, Beijing, 100049, China}
\author{En-Wei Liang}
\affiliation{School of Physical Science and Technology, Guangxi University, Nanning 530004, China}
\author{Yang-Heng Zheng}
\affiliation{School of Physical Science, University of Chinese Academy of Sciences, Beijing, 100049, China}

\begin{abstract}
 Gaseous X-ray polarimetry refers to a class of detectors used for measuring the polarization of soft X-rays. The systematic effects of such detectors introduce residual modulation, leading to systematic biases in the polarization detection results of the source. This paper discusses the systematic effects and their calibration and correction using the Gas Microchannel Plate–Pixel Detector (GMPD) prototype for POLAR-2/Low-Energy X-ray Polarization Detector (LPD). Additionally, we propose an algorithm that combines parameterization with Monte Carlo simulation and Bayesian iteration to eliminate residual modulation. The residual modulation after data correction at different energy points has been reduced to below 1\%, and a good linear relationship is observed between the polarization degree and modulation degree. The improvement in modulation degree after correction ranges from 2\% to 15\%, and the results exceed those of the Imaging X-Ray Polarimetry Explorer (IXPE) above 5 keV.
\end{abstract}


\maketitle

\section{Introduction}
In recent decades, there has been a growing interest in the field of gamma-ray astronomy, particularly in the study of Gamma-ray bursts (GRBs) \cite{De_Angelis_2018}, which are highly energetic cosmic events. While satellite observations from missions like Swift \cite{Burrows2005} and Fermi \cite{Atwood_2009} have yielded valuable insights into the energy spectra and timing properties of GRBs, numerous fundamental questions remain unanswered. These include understanding the mechanisms that propel the energetic jets, elucidating the processes responsible for energy dissipation, determining the composition of the jets, investigating the configurations of magnetic fields, and unraveling the mechanisms behind particle acceleration and radiation \cite{KUMAR20151,wang2015gamma,Zhang_2002,Michaelpaper1,8e185ab94c424cab929ed60022fb4416}. The detection of polarization in GRBs plays a crucial role in providing important clues for addressing the aforementioned issues \cite{Toma_2009,10.1093/mnras/stz2976,10.1093/mnras/stu457,10.1111/j.1365-2966.2004.07387.x,Granot_2003,Zhang2019}.

Scheduled for deployment in 2026 as an external payload on the China Space Station, POLAR-2 \cite{de2021development} is the successor experiment to POLAR \cite{PRODUIT2018259}. Its main goal is to conduct high-precision measurements of polarization across the spectrum from soft X-rays to gamma rays. The GMPD \cite{Feng_2023} is an innovative gaseous pixel detector \cite{feng2024gas} developed to validate the design of the POLAR-2/LPD \cite{feng2023orbit} payload.

Recently launched polarimetric detectors such as PolarLight \cite{Feng2019,LI2015155}, IXPE \cite{10.1117/12.2275485}, and the under-development eXTP \cite{Zhang2018extp,shen2018current}, CATCH type-A \cite{li2023catch} and POLAR-2/LPD all utilize gaseous pixel polarimetric detector structures. This type of detector has high spatial resolution, capable of imaging electron tracks at the level of hundreds of micrometers, thus providing excellent sensitivity in polarimetric detection. However, due to the complex and sophisticated structure of the gaseous pixel detector, as well as its high spatial resolution sensitivity, operational state of the instrument, various components, and electronic devices may introduce systematic effects on the polarimetric detection results. These systematic effects can result in non-zero modulation named residual modulation when detecting unpolarized sources, leading to systematic biases in the measurement of polarized sources. Since low-energy electron tracks are relatively short, the residual modulation effects produced by these systematic effects are more significant in low-energy events and cannot be ignored.

For residual modulation, IXPE employs two methods for correction \cite{Rankin_2022}: the first involves oscillating the detector during the detection process to integrate and eliminate some of the systematic effects. The second method involves calibrating the corresponding Stokes parameters \cite{mcmaster1954polarization} $q$ and $u$ for systematic effects in different regions and energy points, and then subtracting them on a event-by-event basis to eliminate the systematic effects. In this reasearch, We listed some of the known causes of residual modulation and corrected some of these effects based on their generation mechanisms. For another part of the residual modulation where the specific causes are currently unclear, we proposed a correction algorithm and obtained favorable outcomes.

In this paper, we introduce basic structure and polarization detection principles of the POLAR-2/LPD detector in Sect.\ref{sec:Geometric_structure}. We then discuss the residual modulation caused by signal response and its correction methods in Sect.\ref{sec:Complete}. In Sect.\ref{sec:Incomplete}, we discuss the residual modulation caused by geometric effects and proposed a modulation curve correction method based on the parameterization of scale ratios, combined with Monte Carlo simulation and Bayesian iteration \cite{DAGOSTINI1995487} (see in Appendix \ref{sec::Bayesian}), and provide errors of this algorithm. Subsequently, we compare various data reconstruction characteristics before and after algorithm correction and compared them with the modulation calibrated by the IXPE detector. Finally, in Sect.\ref{sec:summary}, we discuss the performance of the GMPD after correction, emphasizing the performance and scalability of the correction algorithm, and outline prospects future work.

\section{Geometric structure and Working principle of LPD}
\label{sec:Geometric_structure}
\begin{figure*}[htbp]
\includegraphics
  [width=0.9\hsize]
  {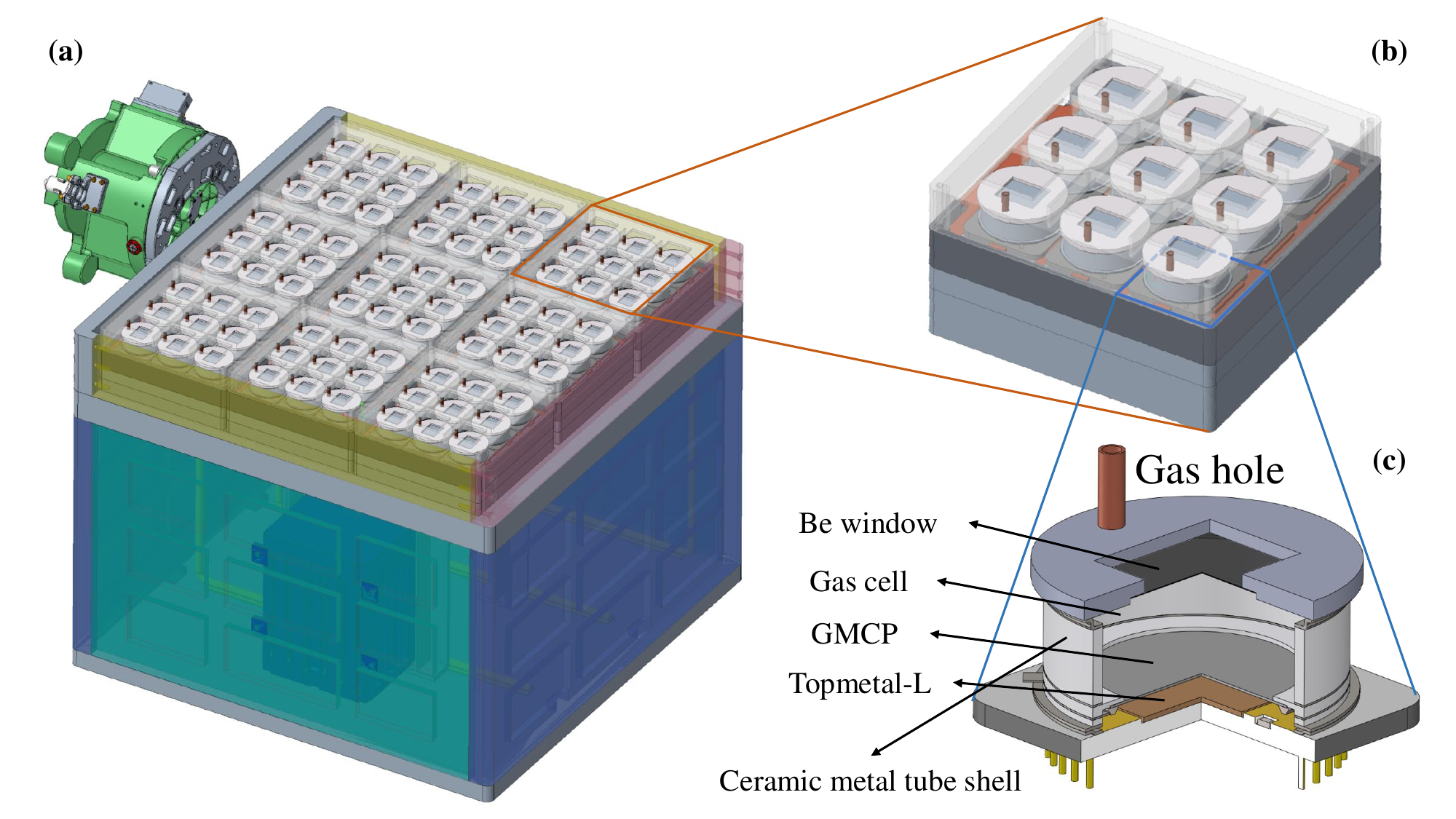}
\caption{Mechanical structures of LPD. (a) The LPD payload. (b) The detector array module. (c) The detector unit.}
\label{LPD Structure}
\end{figure*}

The LPD system shown in Fig.\ref{LPD Structure} is composed of a total of 9 detector modules, arranged in a 3×3 array configuration. Each detector module consists of 9 detection units with 90° field of view (FoV), resulting in a total of 81 detection units. A detection unit of LPD consists of working gas, a Gas Micro-Channel Plate (GMCP) \cite{FENG2023168499}, a pixel readout chip, and frame structure. Working gas: 60\% volume ratio of dimethyl ether (DME), 40\% helium (He), at 0.8\,atmospheres pressure, serving the purpose of photoelectric effects and formation of ionization tracks. The upper end of the gas chamber is sealed with a 50\,$\mu$m beryllium window, which prevents entry of lower energy photons and ensures gas containment to prevent leakage. GMCP layer is positioned near the bottom plane at the chamber, for the purpose of electron avalanche multiplication. At the bottom of the chamber is the Topmetal-L chip, specifically designed for LPD within the Topmetal chip series\cite{gao2016topmetal, AN2016144, REN2020164557, Zhou2024}. It features a 356$\times$ 512 pixel array with a pixel size of 45 $\mu$m and supports readout modes in Rolling Shutter and Region of Interest. Additionally, it operates with low power consumption (0.8\,W).

\begin{figure}[htbp]
\includegraphics
  [width=0.9\hsize]
  {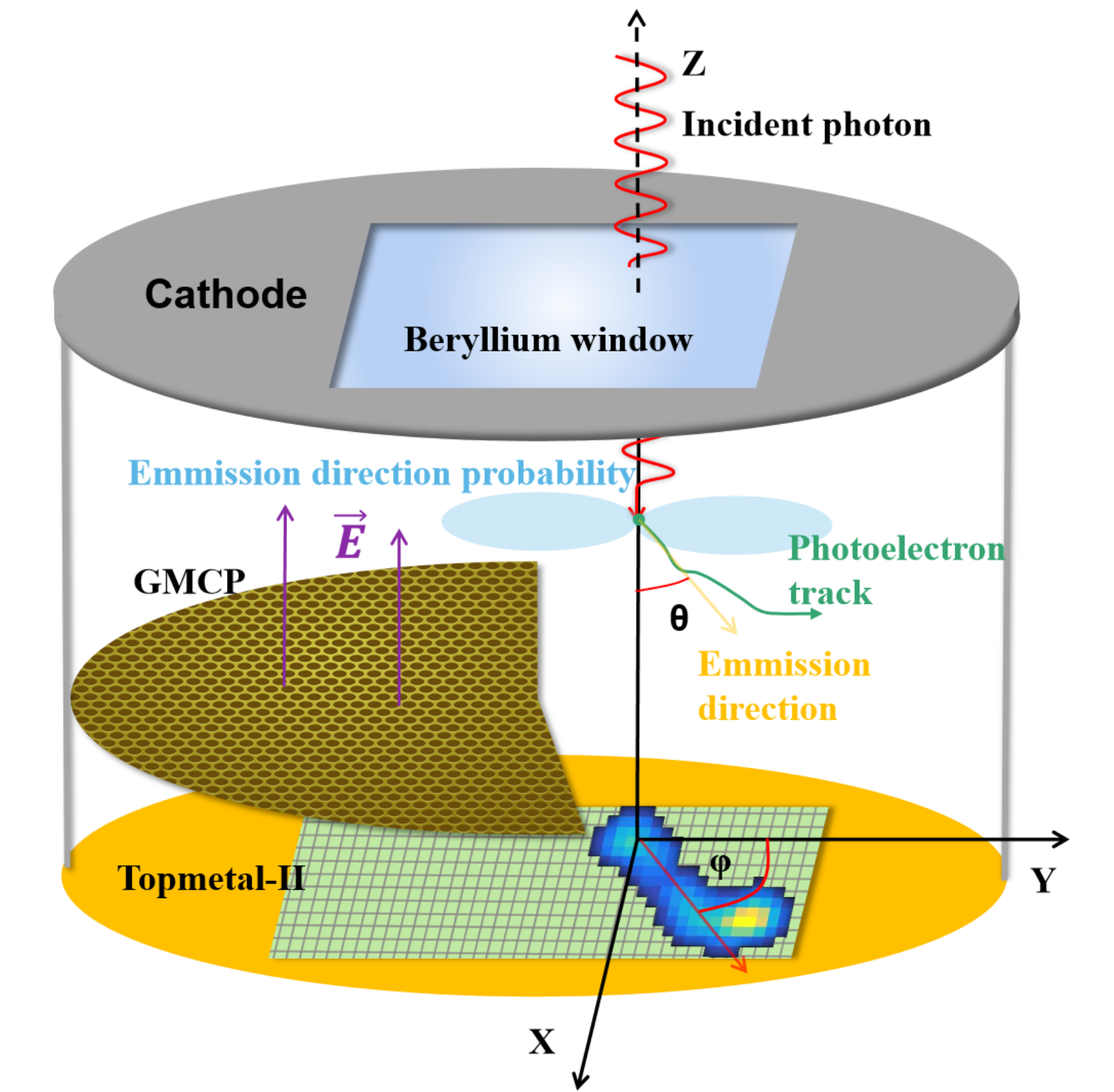}
\caption{Polarization detection principle of GMPD.}
\label{detect principle}
\end{figure}

The soft X-rays, as shown in Fig.\ref{detect principle}, pass through the beryllium window of the detector unit and have a certain probability of undergoing photoelectric effects within the drift region, resulting in the generation of photoelectrons. These photoelectrons carry the polarization information of the incident photons. Photoelectrons deposit ionization energy within the gas and generate secondary ionization electrons until they come to a complete stop. The secondary electrons are multiplied by transferring to the holes inside the GMCP under a drift electric field of 1\,keV/cm in the drift region. Within the induction region, an upward-directed electric field is applied, causing some of the secondary ionization electrons drift downward onto the surface of the GMCP. A portion of these electrons enters the micro-channels and undergoes avalanche multiplication. The multiplied electrons then emerge from the lower surface of the GMCP, where some of them are absorbed, resulting in the production of a pulse signal. The remaining multiplied electrons continue to drift towards the Topmetal chip, inducing signals in the corresponding pixel positions. This process projects the photoelectron tracks onto the 2D plane of the Topmetal chip, ultimately outputting an energy deposition projection image of the track.

\begin{figure}[htbp]
  \centering
  \subfigure[]{\includegraphics[width=0.23\textwidth]{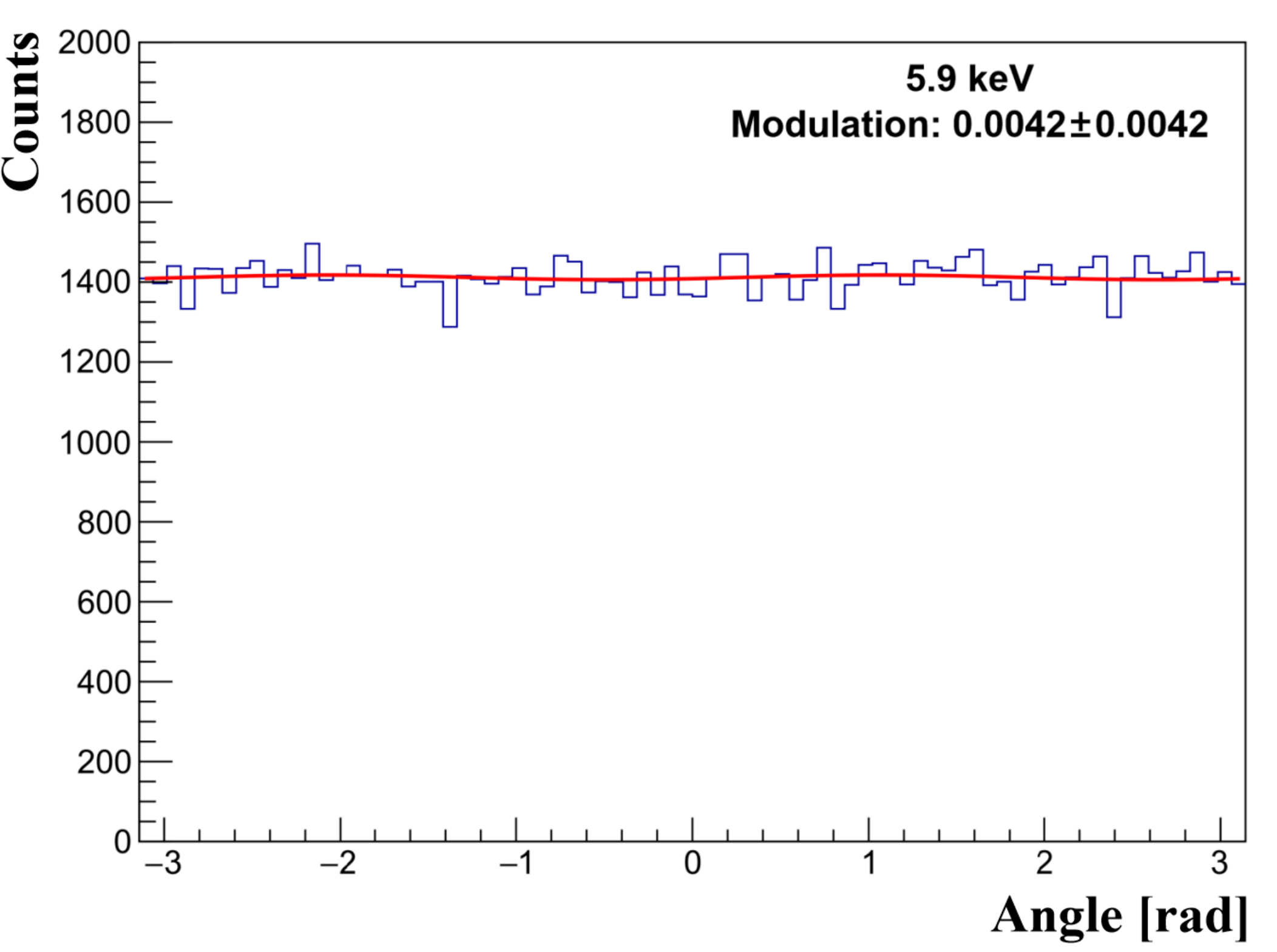}}
  \hfill
  \subfigure[]{\includegraphics[width=0.23\textwidth]{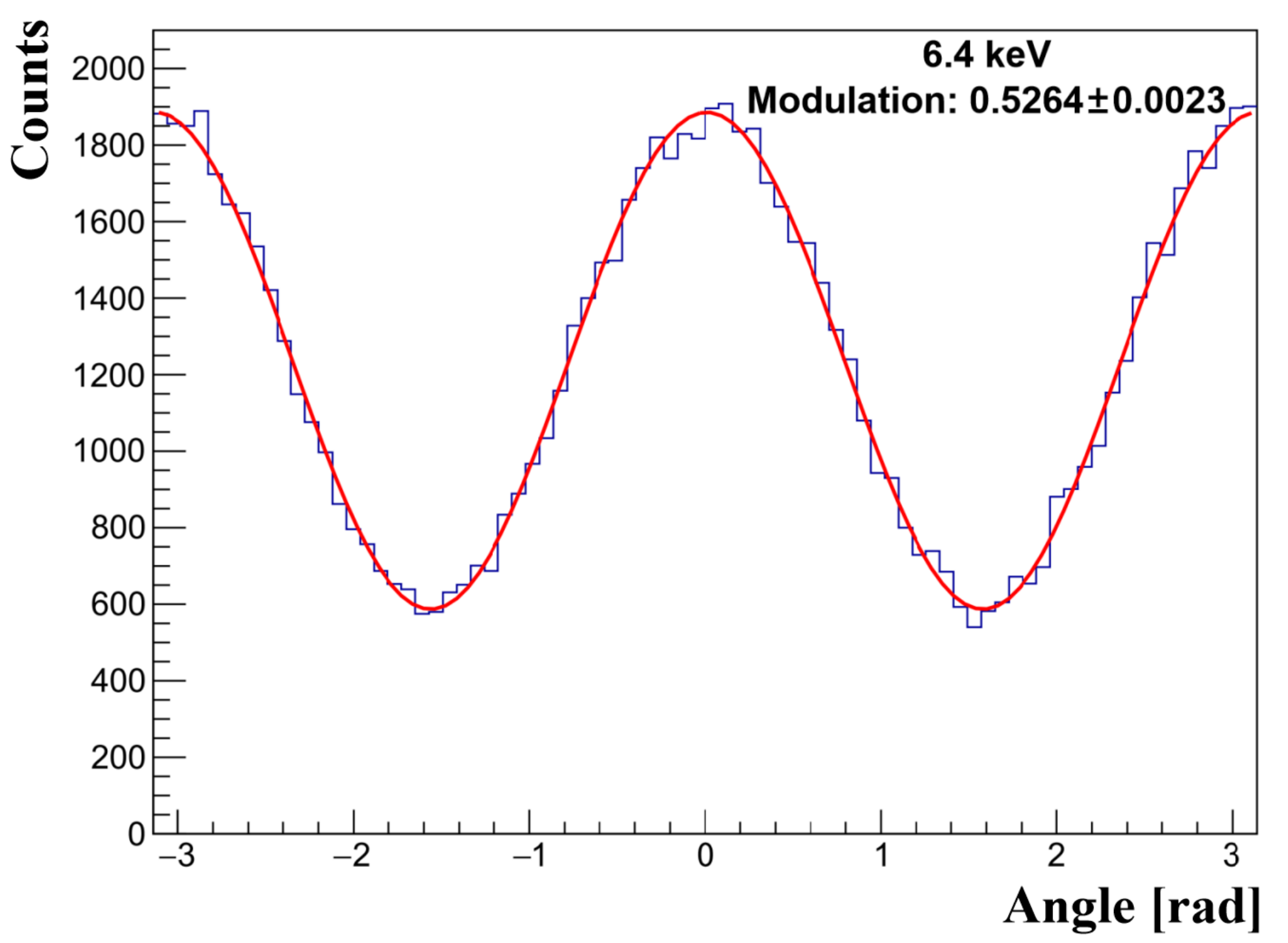}}
  \caption{(a) The modulation curves for unpolarized 5.9\,keV X-rays. (b) polarized 6.40\,keV X-rays with polarization angle at 0°.}
  \label{fig:Recontruction}
\end{figure}

In general, the angular distribution of photoelectrons detected by the gaseous pixel detector is modulated by polarized X-rays. For gaseous pixel detectors, photons primarily interact with the K-shell electrons of gas molecules through photoelectric interactions, and the direction of emitted electron is described by the differential cross-section according to the following formula \cite{Grandy1991}:
\begin{equation} \label{eq:cross-section} \frac{{\rm d}\sigma^{k}}{{\rm d}\Omega} \propto \frac{{{\rm sin}^{2}\theta {\rm cos}^{2}\phi}}{(1+\beta{\rm cos}\theta)^2}. \end{equation}

where $\beta$ is the emission velocity of the photoelectron in units of the speed of light $c$, $\theta$ and $\phi$ are the latitude and azimuth angles, respectively. Due to the lack of resolution in the $Z$ direction for the LPD detector, we reconstruct the 2D projection angular distribution of the azimuthal angle of the emitted photoelectron , corresponding to the integration of $\theta$ in the formula. Therefore, the reconstructed angular distribution is modulated by the ${\rm cos}^2$ factor. In theory, for 100\% polarized X-rays, the minimum value of the true emission distribution in phi should be 0. However, due to limitations in instrument resolution, system effects, and reconstruction algorithm accuracy, there will be a certain proportion of unmodulated components in the angular distribution, as shown in Fig.\ref{fig:Recontruction} (a), (b). Therefore, the modulation function of $\phi$, $M(\phi)$ can be written as:
\begin{equation} \label{eq:Mphi} M(\phi) = A + B{\rm cos}^{2}(\phi -\phi_{0}). \end{equation}
Where $\phi_{0}$ is the polarization phase of source. Modulation factor $\mu$ is defined as the ratio of the area occupied by the modulation component in the distribution:
\begin{equation} \label{eq:Modulatoin} \mu = \frac{B}{2A+B}. \end{equation}

\section{Calibration and Correction of Signal Response}
\label{sec:Complete}

The structural design and operational principles of GMPD result in variations in the response between pixels, which can impact the energy resolution of the detector. More importantly, some of these factors can introduce anisotropic differences, leading to residual modulation. This section primarily discusses the impact and calibration of these factors.

\subsection{Pixel response differences}
\begin{figure}[htbp]
\includegraphics
  [width=0.9\hsize]
  {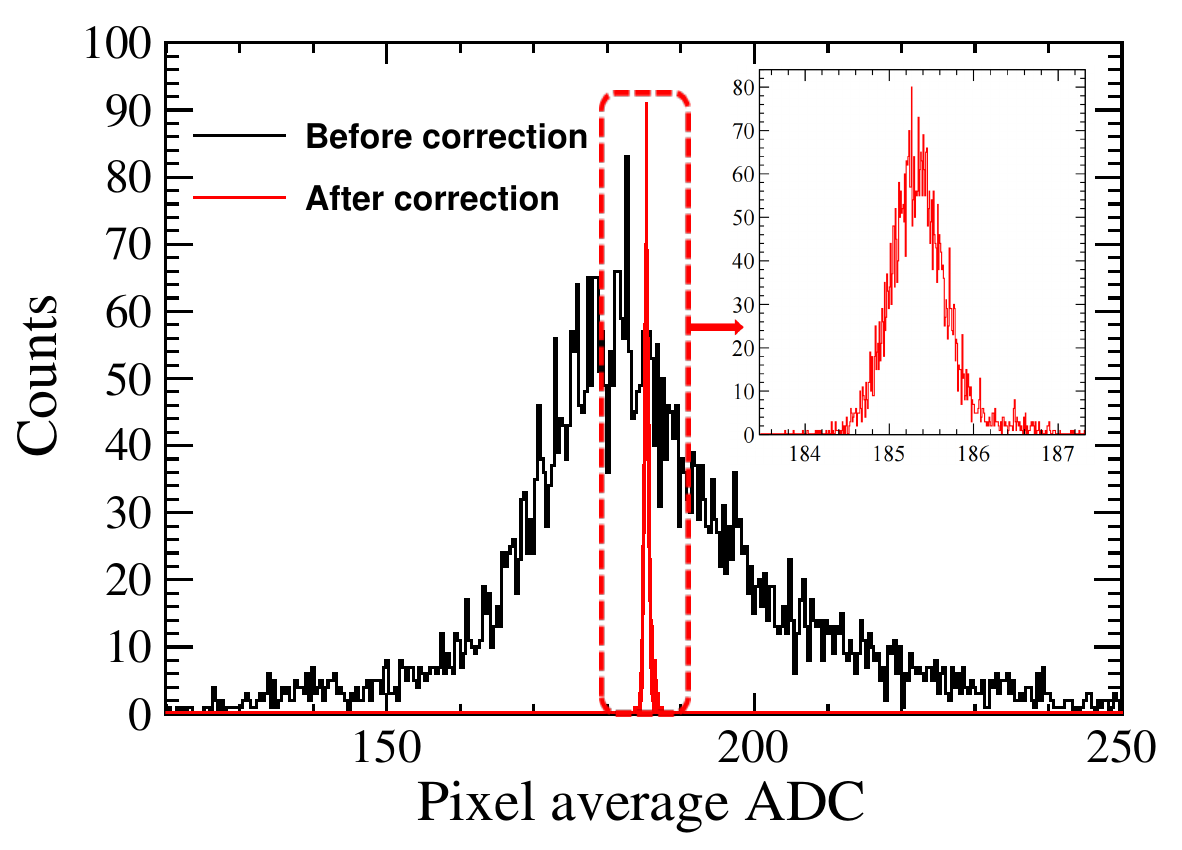}
\caption{The distribution of pixel average ADC values before and after correction, with blue representing the pre-correction values and red representing the post-correction values.}
\label{fig:ADC_Response}
\end{figure}

Due to the subtle structural differences between each pixel, the uniformity of the electric field, and the uniformity of GMCP gain, the signal induction intensity of drift charge varies among different pixels. It is necessary to calibrate the relative signal induction intensity on each pixel. We uniformly irradiate with a 4.51\,keV flat source and statistically record the signal distribution received by each pixel. As the response curve of the pixels exhibits good linearity \cite{10111087}, we can characterize the relative signal induction intensity of a pixel by the mean of the signal distribution received on that pixel. In the process of calculating the mean, we only selected the part of the signal intensity greater than 50 in order to eliminate the interference of noise. Fig.\ref{fig:ADC_Response} illustrates the average distribution of pixel ADC values before and after correction.

\subsection{Rolling-Shutter and Signal Decay}

\begin{figure*}[htbp]
  \centering
  \subfigure[]{\includegraphics[width=0.296\textwidth]{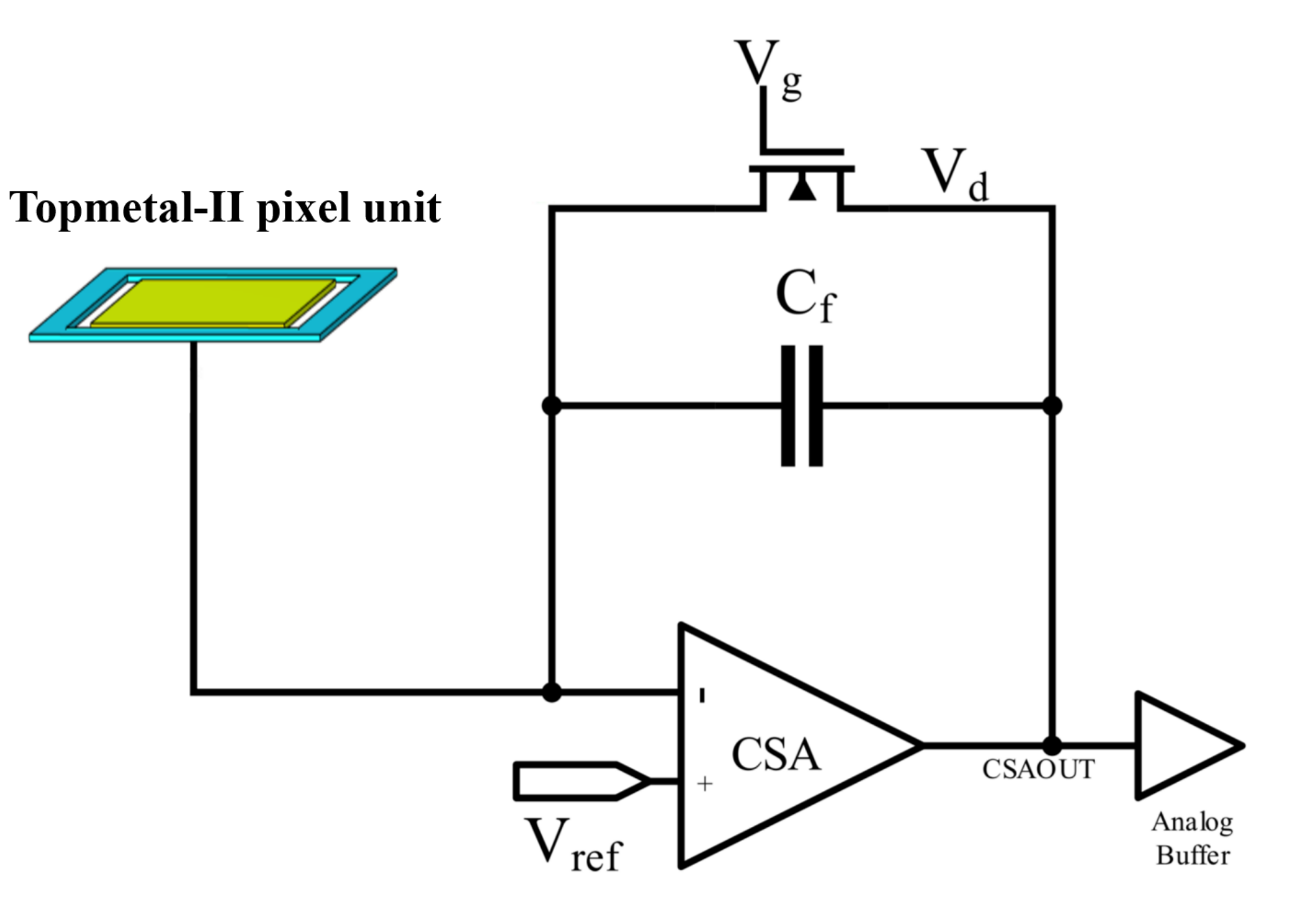}}
  \hfill
  \subfigure[]{\includegraphics[width=0.292\textwidth]{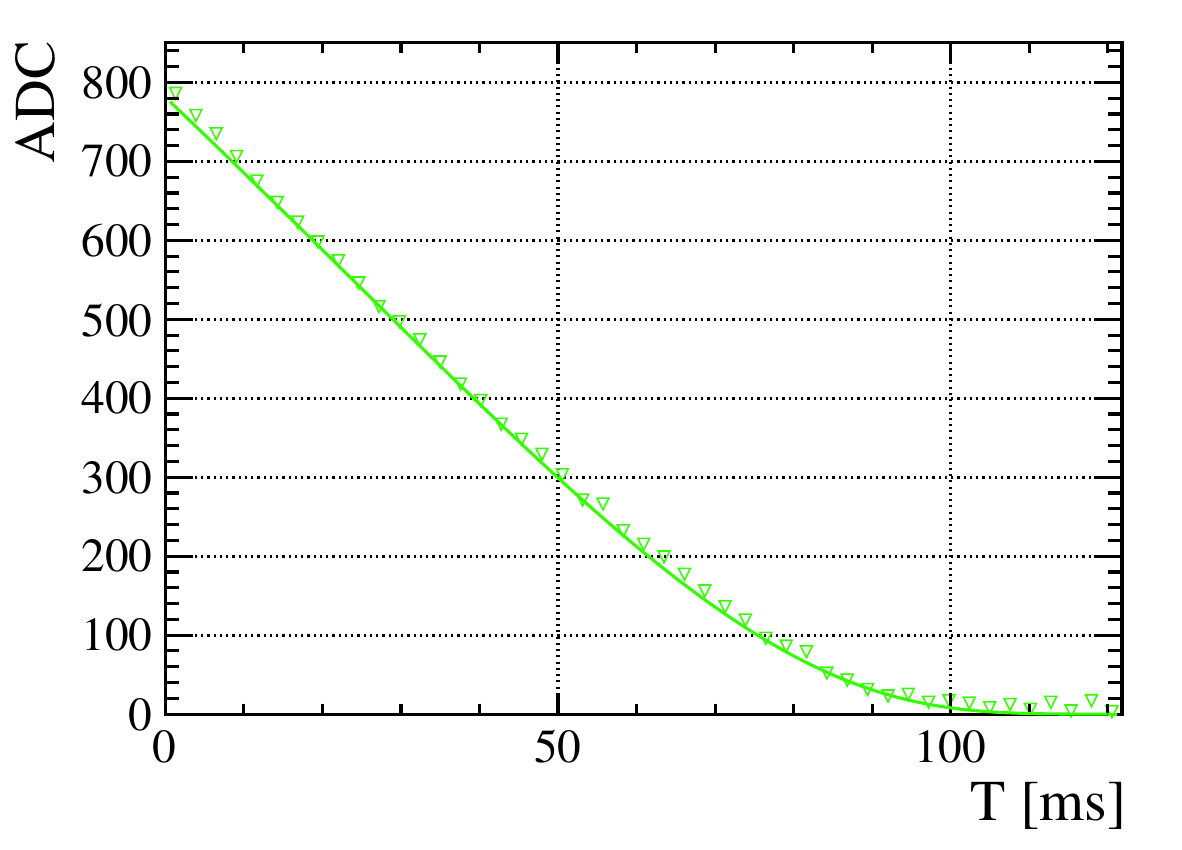}}
  \hfill
  \subfigure[]{\includegraphics[width=0.292\textwidth]{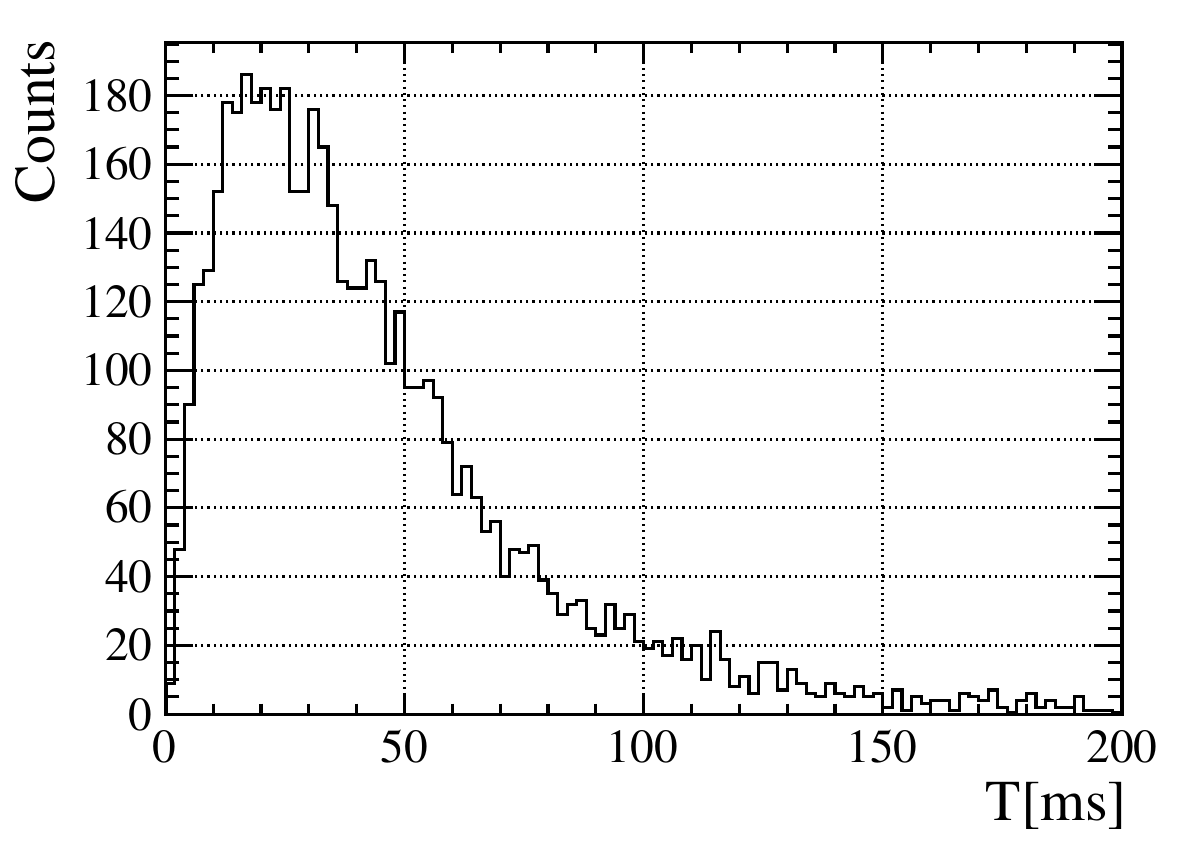}}
  \caption{(a) Topmetal-II$^-$ pixel CSA structure. The green portion represents the top metal, which is used to induce the drift charge signal. C$_\text{f}$ denotes the feedback capacitor. V$_\text{g}$ represents the gate voltage, V$_\text{d}$ represents the drain voltage, and V$_\text{ref}$ represents the amplifier's reference voltage. (b) Trend and fitting curve of pixel ADC value decay over time. (c) Distribution of decay times for all pixels, where the decay time of a pixel is defined as the time interval for the pixel signal value to decay from $a$ to $a/2$.}
  \label{fig:photo-electron-track}
\end{figure*}

Another source of residual modulation is the attenuation of pixel signal amplitude caused by signal readout time delay. Since Topmetal adopts a rolling-shutter readout method to read each frame of the image, pixel signals are read out in sequence, which means there is a certain delay from the triggering of the induction signal to the readout, and there is also a delay in the readout time of different pixels on the same track. The charge-sensitive
preamplifier(CSA) structure of Topmetal-II is shown in the Fig.\ref{fig:photo-electron-track}(a). A CSA  includes a differential amplifier, a sub-threshold nMOS resistor, and a feedback capacitor, where the pixel controls the discharge of induced charge through the drain voltage. Therefore, the scanned readout signal will be attenuated compared to the true signal amplitude at the triggering moment due to the time delay. The scanning time for one frame of Topmetal-II is $\tau_{\text{frame}}=$2.59\,ms, and the scanning time interval for each pixel is $\tau_{\text{pixel}}=$ 500\,ns. Due to the rolling-shutter method of scanning the chip column by column along the 0° direction, the scanning time interval between adjacent pixels in the 90° direction is 35\,$\mu$s, while the scanning time interval between adjacent pixels in the 0° direction is 500 ns, with $\Delta T_{0^\circ} \ll \Delta T_{90^\circ}$ . The difference in scanning time intervals between 0° and 90° can result in inconsistent signal attenuation gradients in these two directions, introducing a vertical bias, namely, residual modulation in the 90° direction.

In order to calibrate the systematic errors caused during the scanning process, it is first necessary to calibrate the signal attenuation behavior of each pixel, and secondly to determine the time difference between each triggered and readout pixel. We input square wave signals to the chip and record the output results of the pixel readout signals for multiple consecutive frames, in order to obtain the decay characteristics of each pixel and perform parameter fitting. The theoretical formula for pixel decay is given by Equation \ref{eq:pixel decay}:
\begin{equation} \label{eq:pixel decay} A(t) = a \cdot \exp\left(-\frac{t}{b \cdot t + c} \right). \end{equation}

Fig.\ref{fig:photo-electron-track}(b) illustrates the decay pattern of signal intensity over time on a pixel and the fitting result. Fig.\ref{fig:photo-electron-track}(c) shows the decay time distribution of all pixels on Topmetal-II, indicating that the most likely decay time for the pixel is 20 ms.

The time precision of GMCP is 10\,ns\cite{10111087}, significantly smaller than the most likely decay time scale of the pixel. We can obtain the arrival time of the event signal at GMCP by comparing the GMCP trigger signal with the Topmetal-II trigger signal. Since the distance between GMCP and Topmetal-II chips is only 3.4\,mm, the typical time scale for the electron multiplied by GMCP to traverse this distance is on the order of tens of nanoseconds, which can be neglected compared to the characteristic time scale of the pixel decay. Therefore, it can be assumed that the time of arrival of the multiplied electron at Topmetal-II is $\text{T}_{\text{Arrival}} = \text{T}_{\text{GMCP}}$. Denoting the time corresponding to the Topmetal-II trigger frame as $\text{T}_{\text{Top}}$, if $\text{T}_{\text{Top}}$ > $\text{T}_{\text{GMCP}}$, it indicates that the position scanned has already passed through the region reached by the photoelectrons when the signal arrived. In this case, for the k-th fired pixel, the time difference between the signal triggering and readout is $\Delta t = \text{T}_{\text{Top}}-\text{T}_{\text{GMCP}}+\tau_{\text{pixel}}\times\text{ID}_{k}$, where $\text{ID}_{k}$ is the index of the k-th fired pixel. If $\text{T}_{\text{Top}}$ < $\text{T}_{\text{GMCP}}$, it indicates that the position scanned has not yet passed through the region reached by the photoelectrons when the signal arrived. In this case, the time difference is: $\Delta t = \text{T}_{\text{GMCP}}-\text{T}_{\text{Top}}+\tau_{\text{pixel}}\times\text{ID}_{k}$. Based on the time difference, we can then correct the decay signal for each pixel using the following formulas:
\begin{equation} \label{eq:decay_fix} A_{\text{truth}} = a \cdot \exp\left(-\frac{t_{0}-\Delta t}{b \cdot (t_{0}-\Delta t) + c} \right), \end{equation}
\begin{equation} \label{eq:decay_t0} t_{0} = \frac{c}{1-b\cdot \log(\frac{a}{A_{out}})}. \end{equation}

\subsection{Charge pile up effect}
\begin{figure*}[htbp]
  \centering
  \subfigure[]{\includegraphics[width=0.5\textwidth]{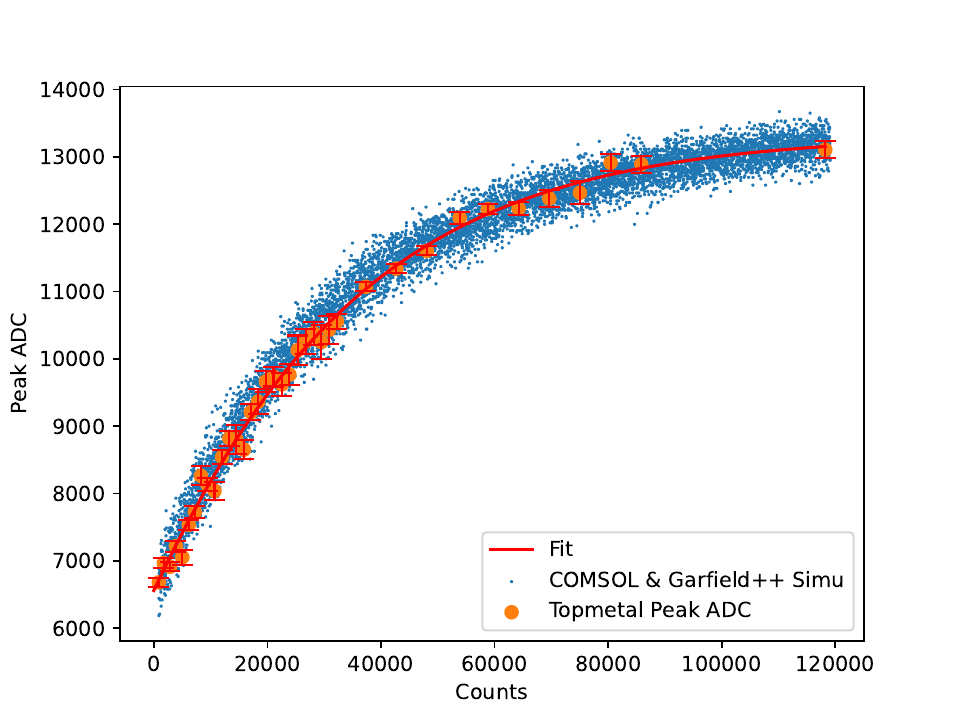}}
  \hfill
  \subfigure[]{\includegraphics[width=0.4\textwidth]{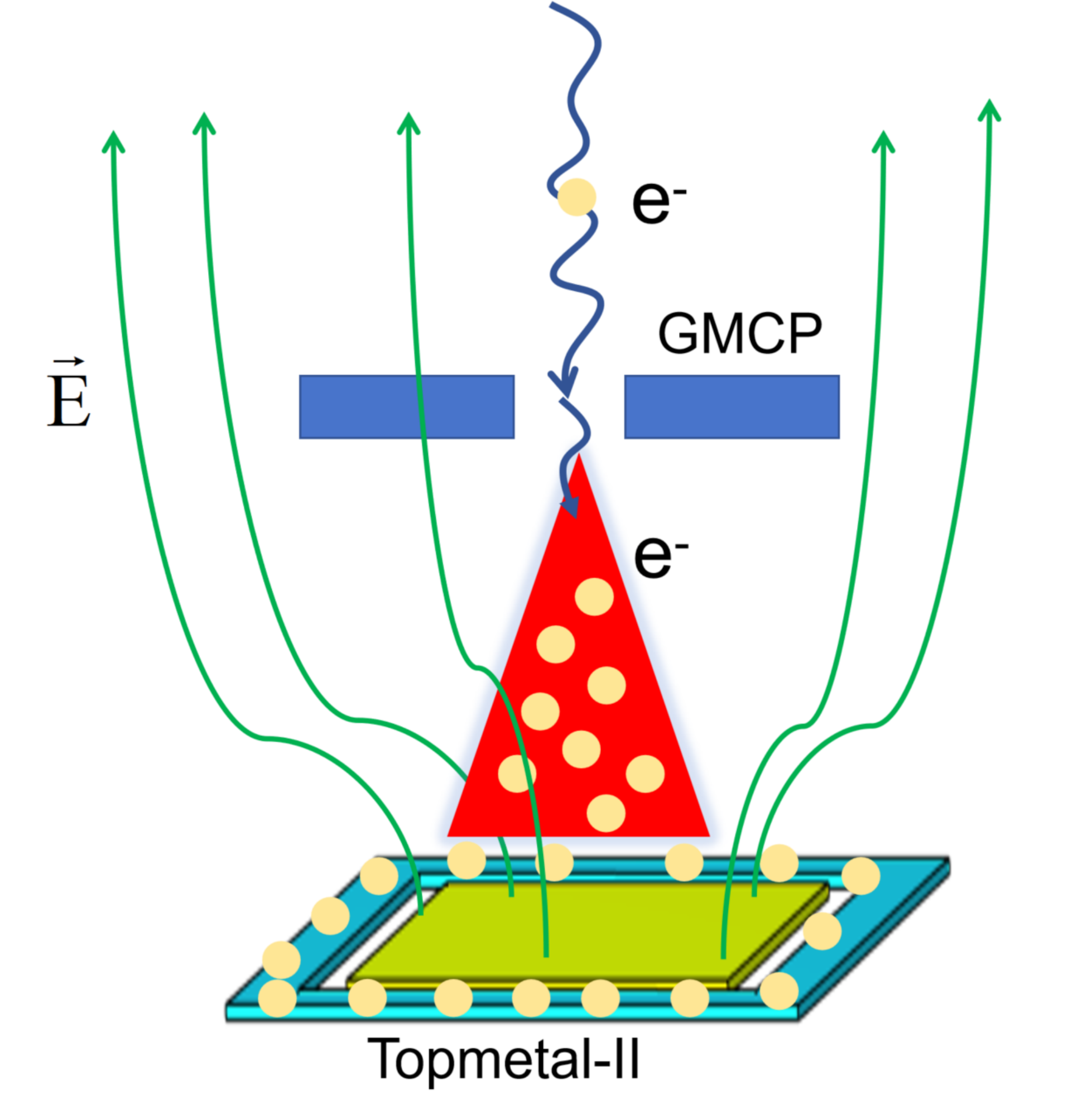}}
  \caption{(a) The variation of chip collection efficiency with the accumulation of events. The orange points represent experimental data results of 4.51\,keV photoelectron track energy deposition, characterized by fitting the peak ADC of the data spectrum to represent the variation in chip charge collection efficiency. The blue points represent the trend of collection efficiency variation obtained from the joint simulation using COMSOL and GARFIELD++. (b) Charge Accumulation Schematic: Electrons accumulate on the insulating layer, forming a low potential region around the top metal. This leads to the formation of a funnel-shaped electric field above the top metal, thereby enhancing the charge collection efficiency of the pixel.}
  \label{fig:Accum}
\end{figure*}

The encapsulated detector exhibits an initial stage where the gain increases with the accumulated number of events, as illustrated in Fig.\ref{fig:Accum}(a). This effect is attributed to charge accumulation. The surface of the Topmetal-II utilized in the detector features a grid-like insulating layer, causing electrons that fall and become adsorbed on this layer to have limited mobility. As the accumulation of avalanche multiplied electrons rises, the potential on the chip surface gradually alters, impacting the charge collection efficiency and modifying the gain, as depicted in Fig.\ref{fig:Accum}(b).

\begin{figure*}[htbp]
  \centering
  \subfigure[]{\includegraphics[width=0.45\textwidth]{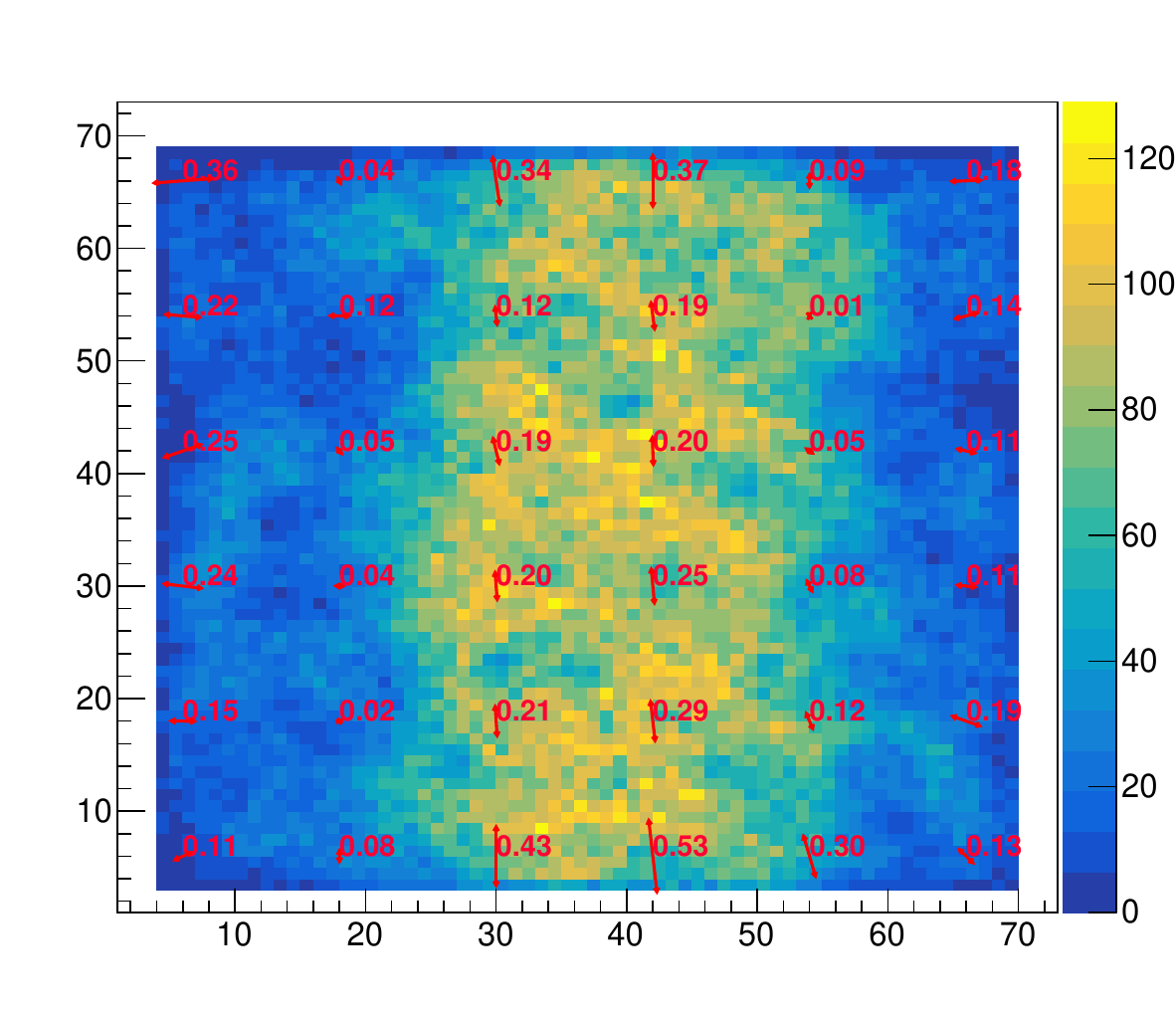}}
  \hfill
  \subfigure[]{\includegraphics[width=0.45\textwidth]{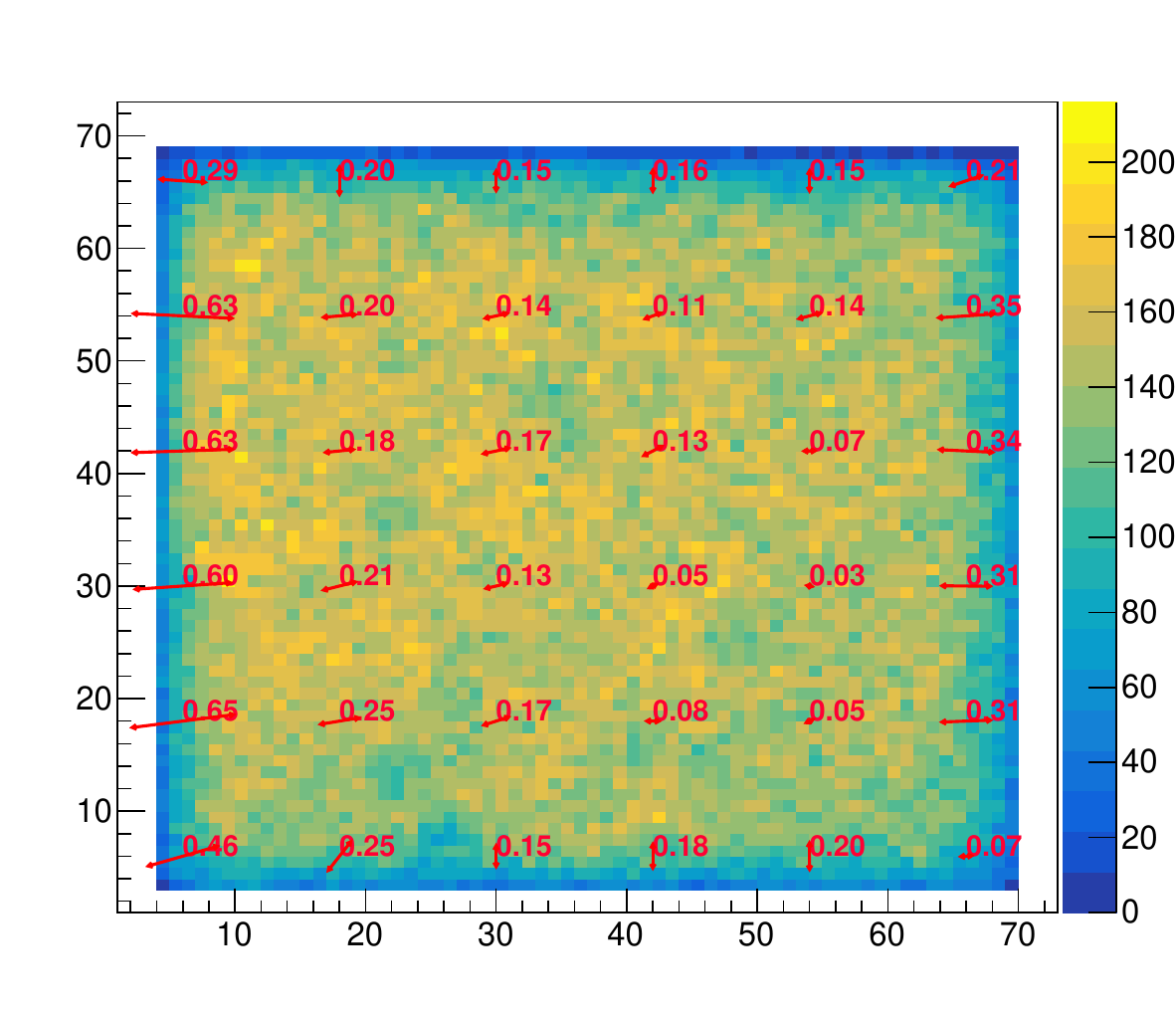}}
  \caption{The residual modulation distribution in different regions of the detector, with each small region composed of 12$\times$12 pixels. (a) The residual modulation distribution of 5.9\,keV Fe$^{55}$ tracks after uneven charge accumulation due to narrow gap obstruction. (b) The residual modulation distribution of 5.9\,keV Fe$^{55}$ tracks after uniform charge accumulation following the removal of the narrow gap obstruction. The heatmap represents the distribution of the reconstructed photoelectron emission positions of the tracks, with the direction of the red lines indicating the direction of residual modulation. The length of the line and the adjacent number represent the value of the residual modulation.}
  \label{fig:Uneven_Accum}
\end{figure*}

If the charge accumulation process is unevenly distributed on the surface of the chip, it will lead to a noticeable signal intensity gradient on the chip surface, eventually resulting in the generation of pseudo-modulation perpendicular to the gradient direction. Fig.\ref{fig:Uneven_Accum} illustrates the residual modulation caused by the charge accumulation effect. Initially, a ferrous strip was used to partially obstruct a section of the detector's field of view, leaving a gap of a few millimeters. Following a 2-hour exposure to an X-ray flat source, the obstruction was removed, and a 5.9\,keV unpolarized Fe$^{55}$ source was used to irradiate and collect the photoelectron tracks. Upon reconstruction, it was observed in Fig.\ref{fig:Uneven_Accum}(a) that the signal gain at the previous narrow gap position was significantly higher than the shaded area, and the residual modulation in the narrow gap area was higher than in the shaded area, with the modulation direction parallel to the gap. Subsequently, without any obstruction, the X-ray flat source was used again for 4 hours to accumulate charges on the entire surface of the chip to near saturation. The detector was then irradiated with the 5.9\,keV unpolarized Fe$^{55}$, and the tracks were reconstructed in Fig.\ref{fig:Uneven_Accum}(b). Comparing the results of the Fe$^{55}$ measurements before and after charge accumulation reached saturation, it was found that the residual modulation caused by the uneven gain due to charge accumulation significantly decreased. Therefore, it is possible to mitigate the impact of the charge accumulation effect by calibrating or measuring the detector after saturating the charge accumulation before conducting experiments. Since the accumulated charge is unlikely to naturally dissipate, once the detector is encapsulated, only one thorough charge accumulation is required.

By employing Garfield++ and COMSOL for charge drift accumulation iteration and updating of the drift electric field, we successfully replicated this effect in simulations, as indicated by the blue data points in Fig.\ref{fig:Accum}(a), which align with the experimentally observed gain variation results. The process of charge accumulation can be described by a simplified Eq.\ref{eq:charge_Accum}. Where $n$ is the number of events, $q$ is the accumulated charge on the chip, and $q_{\text{max}}$ is the maximum saturated accumulated charge, and $a_c$ is the charge adsorption coefficient. Therefore the change in the accumulated charge quantity with respect to the detector counts, q(n), can be expressed in a parametric form as given in Eq.\ref{eq:qn}.

\begin{equation} \label{eq:charge_Accum} \frac{d q(n)}{d n}=\alpha_c\left(1-\frac{q(n)}{q_{\max }}\right), \end{equation}
\begin{equation} \label{eq:qn} q(n) = x_{0}+x_{1}\exp(x_{2}(n+x_{3})). \end{equation}

Both experimental and simulation results indicate that the charge accumulation process gradually reaches saturation, leading to a stable final gain. Additionally, the non-focusing observation mode of the LPD can prevent the uneven accumulation of charge on the chip surface. Therefore, after a sufficient number of accumulated events, the impact of the pseudo-modulation caused by the charge accumulation effect in the encapsulated detection unit can be reduced to a negligible level.

\section{Calibration and Correction of Geometrical Effects}
\label{sec:Incomplete}

\subsection{Pixelization influence}

\begin{figure}[htbp]
  \centering
  \subfigure[]{\includegraphics[width=0.22\textwidth]{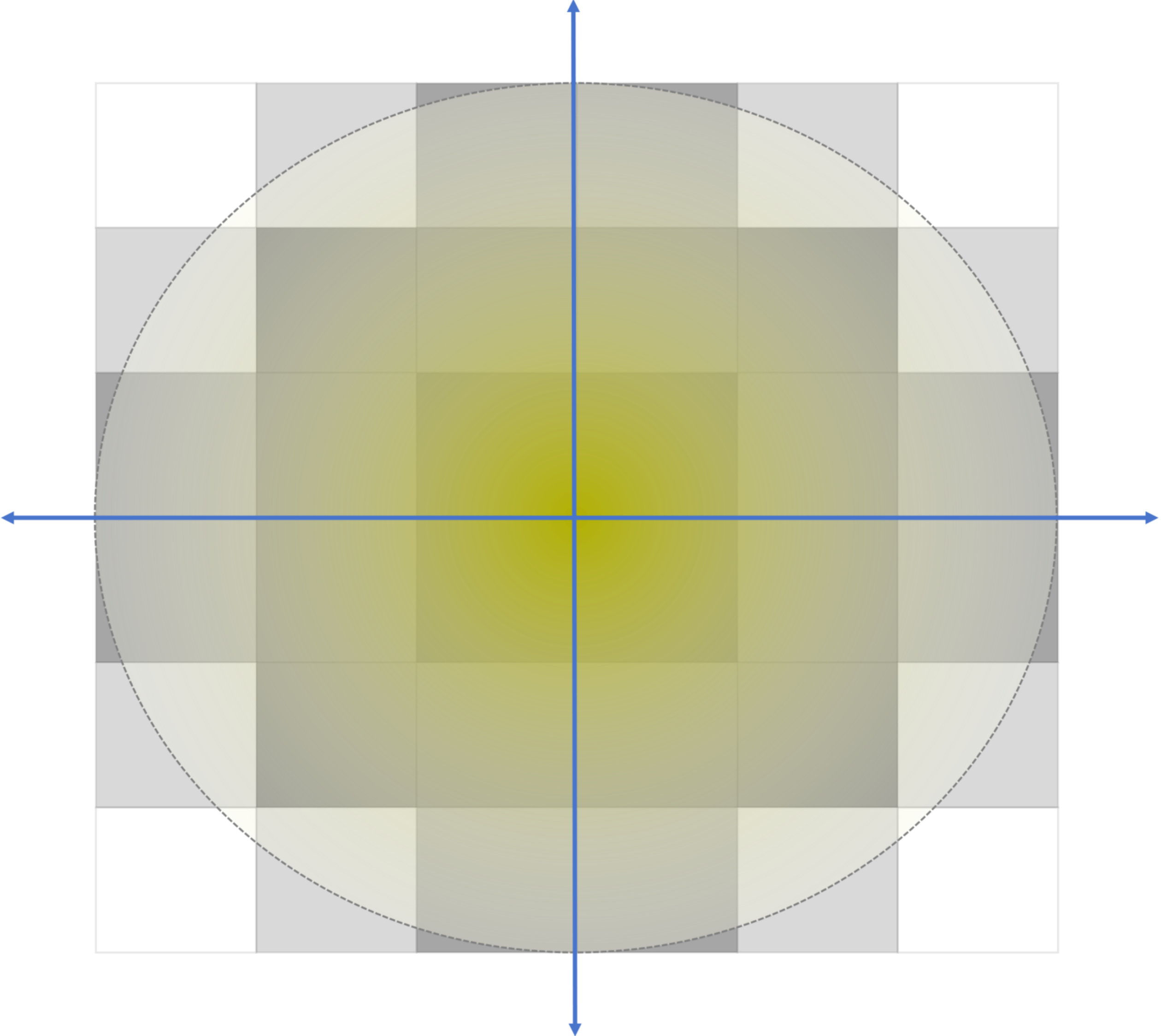}}
  \hfill
  \subfigure[]{\includegraphics[width=0.22\textwidth]{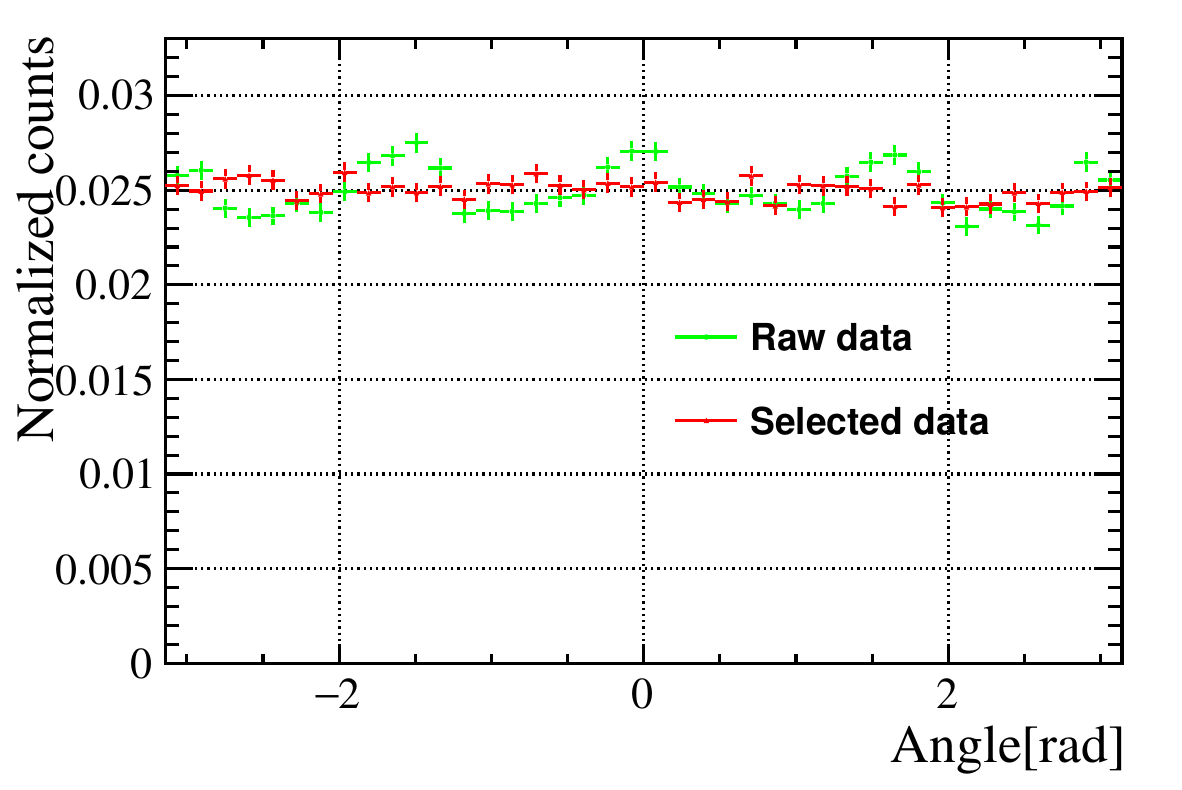}}
  \caption{(a)The shorter circular tracks are projected onto the readout Topmetal-II chip, where the pixel grayscale represents the signal intensity, with darker colors indicating stronger signals. The axis represents the direction obtained from track reconstruction. The pixel size is 83\,$\mu$m. (b)The angular distribution of reconstructed electron tracks at 2.98\,keV. The green data points represent the reconstruction results without any filtering based on the number of triggered pixels and ellipticity. The red data points represent the filtered results, excluding events with a number of triggered pixels less than 36 and the bottom 20\% of events with smaller ellipticity.}
  \label{fig:PixelArrange}
\end{figure}

As shown in Fig.\ref{fig:PixelArrange}(a), we consider a shorter track with a circular projection. Due to the parallel arrangement of Topmetal-II chips in the X and Y directions, the signal distribution sensed on the chip pixels exhibits anisotropy for such tracks. The symmetry is most pronounced in the directions of 0° and 90°, which are aligned with the pixel arrangement. The commonly used moment analysis algorithm for such shorter tracks calculates the centroid line of the pixel track to determine the direction of electron emission. This can lead to a bias in the reconstruction direction of these tracks towards 0° and 90°.

To mitigate the residual modulation caused by pixel arrangement, we need to exclude events with too few responsive pixels and events that are too short or too circular during event selection. Therefore, during reconstruction, we only select events with a number of hit pixels greater than or equal to 27 and exclude the bottom 20\% of events with smaller ellipticities. Fig.\ref{fig:PixelArrange}(b) below shows the angular distribution of the reconstructed unbiased events before and after the event selection. After the event selection, the residual modulation caused by pixel arrangement is significantly improved.

\subsection{Truncation effect}
Similarly, due to the Rolling-Shutter line-by-line scanning readout of the chip, if an event occurs precisely at the position covered by the pixels being scanned at that moment, the event will be truncated and appear in both the preceding and subsequent frames. If the truncated part in one frame has fewer fired pixels that do not exceed the threshold, we can only obtain an incomplete truncated event. As the edge of the truncated track is always parallel to the scanning direction, it introduces a systematic bias in scan direction.

Therefore, we need to select and filter out these truncated instances. Although most truncated tracks exhibit clearly defined edges, this feature is not a sufficient condition for identifying truncated instances, especially for shorter tracks at lower energies. Thus, we continue to utilize the time information from GMCP and Topmetal-II to determine if an instance is truncated. If the time difference recorded by Topmetal-II and GMCP is:$\Delta T_{\text{Diff}}=T_{\text{Top}}-T_{\text{GMCP}}$.
\begin{itemize}
\item When $\Delta T_{\text{Diff}}$ < 0, the pixel ID scanned when the signal arrives is:
$\text{ID}_{\text{Pixel}} = \frac{\Delta T_{\text{Diff}}}{\tau _{pixel}}$.

\item $\Delta T_{\text{Diff}}$ > 0, the signal actually arrives and is read by Topmetal-II in the second frame. Therefore, the pixel ID scanned when the signal arrives is:$\text{ID}_{\text{Pixel}} = 72\times72 - \frac{\Delta T_{\text{Diff}}}{\tau _{pixel}}$.
\end{itemize}

Considering the combined time resolution of GMCP and Topmetal-II is 262\,ns \cite{10111087} and $\tau _{pixel}$, we determine if a track is truncated by examining whether the pixel scanned when the signal arrives and the positions of the five pixels before and after it precisely overlap with the region covered by the photoelectron track signal.

\begin{figure*}[htbp]
  \centering
  \subfigure[]{\includegraphics[width=0.26\textwidth]{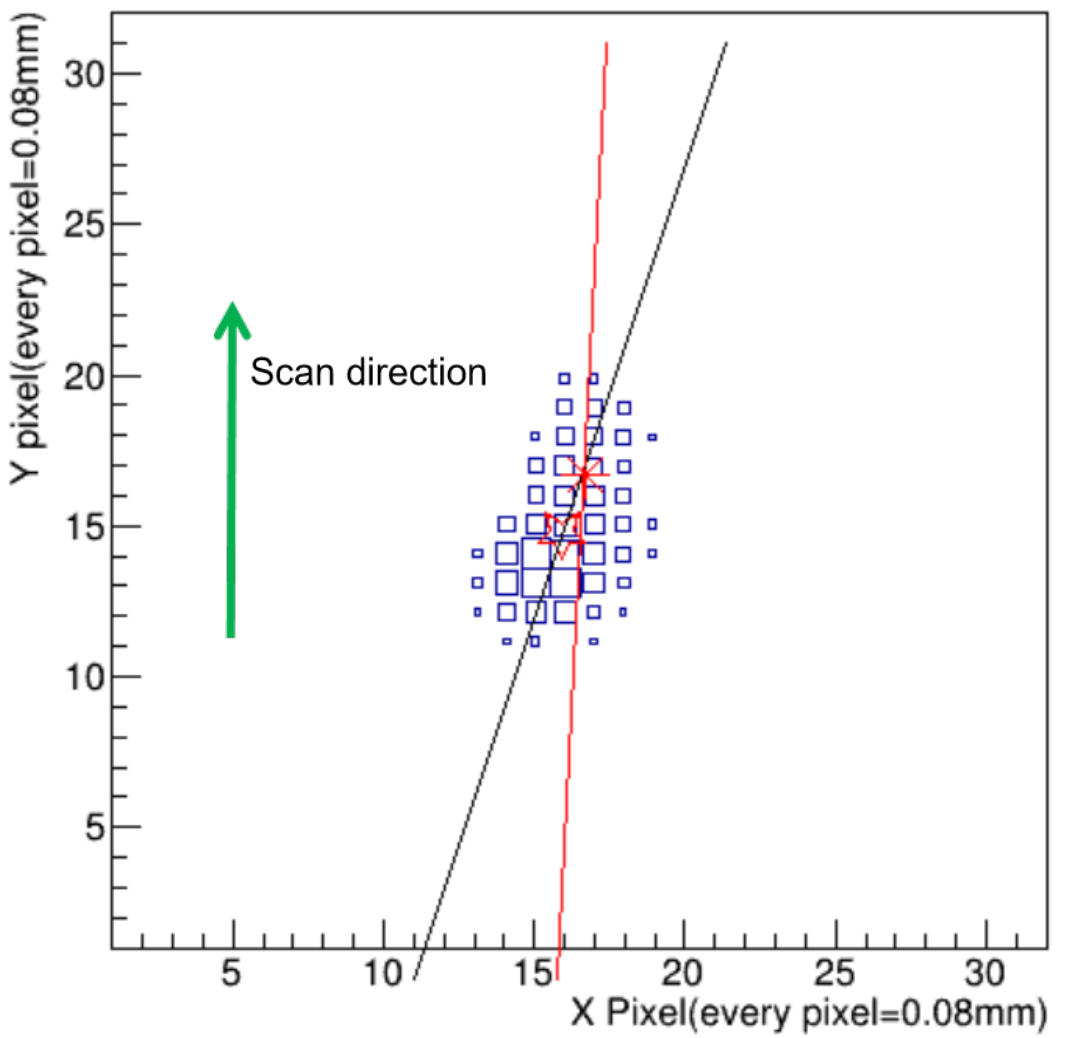}}
  \hfill
  \subfigure[]{\includegraphics[width=0.26\textwidth]{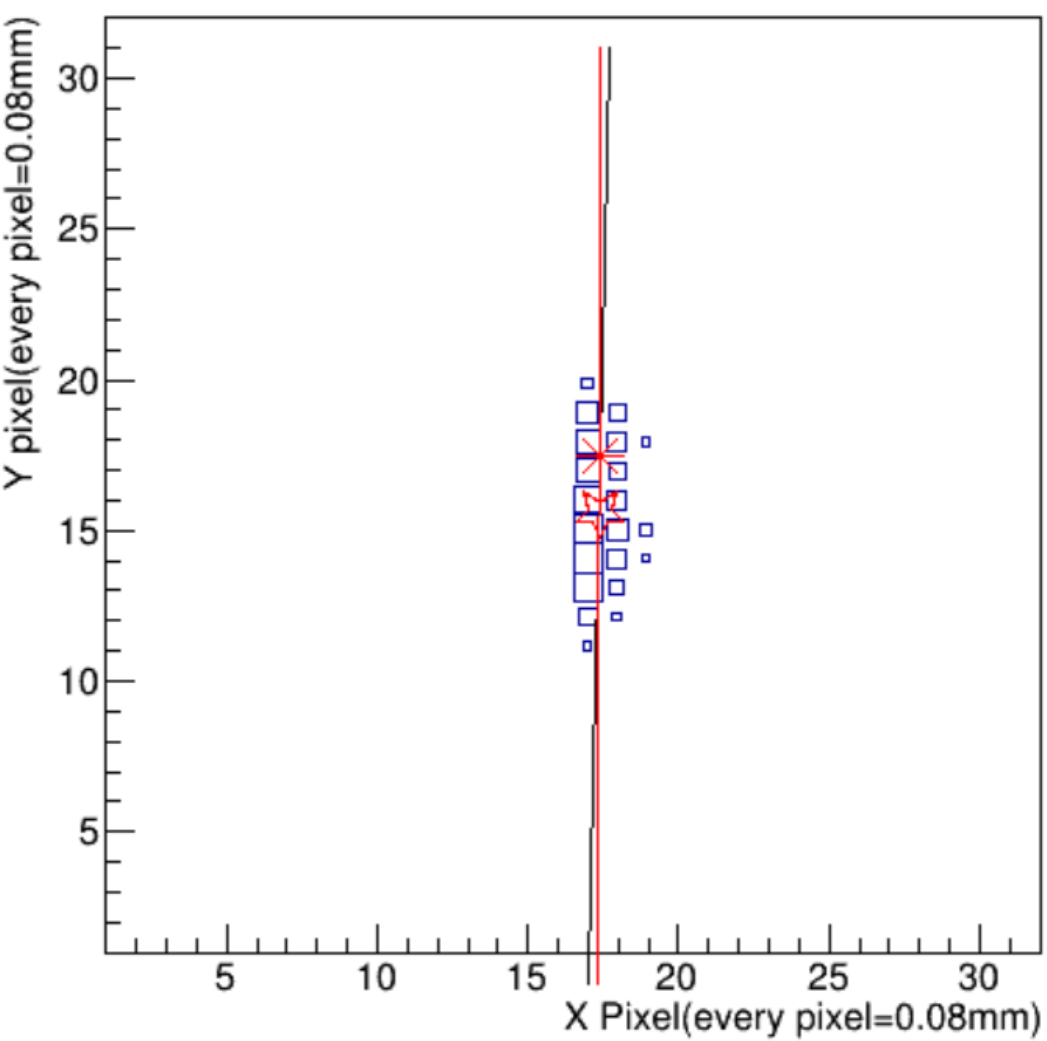}}
  \hfill
  \subfigure[]{\includegraphics[width=0.28\textwidth]{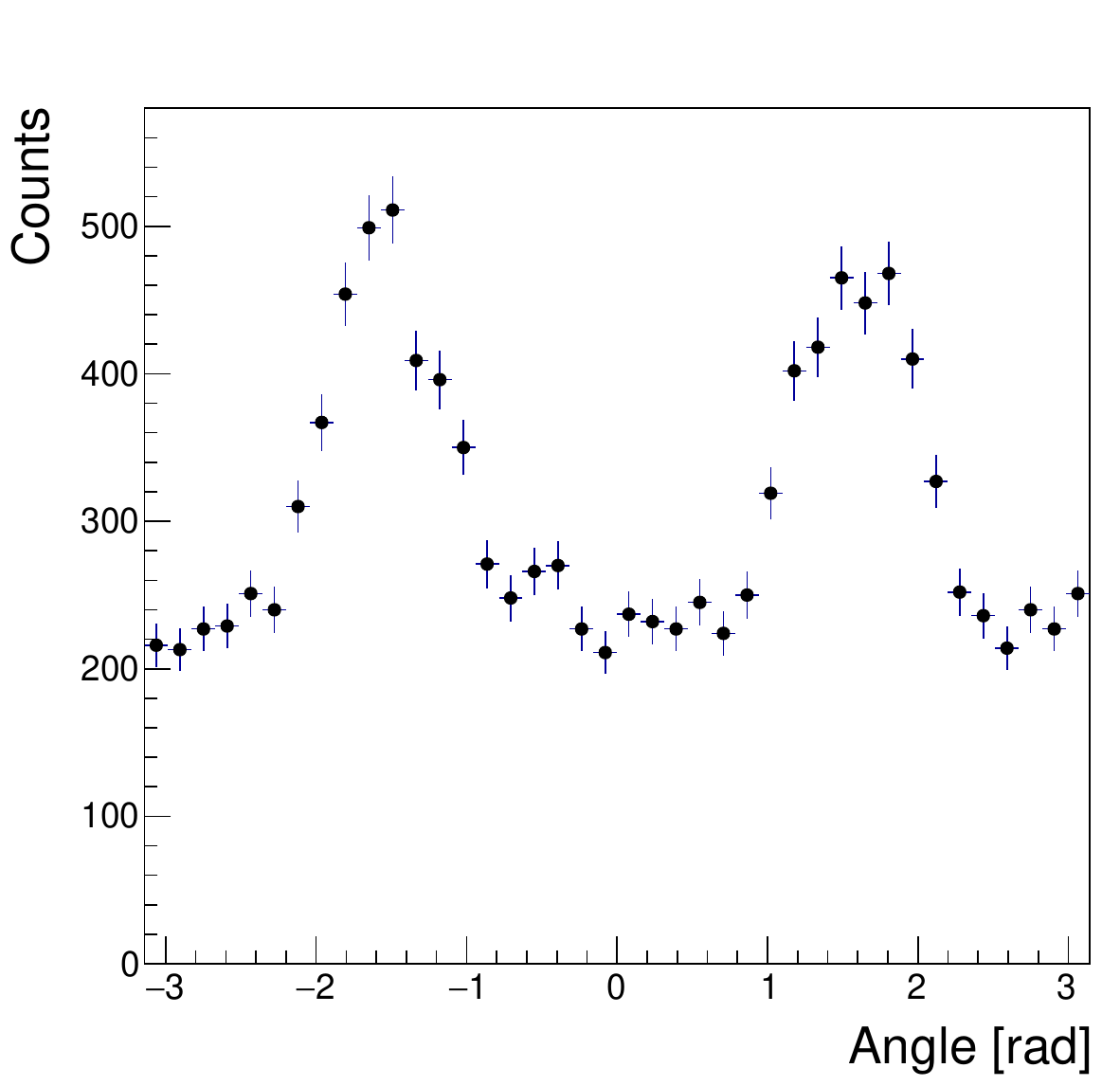}}
  \caption{Track turncation. The scale of the box represents the strength of the signal at that pixel. The red star denotes the reconstructed centroid position of the track, while the black line represents the reconstructed centroid line direction, which corresponds to the reconstructed azimuthal angle of the photoelectron emission. The red line divides the track into the energy deposition part of the Bragg peak and the initial position part of the photoelectric interaction.(a) A complete track. (b) A turncated track. (c) Reconstructed angular distribution of selected truncated events.}
  \label{fig:turncate}
\end{figure*}

Fig.\ref{fig:turncate}(a),(b) depict the reconstructed results of the complete events and the truncated events under scanning, respectively. Fig.\ref{fig:turncate}(c) presents the reconstructed angle distribution of the truncated events in the unpolarized 5.9\,keV dataset selected by the algorithm. It is observed that there is a significant bias in the direction of $\pm$90°.

\subsection{Track image distortion}
\begin{figure}[htbp]
\includegraphics
  [width=1.\hsize]
  {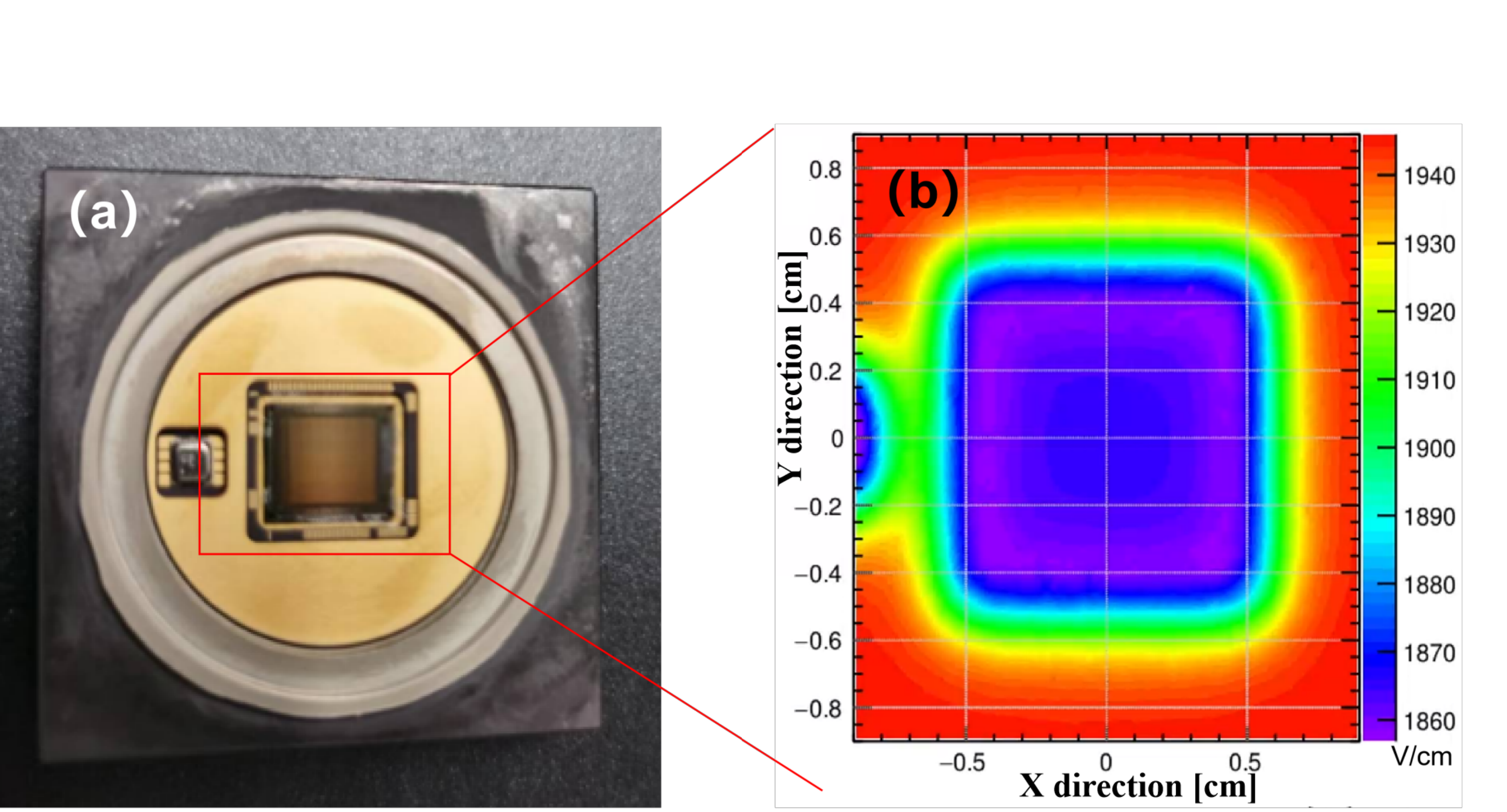}
\caption{(a) Physical diagram of the detector unit base, with the larger central chip being the Topmetal-II charge induction chip, and the smaller chip on the left being the temperature and pressure sensing chip. (b) Simulation results of the electric field near the detector base. Due to the structural characteristics of the pressure sensing chip, the overall geometric structure of the detection unit exhibits a certain degree of asymmetry, leading to some distortion in the electric field.}
\label{fig:FieldMC}
\end{figure}

Excluding the systematic effects and corrections discussed above, the angular reconstruction of track data obtained from the detector still exhibits some residual modulation. This may partly be attributed to the geometry and potential distribution of the detector. The gas cavity of the LPD detection unit is not completely symmetrical. As shown in Fig.\ref{fig:FieldMC}, in addition to the charge induction chip Topmetal, a temperature and pressure sensor chip is also placed nearby. This placement leads to a relatively significant distortion of the electric field near the side of the Topmetal chip adjacent to the sensor chip, resulting in a noticeably higher residual modulation on that side. Furthermore, there is a 1\,mm wide and 0.8\,mm deep groove between the charge induction collection plane of the Topmetal chip and the base plane of the detection unit. Additionally, several to a dozen bonding wires are present around the chip. The geometric structure of the chip's edge and the potential on the bonding wires also cause distortion of the electric field at the edge of the chip. Consequently, it can be observed that the direction of the residual modulation reconstructed in the edge portion of Fig.\ref{fig:Uneven_Accum} is generally perpendicular to the edge of the chip. Therefore, in order to minimize the influence of edge electric field distortion on the reconstruction, we choose to exclude events within 12 pixels of the charge center distance from the edge when selecting valid events.

The residual modulation distribution in different regions near the center of the chip appears to be more random. This variability in residual modulation in certain regions may stem from systematic process issues during chip etching, subtle irregularities during detector installation, and the uneven accumulation of charge resulting in differences in the electric field across different areas of the chip. These issues can all impact the electric field distribution near the chip surface, and the distortion of the electric field can alter the track shape. This alteration is often nonlinear, and the impact on tracks at different positions, heights, and lengths varies. As a result, we lack sufficiently precise information to make pixel-by-pixel or event-by-event corrections from a first-principles perspective for the obtained tracks in the experiment.

\begin{figure}[htbp]
\includegraphics
  [width=1.\hsize]
  {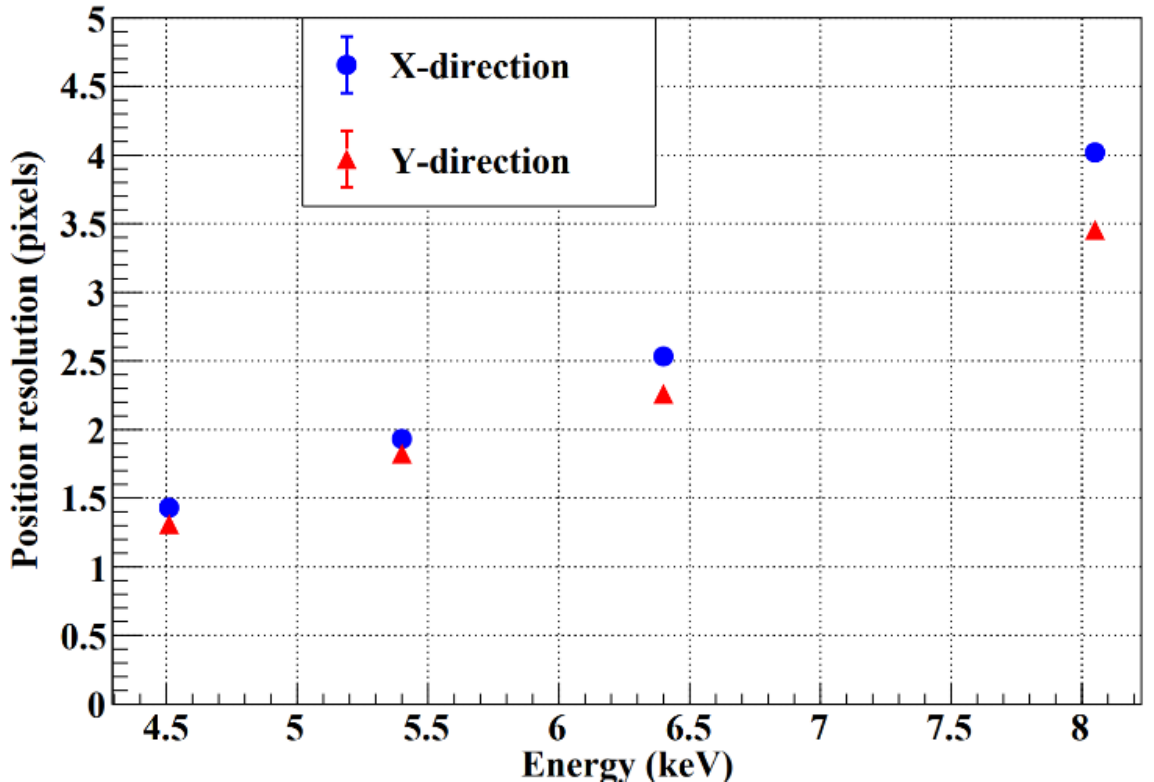}
\caption{The position resolution of GMPD of varying energies. Circular markers depict the results in the X-direction, while triangular markers represent the results in the Y-direction.}
\label{fig:XYResolution}
\end{figure}

The deformation of tracks is reflected in the differences in position resolution in different directions of the detector. As shown in the Fig.\ref{fig:XYResolution}, at different energy points, the position resolution in the X direction of the detector is worse than that in the Y direction. This indicates that the distortion of the track is more severe in the X direction, and this anisotropic deformation of tracks leads to excessive stretching in the X direction, resulting in significant residual modulation.

Similar residual modulation issues also arise in the IXPE detector. The correction scheme for residual modulation provided by IXPE involves calibrating experimental scales for each chip region to correct the Stokes parameters required for event reconstruction. Since the IXPE detector needs to image the observed objects, segmenting and correcting different regions is necessary. However, for the LPD, which lacks imaging capabilities, photons from the source will uniformly fall on the entire chip surface. Therefore, the LPD only needs to consider correcting the distribution of residual modulation integrated over the entire chip surface for events. To address this, we propose a Bayesian method combined with Monte Carlo simulations to correct residual modulation.

\subsubsection{Correction algorithm}
When correcting the data for an energy point, we only need to calibrate a correction parameter $\eta$: the ratio of the pixel size in the Y direction to the pixel size in the X direction. We can phenomenologically explain the need to introduce the parameter $\eta$: the distortion of the electric field will cause the equipotential surfaces to no longer be parallel to the Topmetal chip plane. Therefore, by projecting the chip plane onto the deformed equipotential surface, the linearity in different directions of the chip will have different scaling rates. We select the ratio of the scaling rates calibrated in the two directions parallel and perpendicular to the scanning direction as $\eta$. It should be noted that the $\eta$ value for different regions of the chip is different. However, because the LPD does not have polarized imaging capabilities, the correction parameter we consider is actually the weighted average value $\Bar{\eta}$ of the parameters for different chip regions.

Calibrating $\Bar{\eta}$ requires a comparison of experimental data and simulations. Taking the 5.40\,keV energy point as an example, we calibrated the modulation curves obtained from a 99.9\% polarized source at 0°, 30°, 60°, 90°, 120°, and 150° polarization phases. It can be observed that due to residual modulation, there are significant differences in the modulation at these phases, with a difference of approximately 18\% between the modulation at 0° and 90° as shown in Fig.\ref{fig:DifMod}(g). Since residual modulation is an inherent property of the detector and is independent of the polarization phase of the source, the overall modulation curve measured is a result of the superposition of residual modulation and source modulation. Therefore, the overall modulation curve can be described as:
\begin{equation} 
\label{eq:Obs_Modulation} 
M_{\text{Obs}}(\phi)=M_{\text{Res}}(\phi) \cdot M_{\text{Source}}(\phi,\phi_{0}). 
\end{equation}
Where $M_{\text{Obs}}$ is the modulation curve obtained from reconstructed angular distribution data, $M_{\text{Res}}$ represents the impact of residual modulation, and $M_{\text{Source}}$ is the modulation curve generated by the polarized source. At normal incidence, the form of $M_{\text{Source}}$ is:
\begin{equation} 
\label{eq:Source_Modulation} 
M_{\text{Source}}(\phi,\phi_{0}) = A\cos^2{(\phi - \phi_{0})} + B. 
\end{equation}

\begin{figure}[htbp]
  \centering
  \subfigure[Combining Polarization Data at 0° and 90°]{\includegraphics[width=0.22\textwidth]{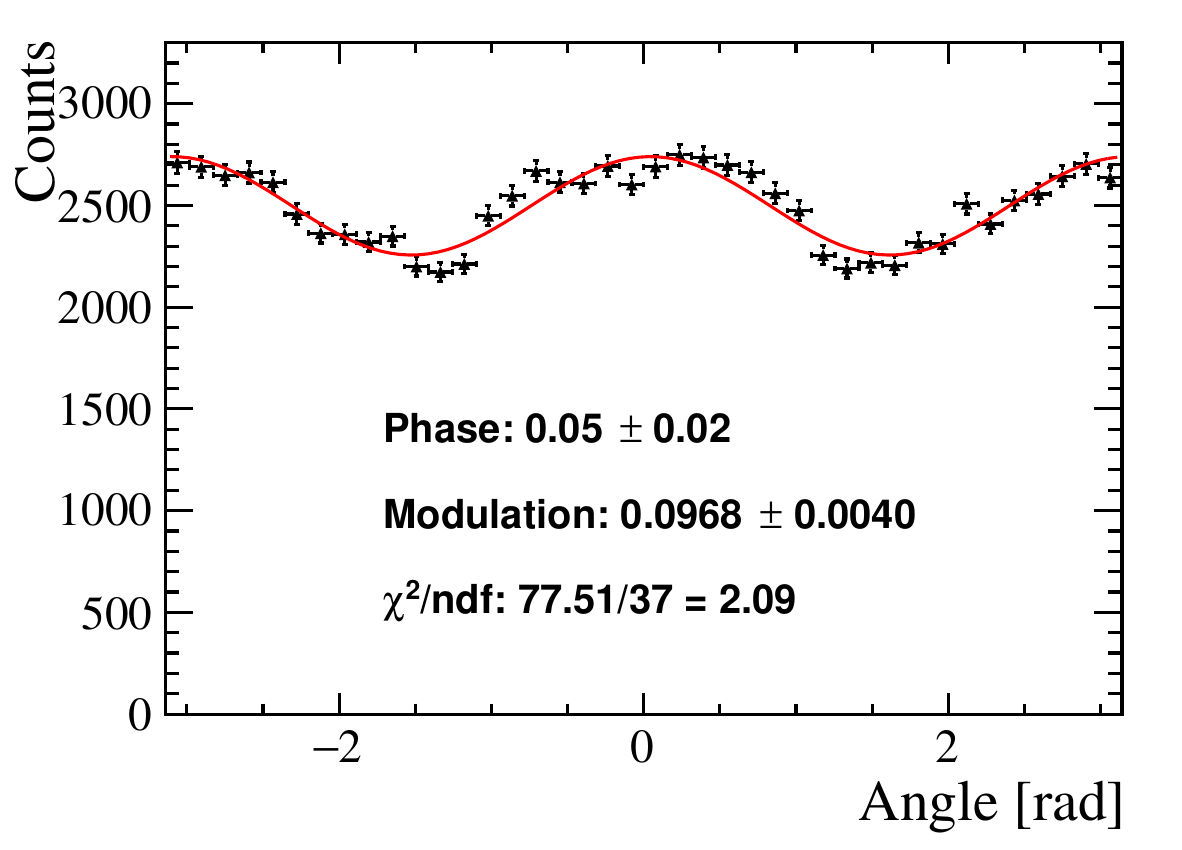}}
  \hfill
  \subfigure[Combining Polarization Data at 30° and 120°]{\includegraphics[width=0.22\textwidth]{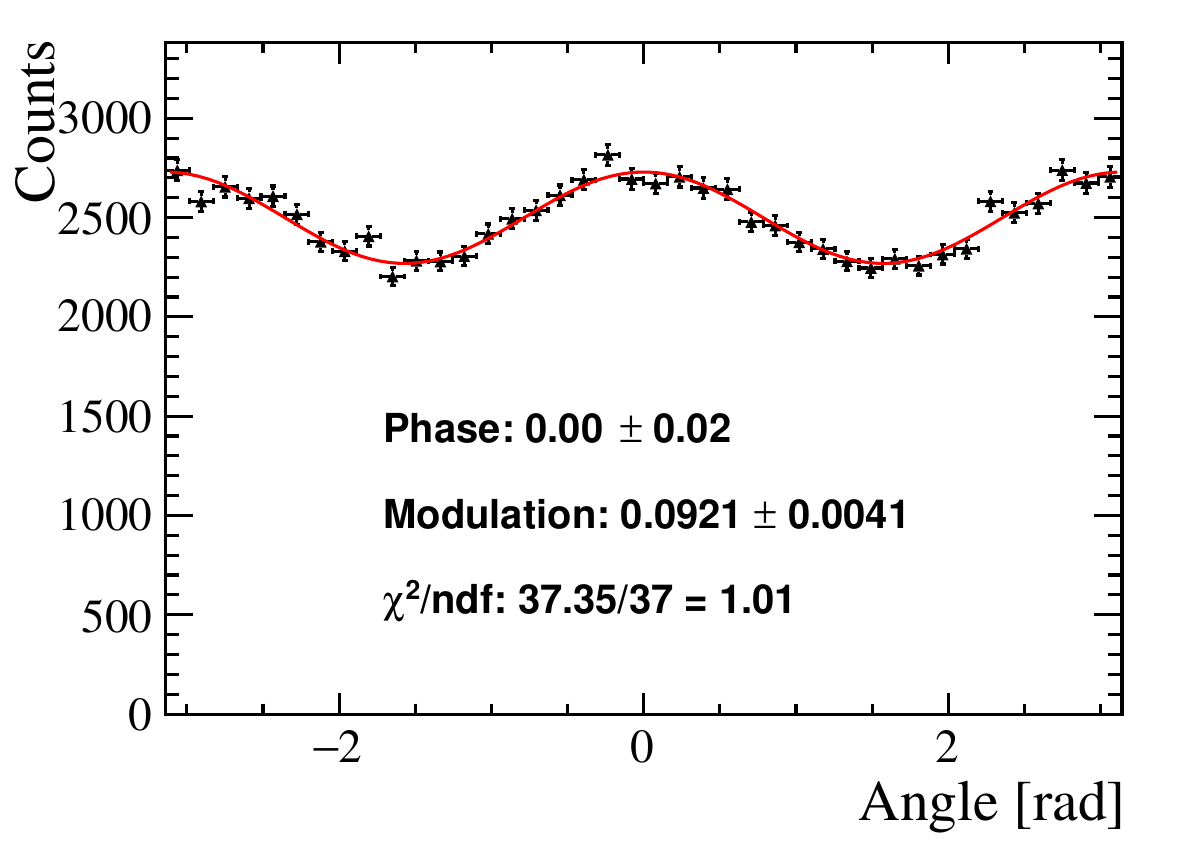}}
  \hfill
  \subfigure[Combining Polarization Data at 60° and 150°]{\includegraphics[width=0.22\textwidth]{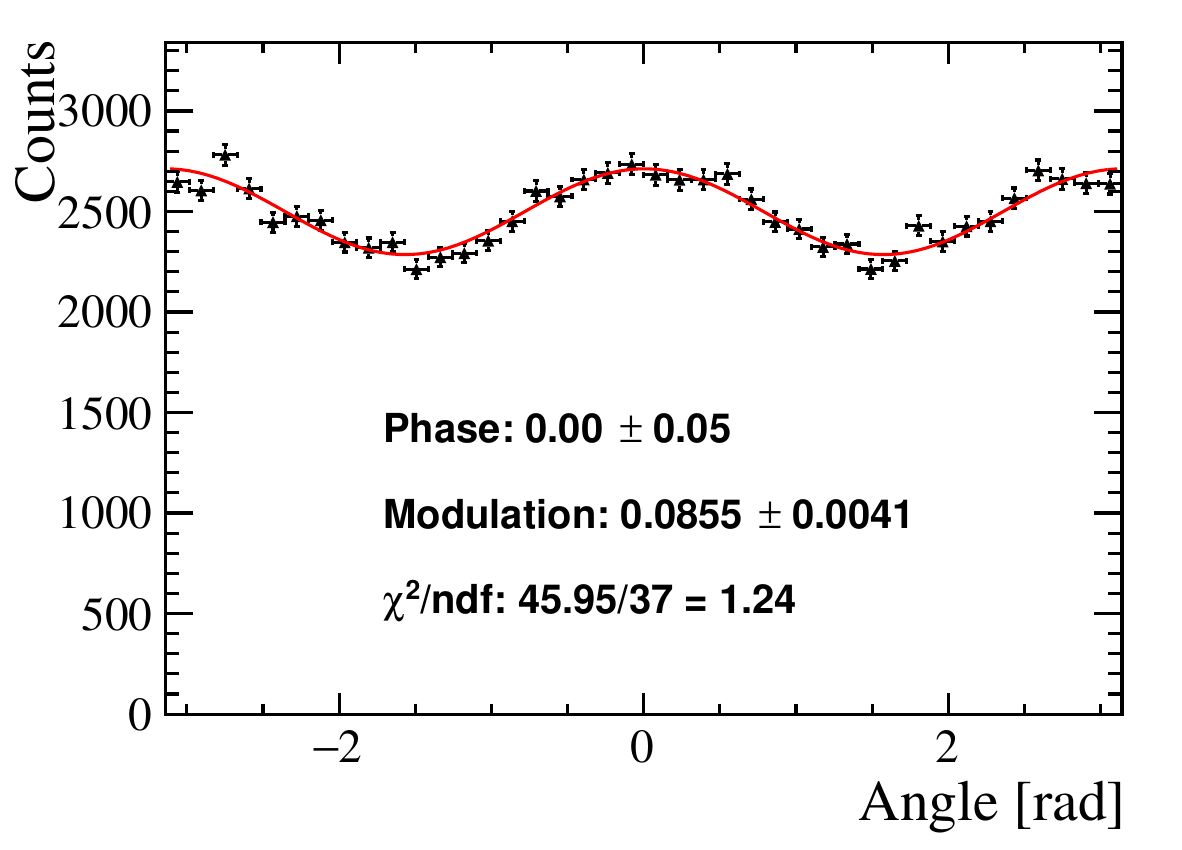}}
  \hfill
  \subfigure[Combining Polarization Data at 0°, 30°, 60°, 90°, 120°, and 150°]{\includegraphics[width=0.22\textwidth]{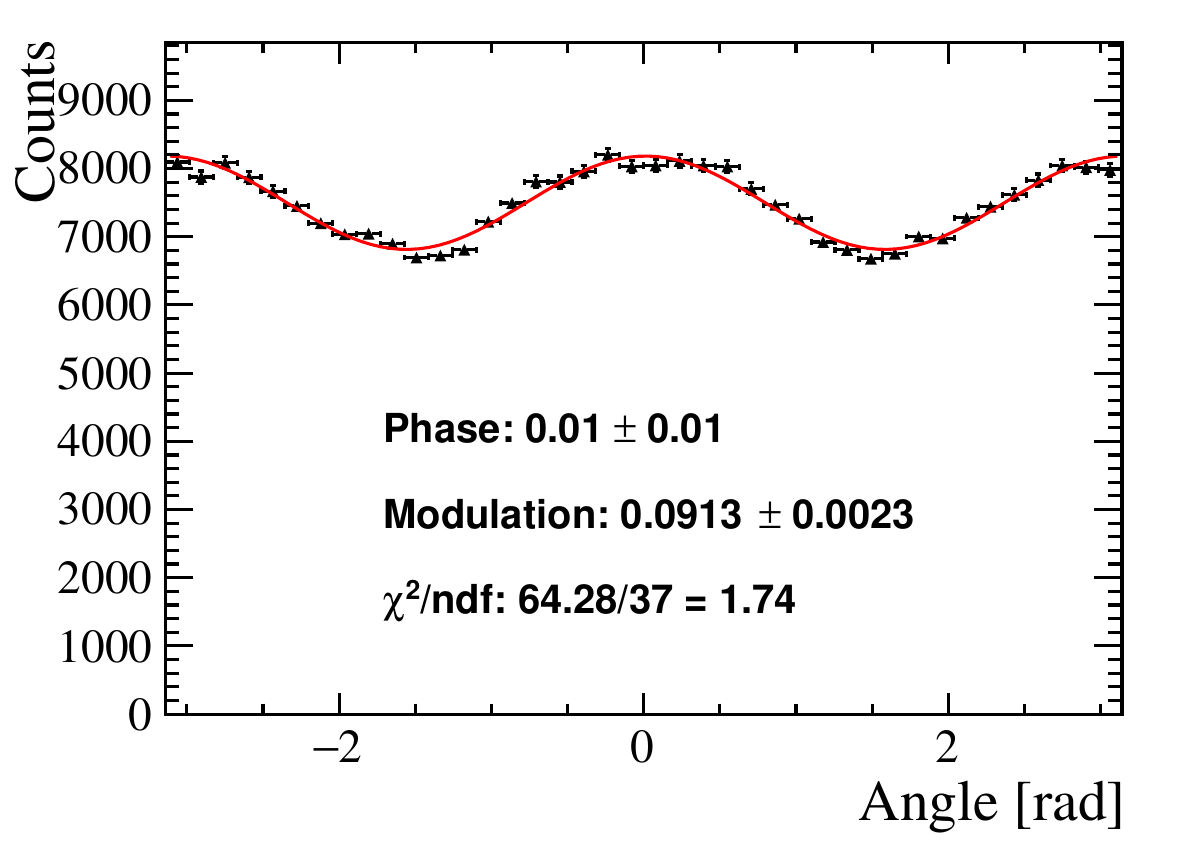}}
  \caption{Combination results of data at different polarization phases for 5.40\,keV.}
  \label{fig:Mod_Combine}
\end{figure}

According to equation \ref{eq:Source_Modulation}, we observe that $M_{\text{Source}}$ is modulated by $\cos^2$. Therefore, by equally combining two sets of data with a 90° difference in polarization phase, the modulation caused by the polarized source can be eliminated. As a result, when equally mixing six sets of data at 0°, 30°, 60°, 90°, 120°, and 150° polarization phases, the modulation curve of the angular distribution $M_{\text{Obs}} \propto M_{\text{Res}}$. The modulation distribution of the combined data is shown in Fig.\ref{fig:Mod_Combine}. The combined results indicate that the residual modulation distribution still follows equation \ref{eq:Source_Modulation}, and fitting different combined data sets within the error range shows that the residual modulation values obtained from different data sets are consistent, with the phase of the residual modulation being 0°.

\begin{figure*}[htbp]
  \centering
  \subfigure[$\Bar{\eta}$=1]{\includegraphics[width=0.29\textwidth]{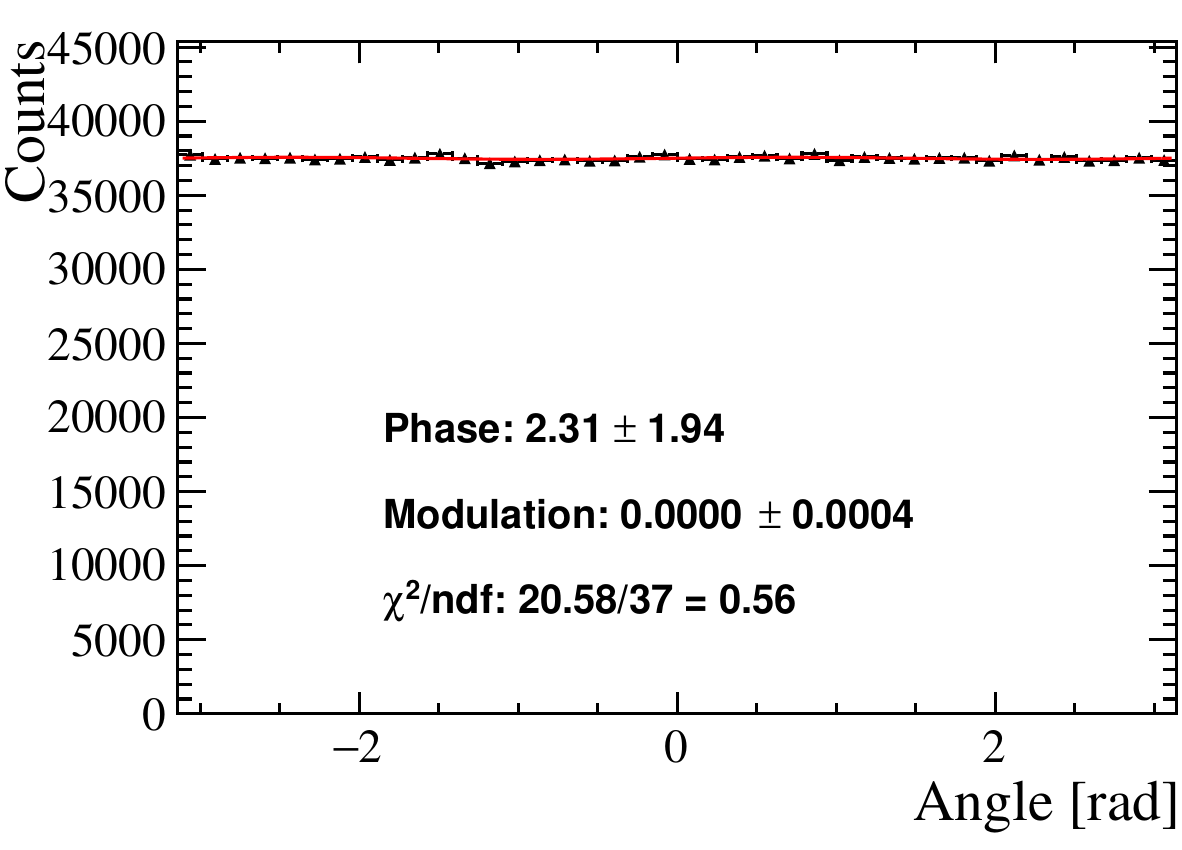}}
  \hfill
  \subfigure[$\Bar{\eta}$=0.981]{\includegraphics[width=0.29\textwidth]{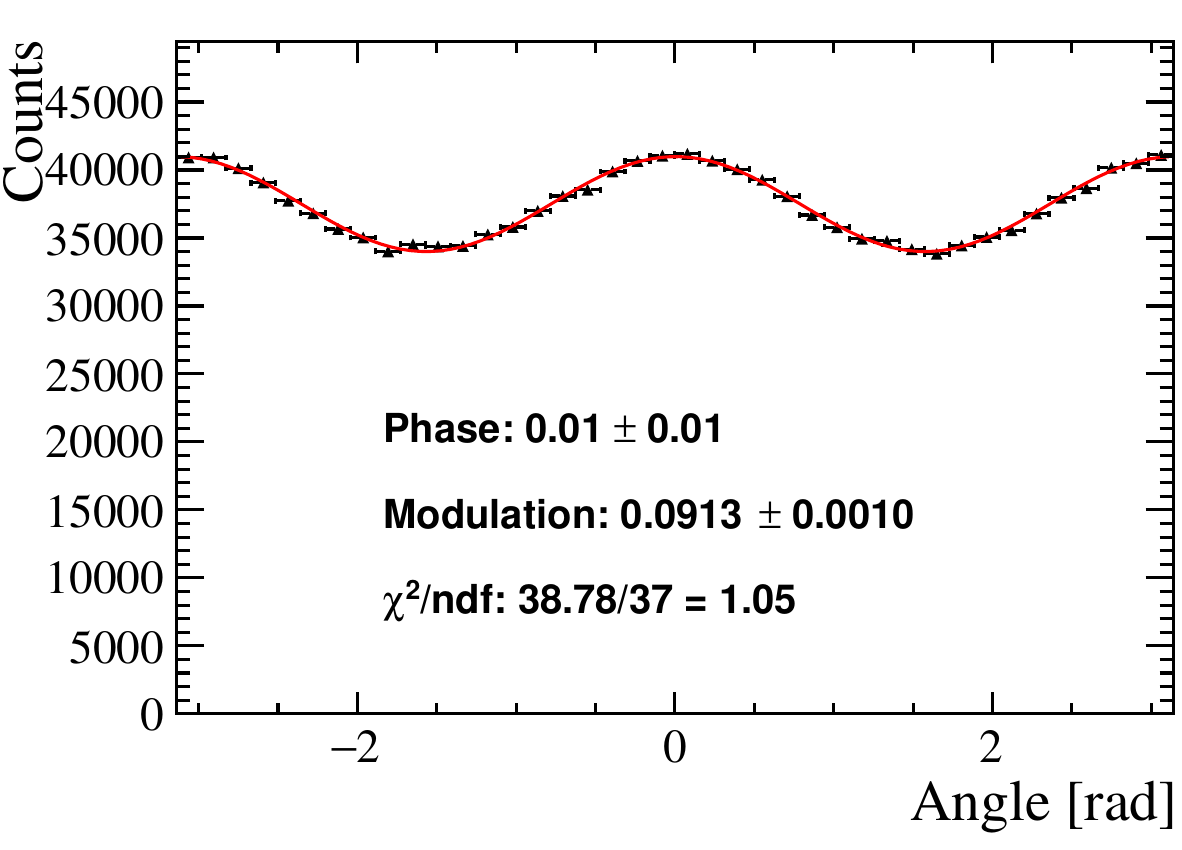}}
  \hfill
  \subfigure[$M_{\text{Distor}}$]{\includegraphics[width=0.29\textwidth]{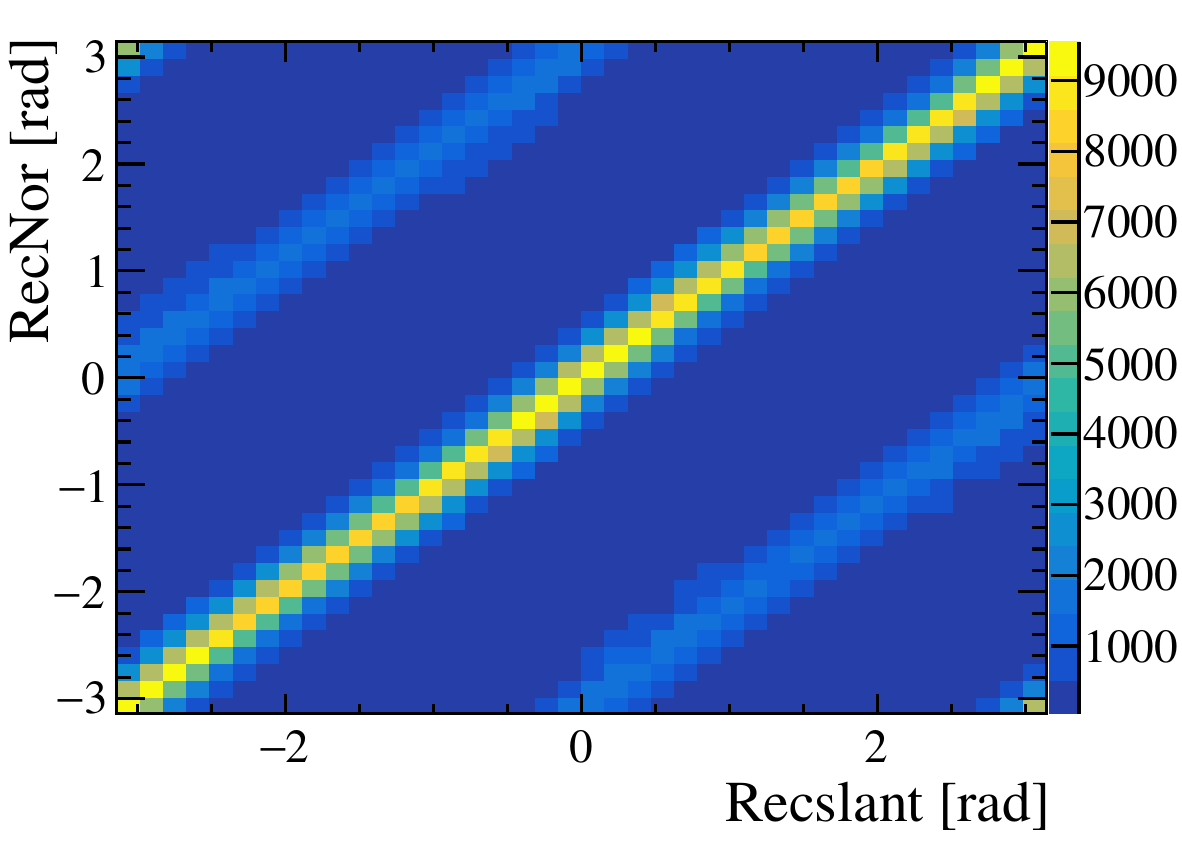}}
  \caption{(a) The distribution of $\text{Angle}_{\text{Truth}}$, with the red line representing the fitted modulation curve, exhibiting a modulation degree of 0. (b) The distribution of $\text{Angle}_{\text{Distor}}$, with the modulation degree adjusted by tuning $\Bar{\eta}$ to match the experimental data in Fig.\ref{fig:Mod_Combine}(d). (c) The response matrix, with the ordinate representing the reconstructed angles of events at $\Bar{\eta}$=1, and the abscissa representing the reconstructed angles of events at $\Bar{\eta}$=0.981.}
  \label{fig:MC}
\end{figure*}

We obtained the residual modulation amplitude at the 5.40\,keV energy point by fitting the residual modulation curve from Fig.\ref{fig:Mod_Combine}(d). Next, we consider using simulations to reproduce the same residual modulation distribution and obtain a response matrix for correcting the residual modulation in the experimental data. We utilized the star-XP software framework \cite{YI2024101626} specifically designed for the LPD detector. The simulation framework meticulously simulates the interaction processes between photoelectrons and the detector, as well as the digitization process. The simulated data output by the framework shows good agreement with the experimental data. Our operational procedure followed the steps below:
\begin{enumerate}
\item In the simulation framework, we simulated the tracks of 1.5 million unpolarized 5.40\,keV X-ray photons and maintained the parameters set in the simulator consistent with the actual operating parameters of the detector.

\item Initially, we set $\Bar{\eta}$=1, representing the state of the detector without electric field distortion, and simulated the two-dimensional image of the photoelectron tracks after digitization. We reconstructed each of the 1,500,000 tracks without distortion to obtain the reconstructed angle information, $\text{Angle}_{\text{Truth}}$. It is important to note that this is not the true value of the emission angle of the photoelectrons provided by the simulation, but rather the angle value obtained from the reconstruction. The distribution of $\text{Angle}_{\text{Truth}}$ is shown in Fig.\ref{fig:MC}(a).

\item In the simulation, we adjusted the value of $\Bar{\eta}$ to deviate from 1, representing the occurrence of electric field distortion in the detector. We used the photoelectron track simulations from Step 1 after digitization, and due to different scaling in the X and Y directions, the reconstructed angle distribution, $\text{Angle}_{\text{Distor}}$, exhibited a non-zero residual modulation. When $\Bar{\eta}$<1, the phase of the residual modulation is 0°, consistent with the experimental data. By adjusting the value of $\Bar{\eta}$, we were able to align the modulation amplitude of the $\text{Angle}_{\text{Distor}}$ distribution with Fig.\ref{fig:Mod_Combine}(d), as shown in Fig.\ref{fig:MC}(b). For 5.40\,keV, the value of $\Bar{\eta}$ was determined to be 0.981.

\item Combining the $\text{Angle}_{\text{Truth}}$ and $\text{Angle}_{\text{Distor}}$ reconstructed step by step in the second and third steps, we can obtain the response matrix $M_{\text{Distor}}$, which arises due to the adjustment of the parameter $\Bar{\eta}$. The physical interpretation of $M_{\text{Distor}}$ is as follows: if we denote $N^{\text{Truth}}_{i}$ as the number of instances for which the reconstructed angle falls within the i-th bin when $\Bar{\eta}$=1, then the number of instances for the same events after adjusting $\Bar{\eta}$ and reconstructed within the j-th bin is given by equation \ref{eq:ResponseMatrix}. In other words, the element (m,n) of $M_{\text{Distor}}$ is proportional to the probability value P(n|m): the probability that an event originally reconstructed in the m-th bin is reconstructed in the n-th bin due to the adjustment of $\Bar{\eta}$.
\end{enumerate}
\begin{equation} 
\label{eq:ResponseMatrix} 
N_{j}^{{\text{Distor}}} = \Sigma_{i} N^{\text{Truth}}_{i}\cdot M_{ij} . 
\end{equation}

After obtaining the parameter $\Bar{\eta}=1$ that describes the system effect and its corresponding response matrix $M_{\text{Distor}}$, we employed a Bayesian iterative process algorithm to decouple the modulation distribution generated by the polarization source, which does not have electric field distortion, from the overall system effects. Many software packages offer Bayesian algorithm capabilities, and in our study, we utilized the RooUnfoldBayes packages integrated within RooUnfold \cite{BogAleLev23}. The RooUnfoldBayes package is capable of iteratively restoring the input angular modulation distribution to its undistorted state, based on the provided response matrix $M_{\text{Distor}}$, and automatically calculates the errors for each bin following the Bayesian iteration process.

\begin{figure*}[htbp]
  \centering
  \subfigure[]{\includegraphics[width=0.29\textwidth]{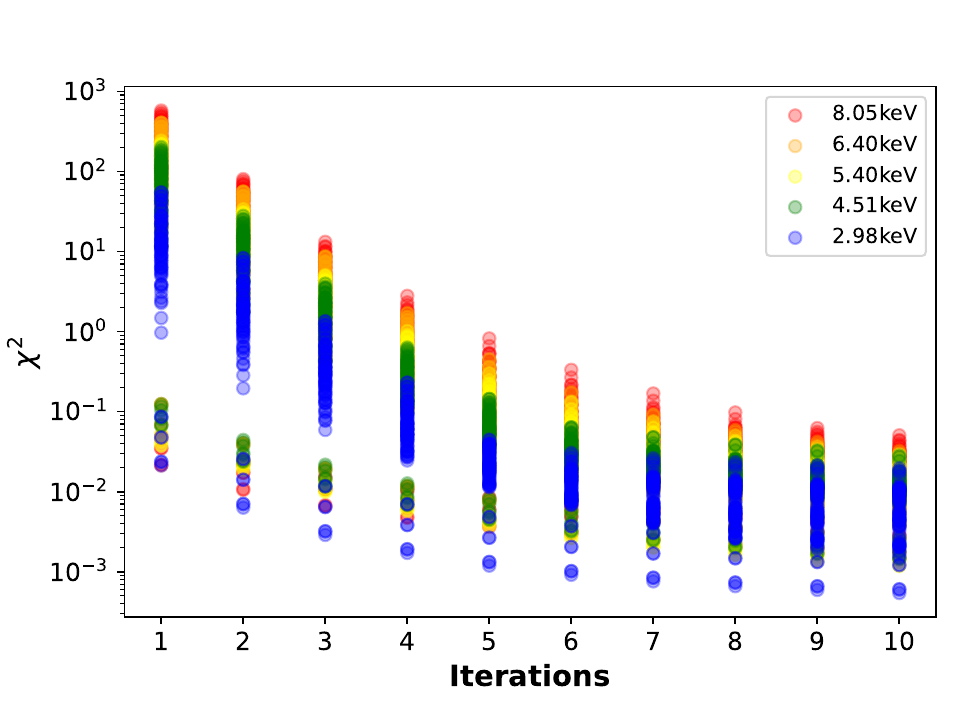}}
  \hfill
  \subfigure[]{\includegraphics[width=0.29\textwidth]{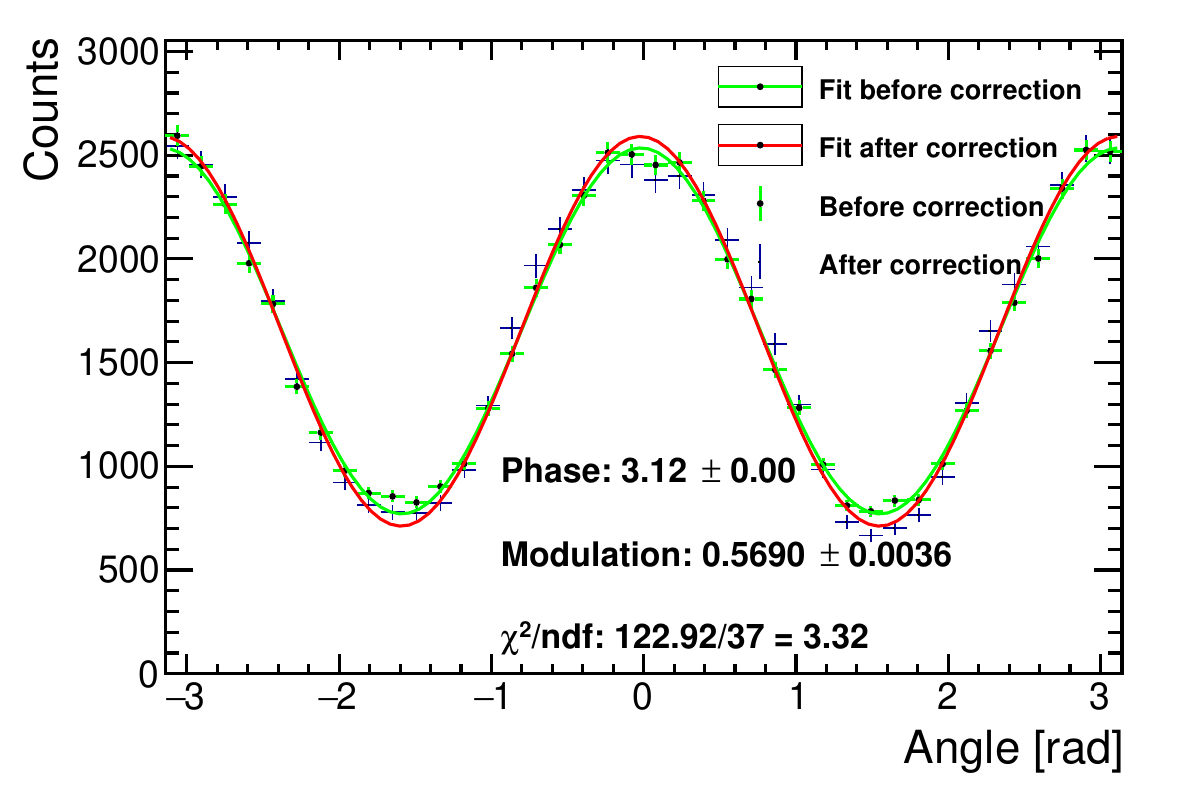}}
  \hfill
  \subfigure[]{\includegraphics[width=0.29\textwidth]{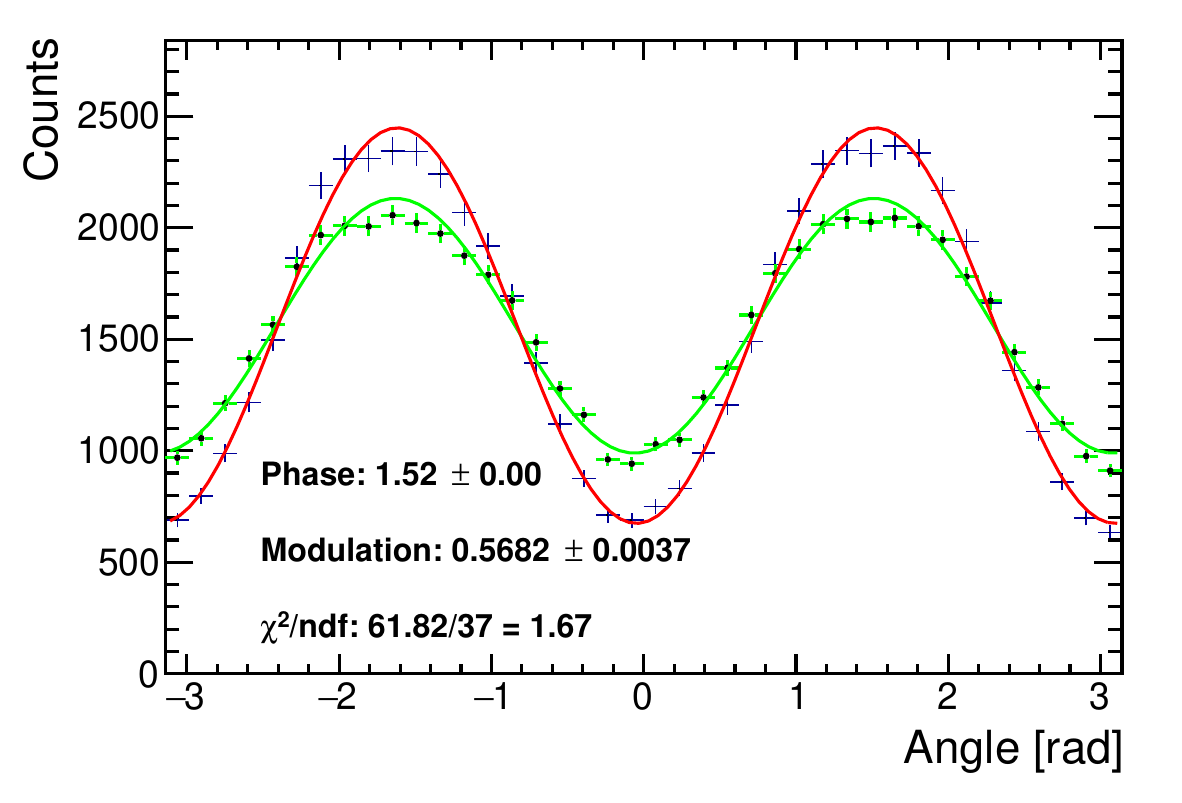}}
  \caption{(a) The $\chi^2$ variation of the distributions after 1-10 Bayesian iterations for different polarization degrees and polarization phases at different energy points. After 10 iterations, the $\chi^2$ values are all less than 0.5, indicating that the Bayesian iterations have essentially converged. (b) and (c) respectively show the completely polarized data at 5.40 keV and 0°, 90° phases. The green curve represents the original measured reconstructed angular distribution, while the blue curve represents the distribution after 10 Bayesian iterations. A comparison between (b) and (c) reveals that the Bayesian iterations have corrected the modulation levels at the two phases to the same level.}
  \label{fig:Bayes}
\end{figure*}

The use of Bayesian methods involves the selection of prior distributions and the adjustment of the number of iterations. Firstly, due to the periodicity of the modulation curves with a period of $\pi$, monotonically increasing or decreasing distributions are not appropriate. Therefore, for simplicity, we set the prior distributions to uniform distributions. Secondly, concerning the number of iterations, we determine the convergence of the iteration process by comparing the $\chi^2$ values of the distributions $M(\phi)_{n+1}$ and $M(\phi)_n$ after the $n+1$ th and $n$ th iterations. We found that when the number of iterations is set to 10, the $\chi^2$ values for different phases, polarizations, and energies are all less than 0.5, which indicates that the iterative process has essentially reached convergence. Additionally, after 10 iterations, the introduced iteration errors in each bin are relatively small. Therefore, we set the number of iterations to 10. Fig.\ref{fig:Bayes}(a) illustrates the variation of the $\chi^2$ values corresponding to different numbers of iterations, while Fig.s \ref{fig:Bayes}(b) and \ref{fig:Bayes}(c) present the corrected results for the 5.40\,keV 99.9\% polarized data at 0° and 90° phases, respectively.

\subsubsection{Result}

\begin{figure*}[htbp]
  \centering
  \subfigure[unpolarized]{\includegraphics[width=0.22\textwidth]{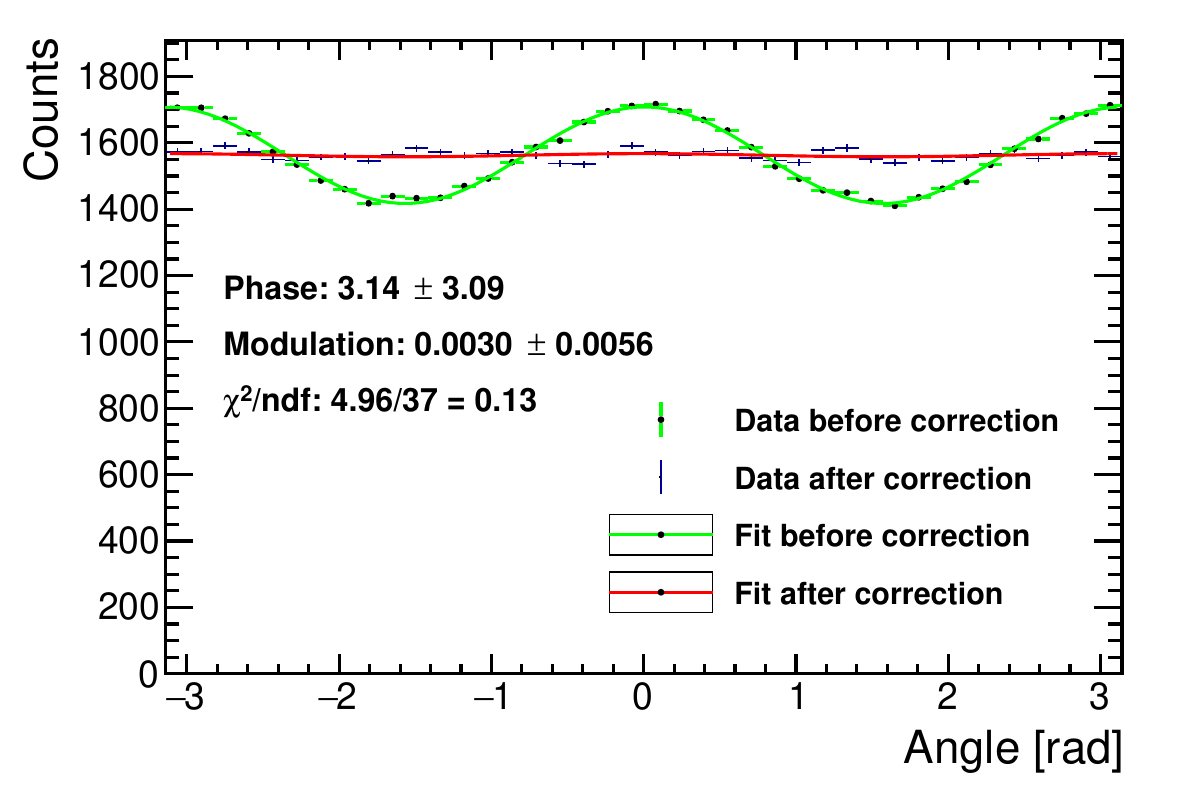}}
  \hfill
  \subfigure[PD = 0.1]{\includegraphics[width=0.22\textwidth]{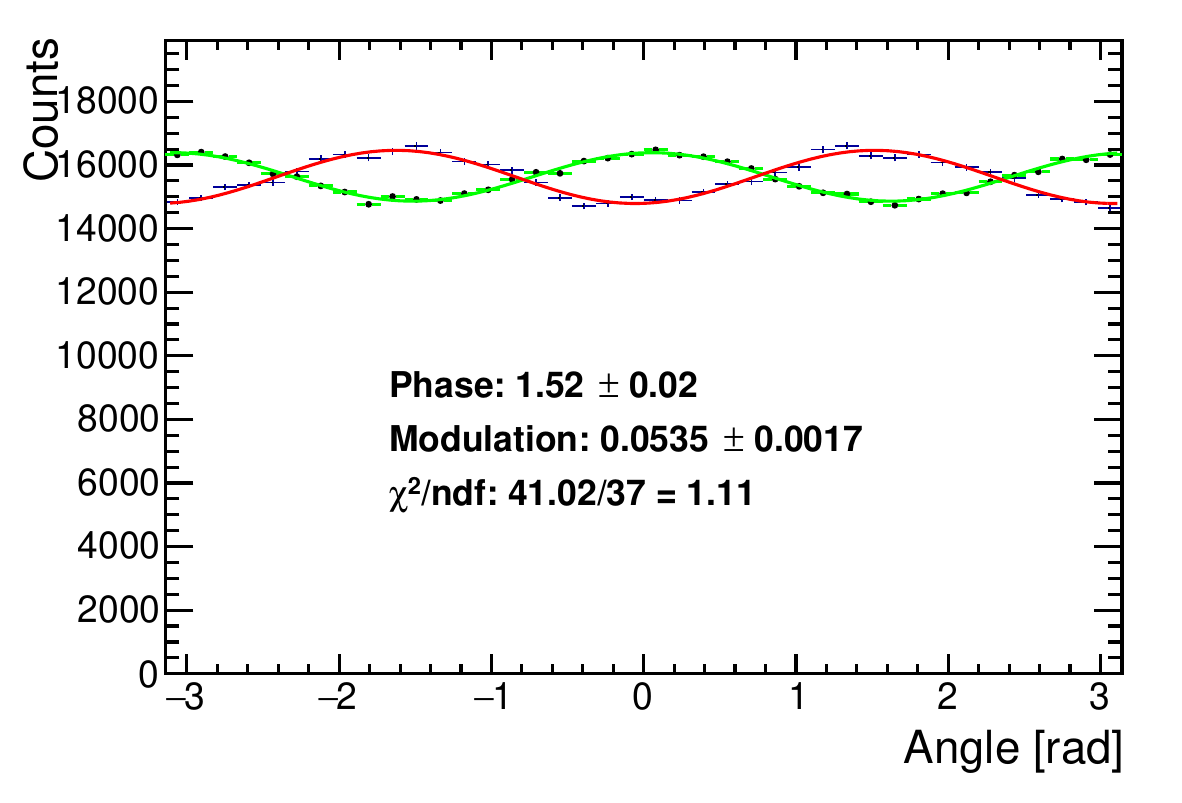}}
  \hfill
  \subfigure[PD = 0.2]{\includegraphics[width=0.22\textwidth]{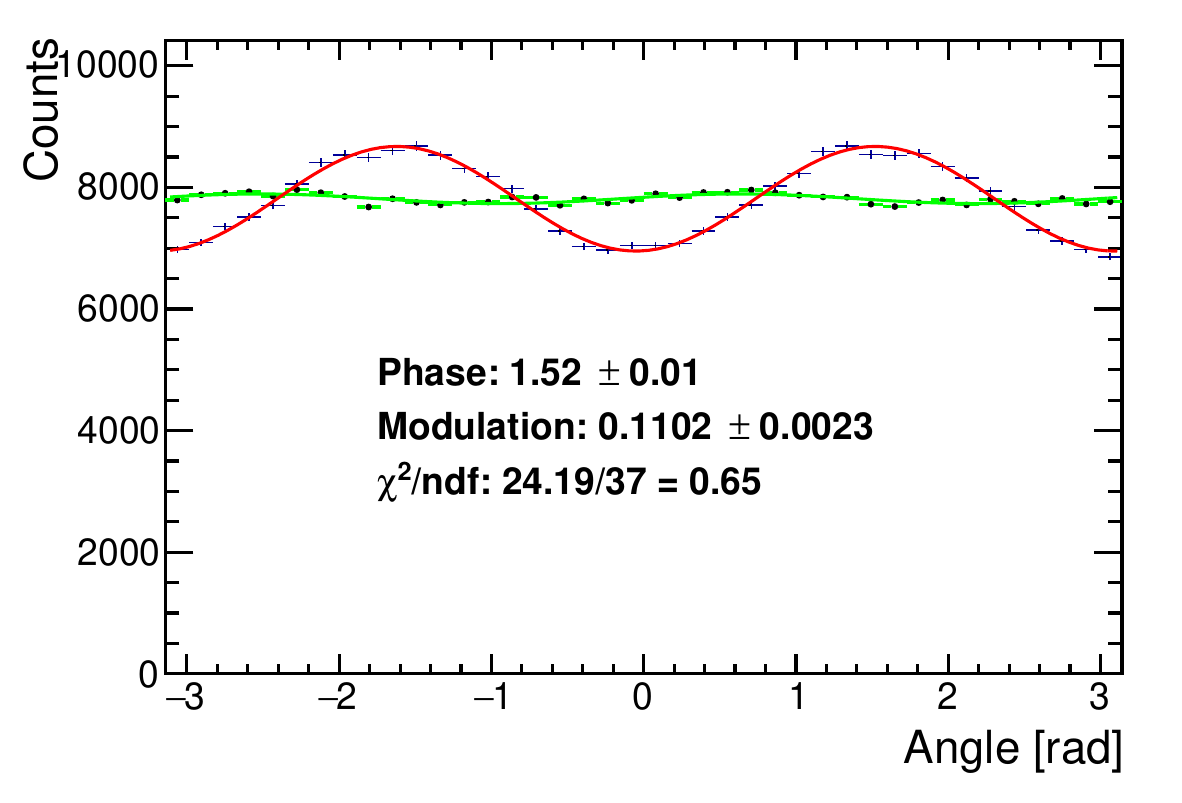}}
  \subfigure[PD = 0.3]{\includegraphics[width=0.22\textwidth]{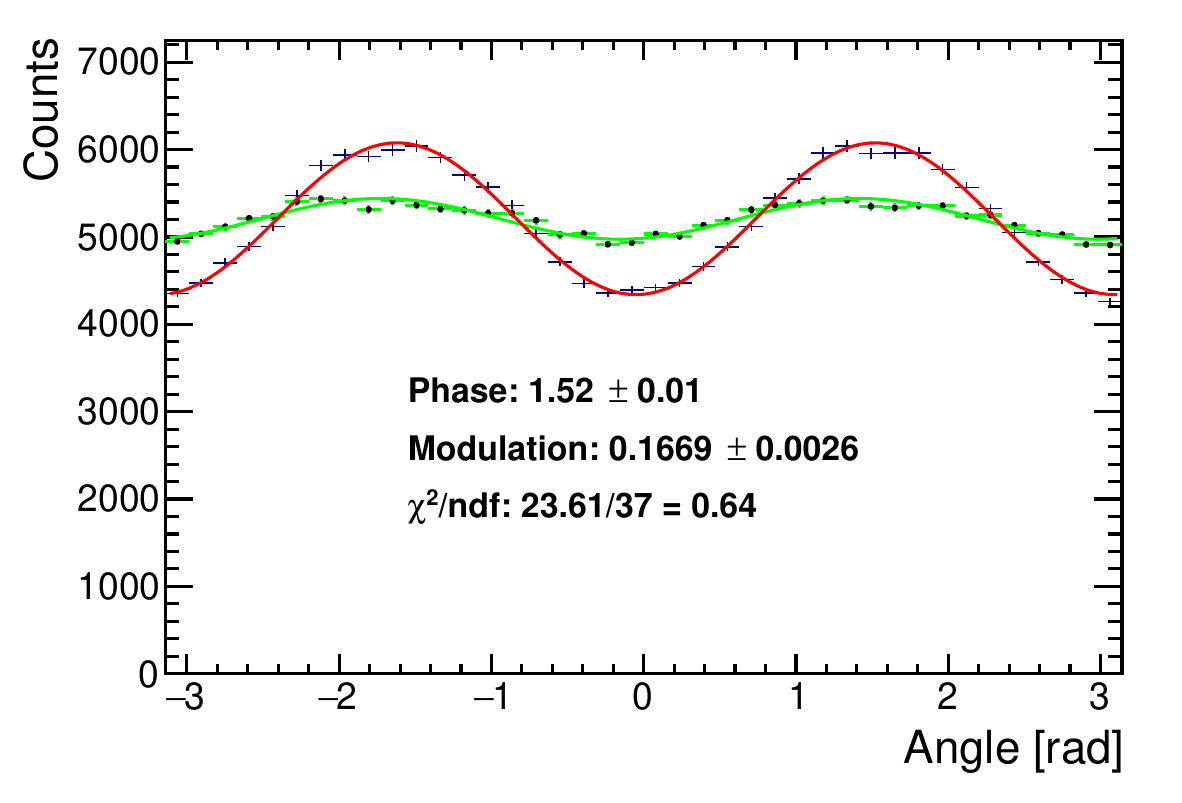}}
  \subfigure[PD = 0.4]{\includegraphics[width=0.22\textwidth]{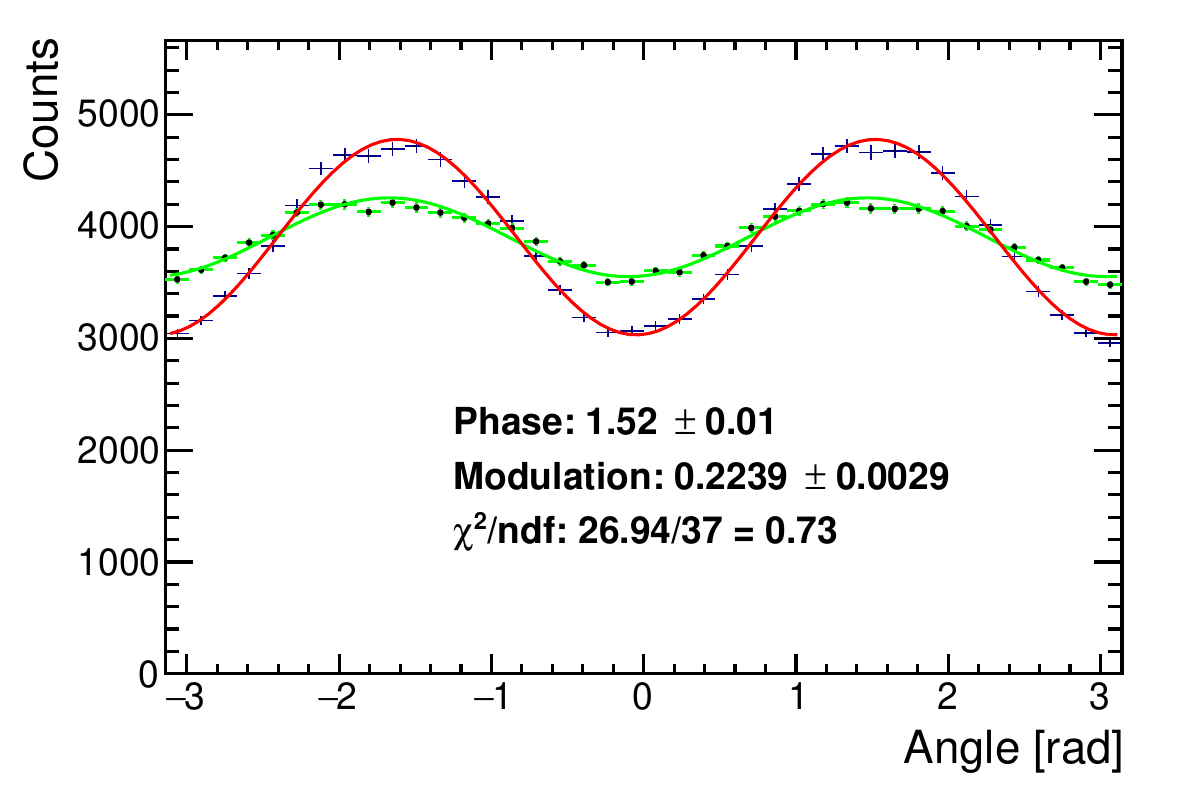}}
  \hfill
  \subfigure[PD = 0.5]{\includegraphics[width=0.22\textwidth]{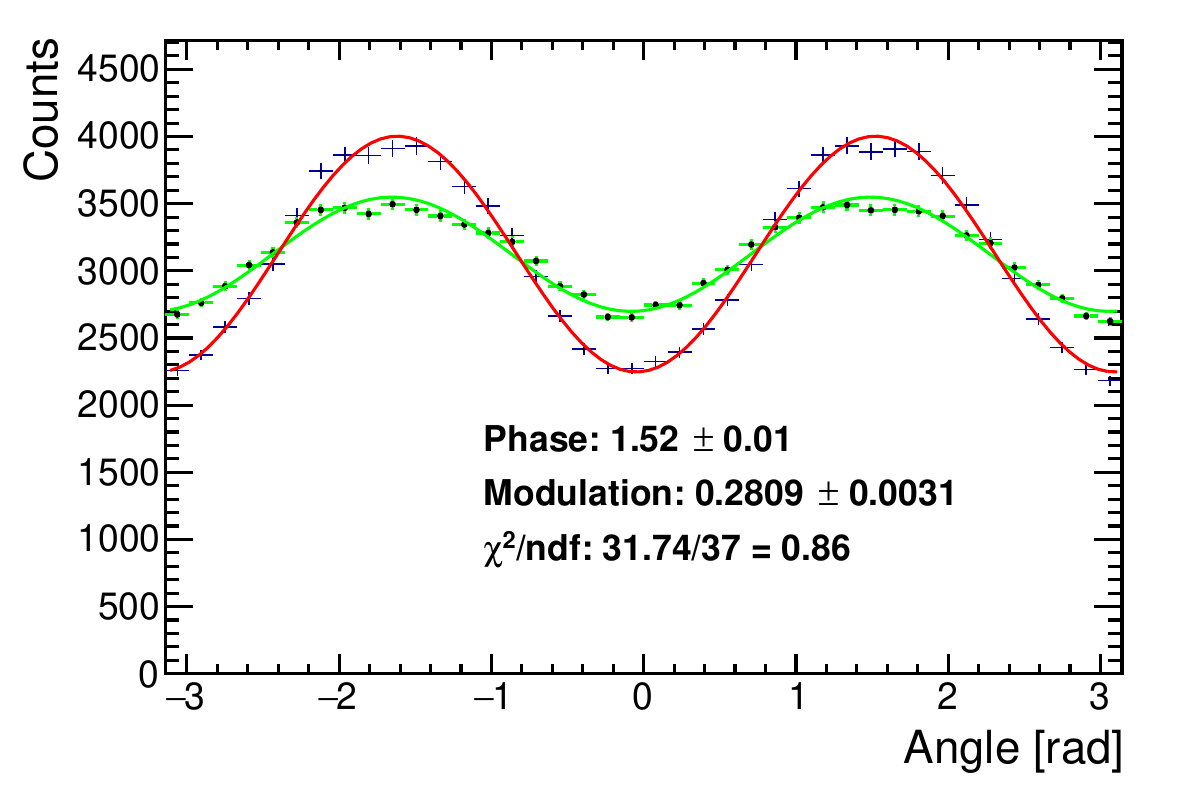}}
  \hfill
  \subfigure[PD = 0.6]{\includegraphics[width=0.22\textwidth]{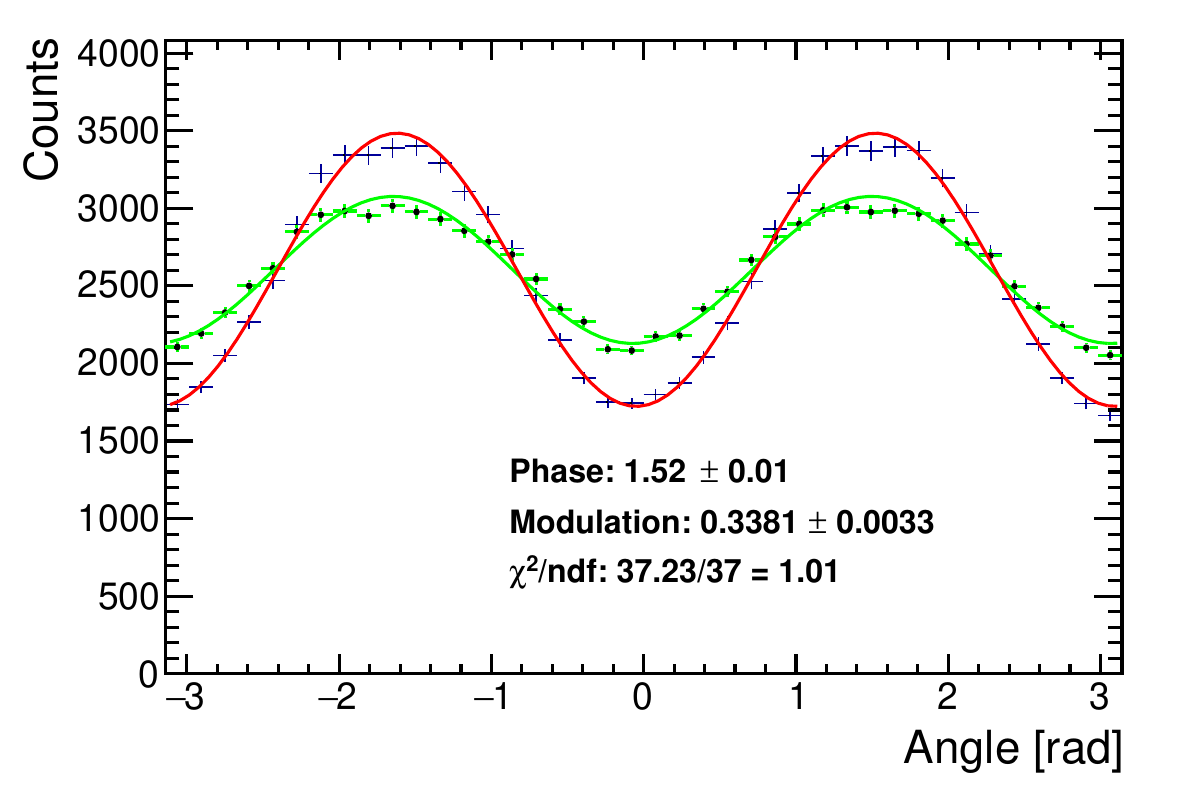}}
  \subfigure[PD = 0.7]{\includegraphics[width=0.22\textwidth]{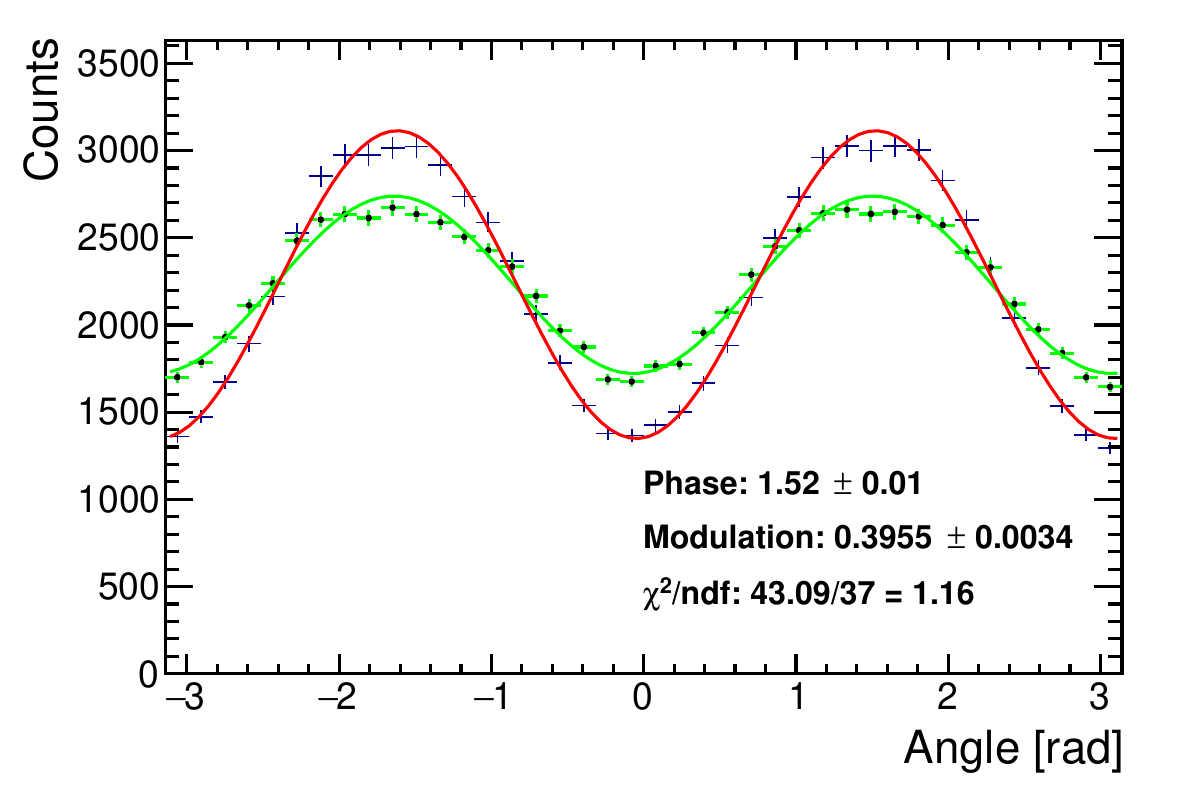}}
  \subfigure[PD = 0.8]{\includegraphics[width=0.22\textwidth]{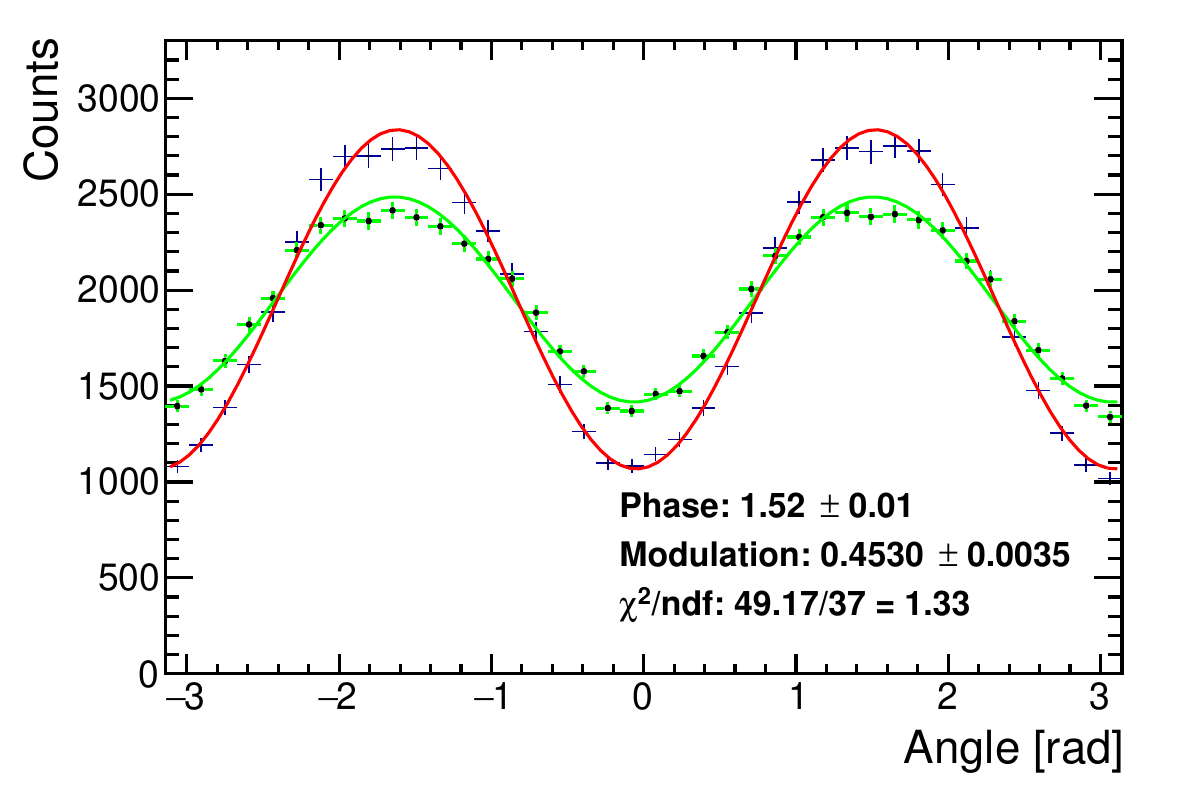}}
  \hfill
  \subfigure[PD = 0.9]{\includegraphics[width=0.22\textwidth]{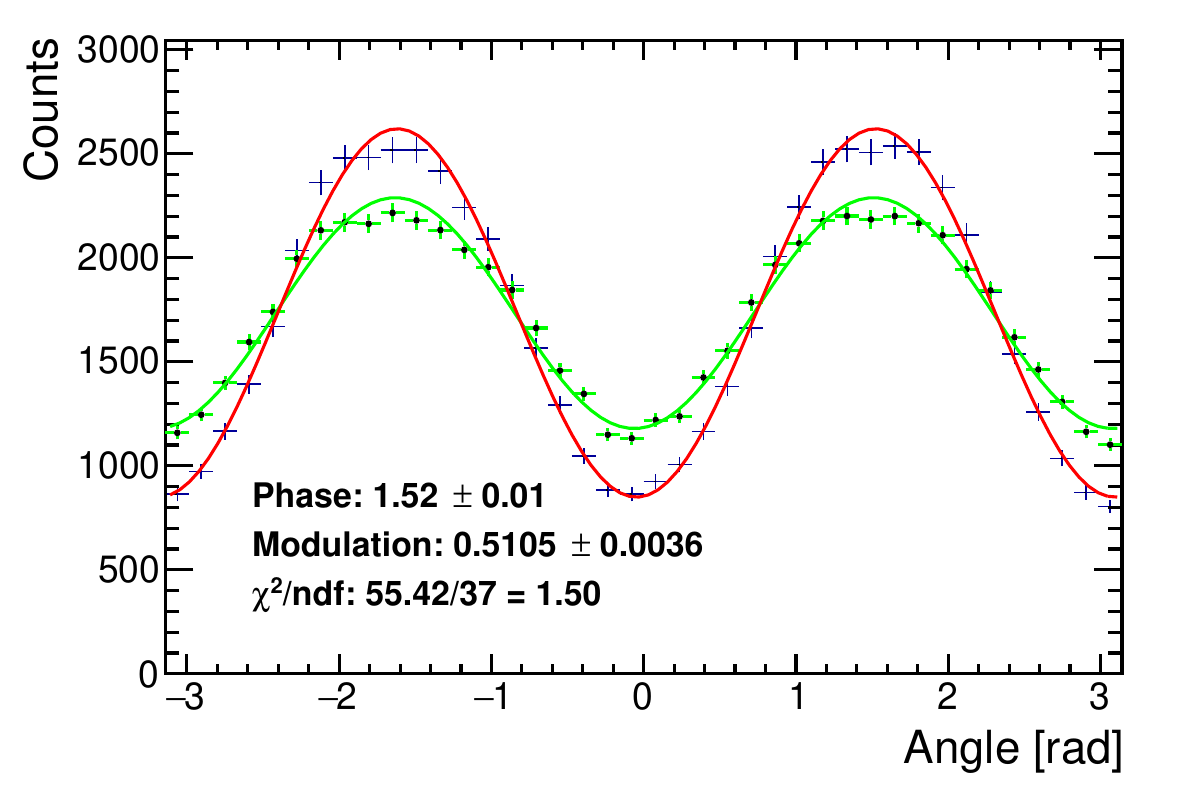}}
  \hfill
  \subfigure[PD = 1.0]{\includegraphics[width=0.22\textwidth]{RooUnfoldExampleEXP_unpolarized_90deg_1.000000.pdf}}
  \hfill
  \subfigure[linear fit]{\includegraphics[width=0.22\textwidth]{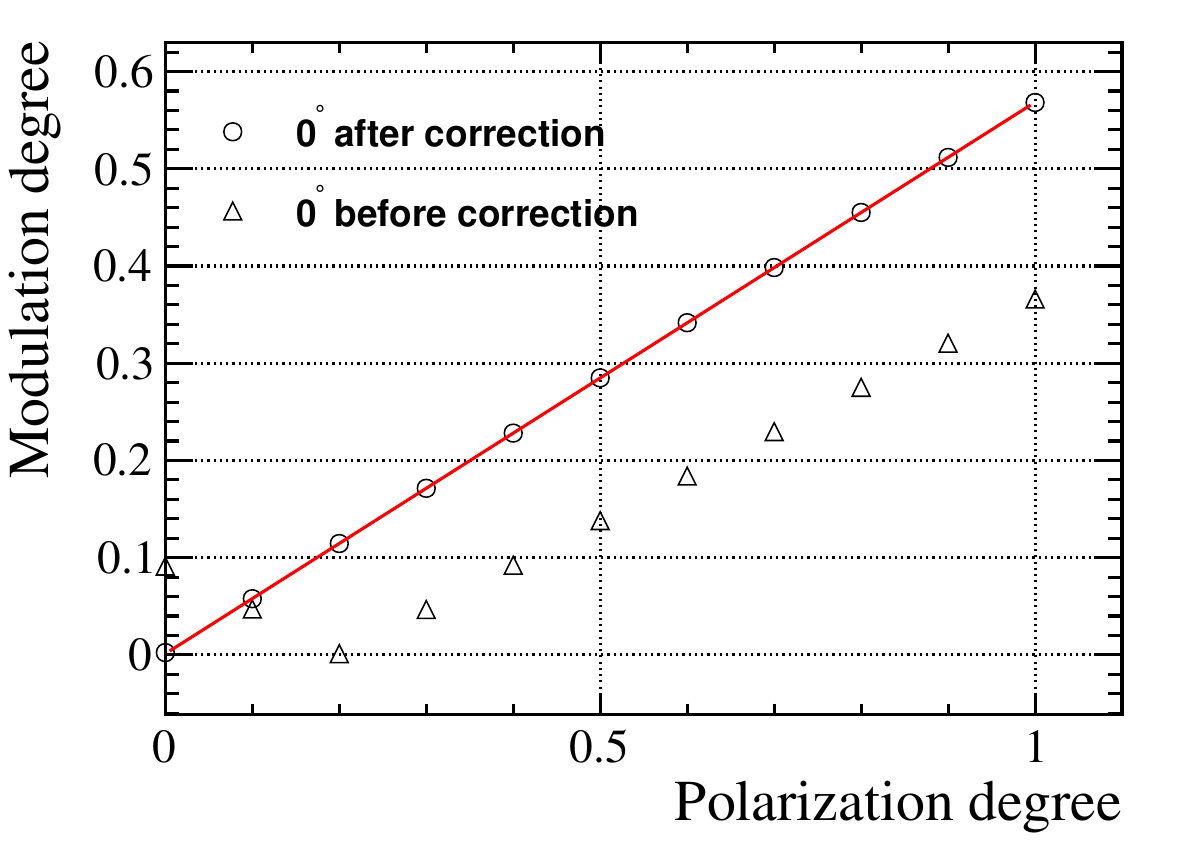}}
  \caption{(a)-(k) Comparison of modulation degree before and after correction for experimental data at 5.40\,keV and 90° polarization phase for different polarization degrees where PD represents the polarization degree. The legend is the same as in Fig.\ref{fig:Bayes}. The unpolarized data were obtained by mixing datasets with a 90° phase difference using the method described at Fig.\ref{fig:Mod_Combine}, and the datasets for different polarization degrees were obtained by proportionally mixing unpolarized and fully polarized data. (l) The triangle represents the relationship between the modulation degree reconstructed from experimental data before correction and the polarization degree, while the circle represents the relationship after correction, and the red line represents the fitted curve of the corrected data points.}
  \label{fig:RecFix_90deg}
\end{figure*}

\begin{figure*}[htbp]
  \centering
  \subfigure[2.98\,keV]{\includegraphics[width=0.29\textwidth]{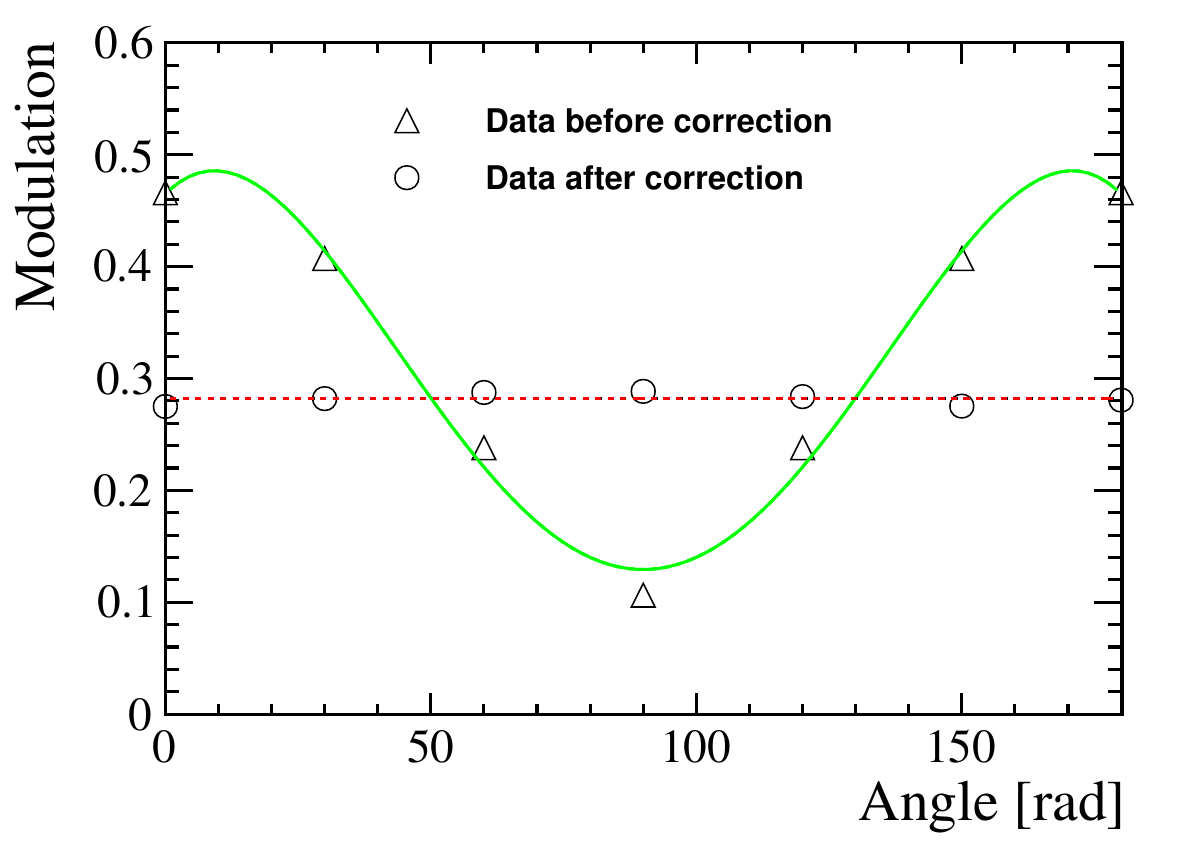}}
  \hfill
  \subfigure[]{\includegraphics[width=0.29\textwidth]{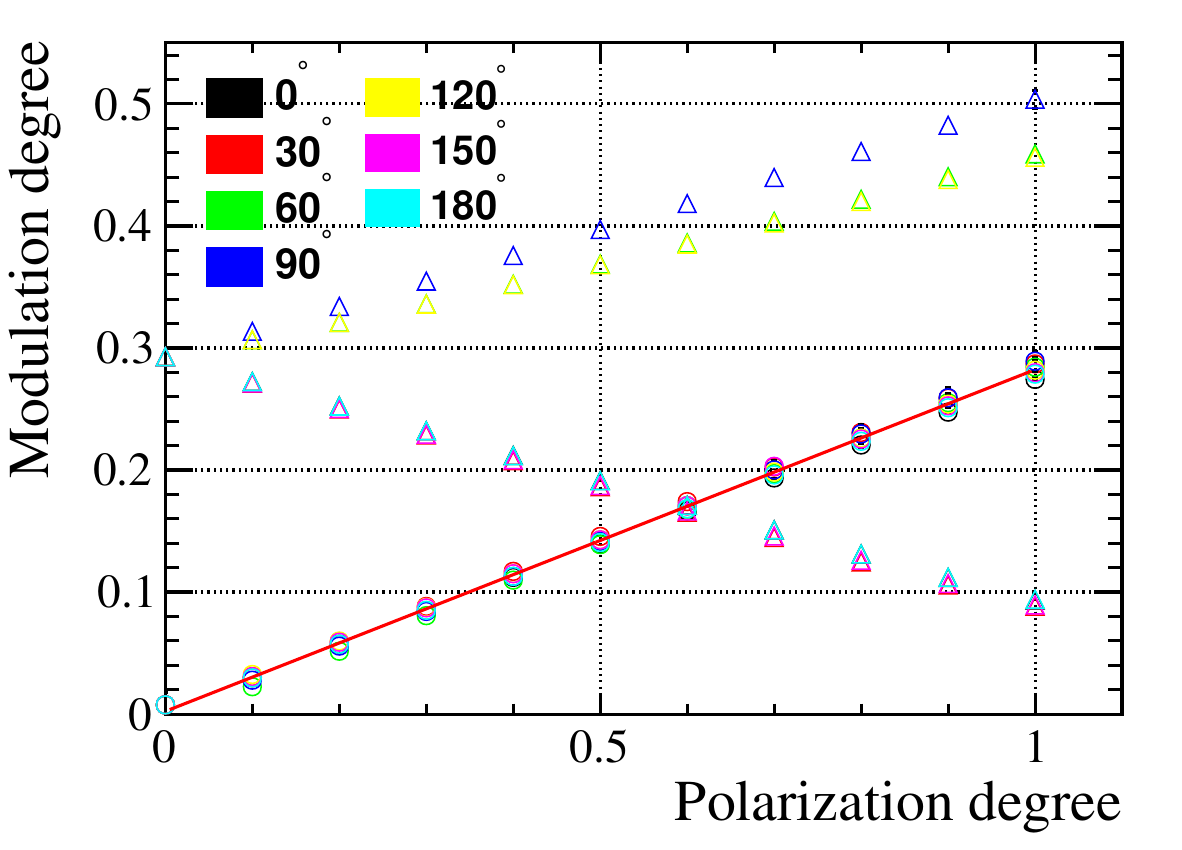}}
  \hfill
  \subfigure[]{\includegraphics[width=0.29\textwidth]{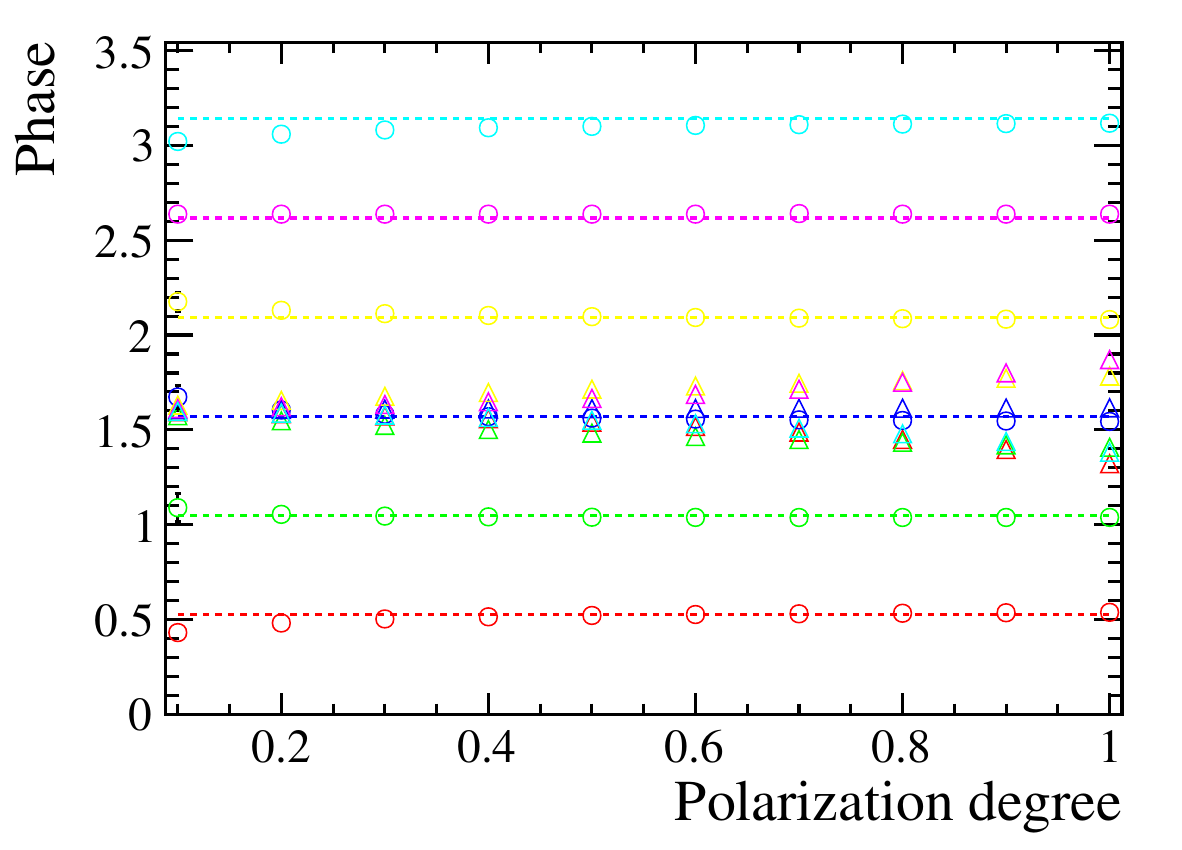}}
  \hfill
  \subfigure[4.51\,keV]{\includegraphics[width=0.29\textwidth]{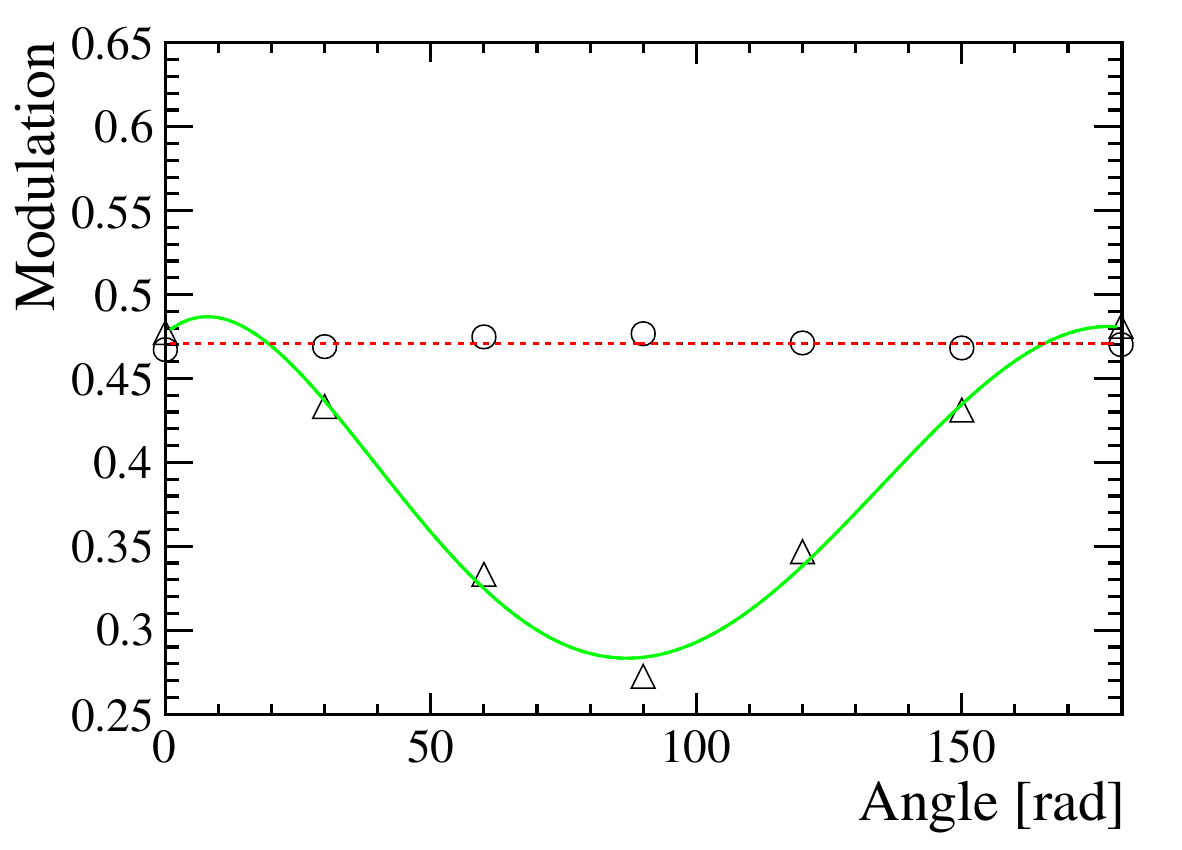}}
  \hfill
  \subfigure[]{\includegraphics[width=0.29\textwidth]{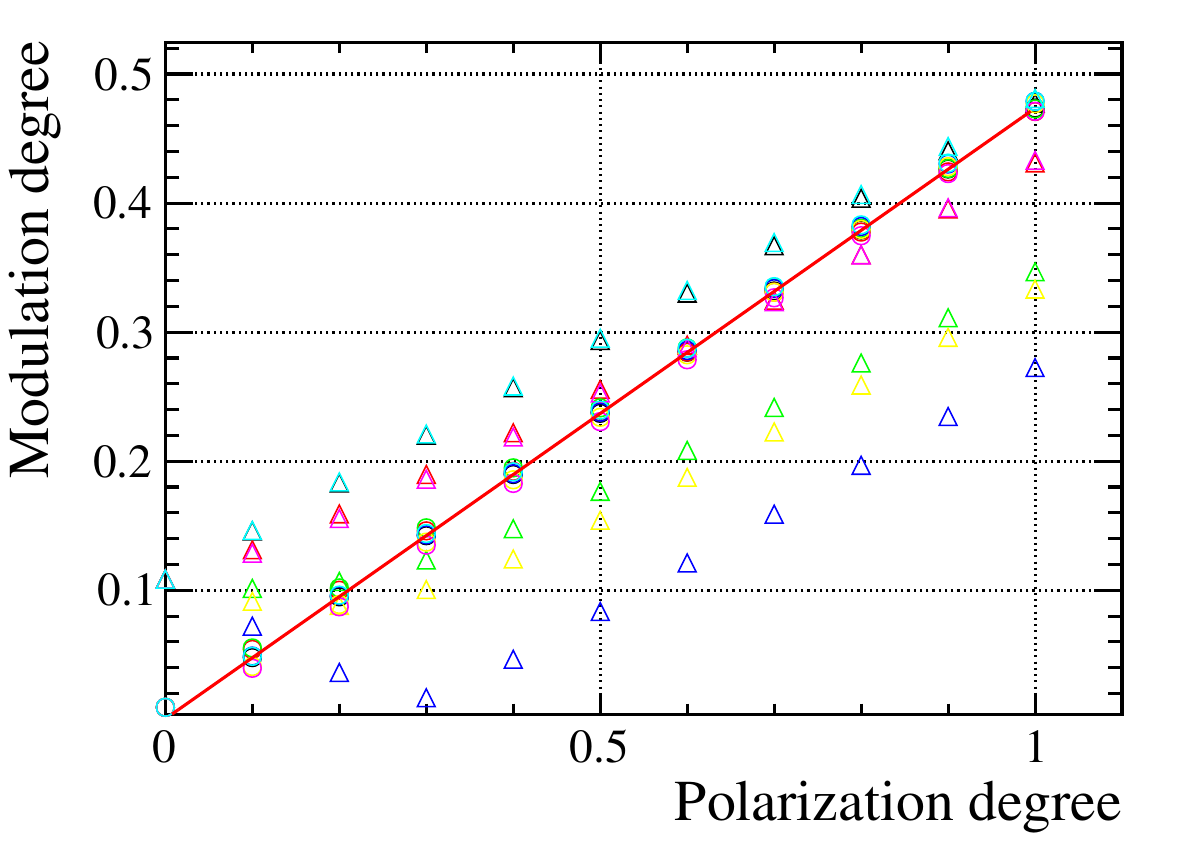}}
  \hfill
  \subfigure[]{\includegraphics[width=0.29\textwidth]{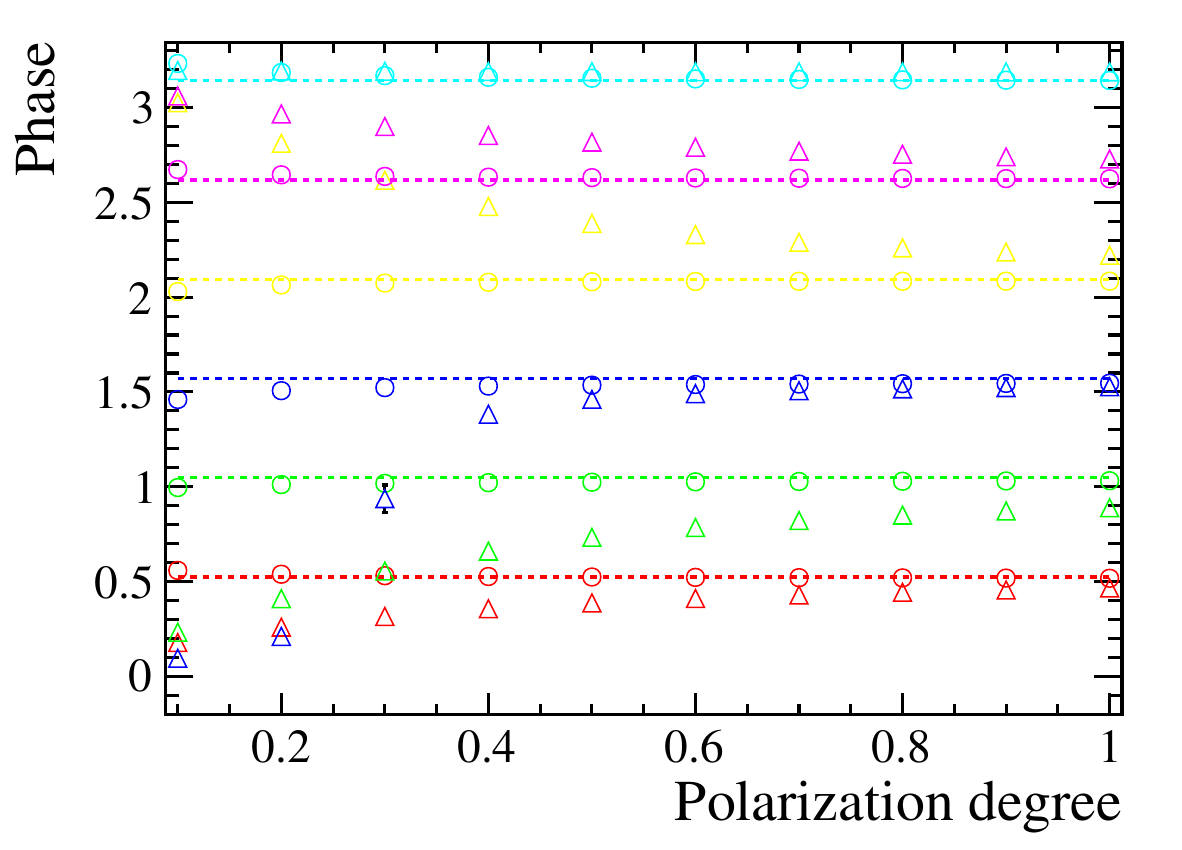}}
  \hfill
  \subfigure[5.40\,keV]{\includegraphics[width=0.29\textwidth]{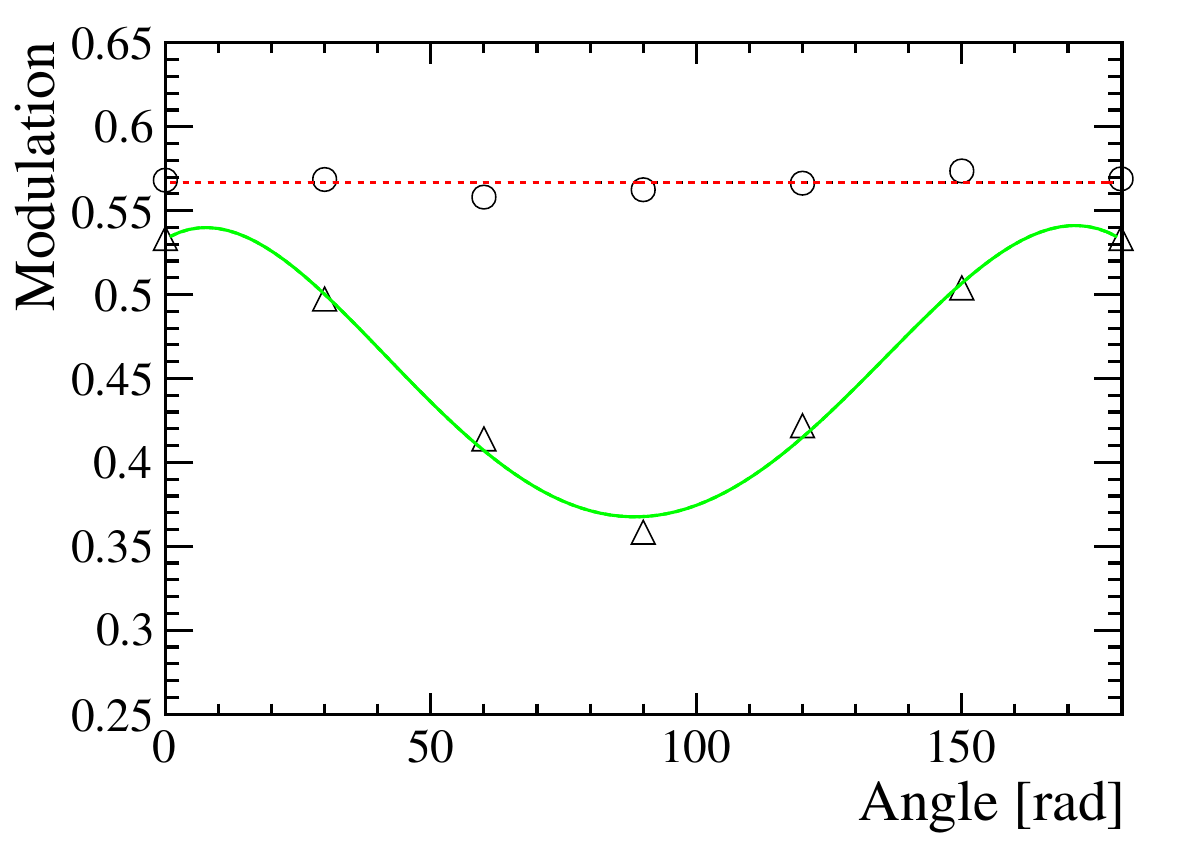}}
  \hfill
  \subfigure[]{\includegraphics[width=0.29\textwidth]{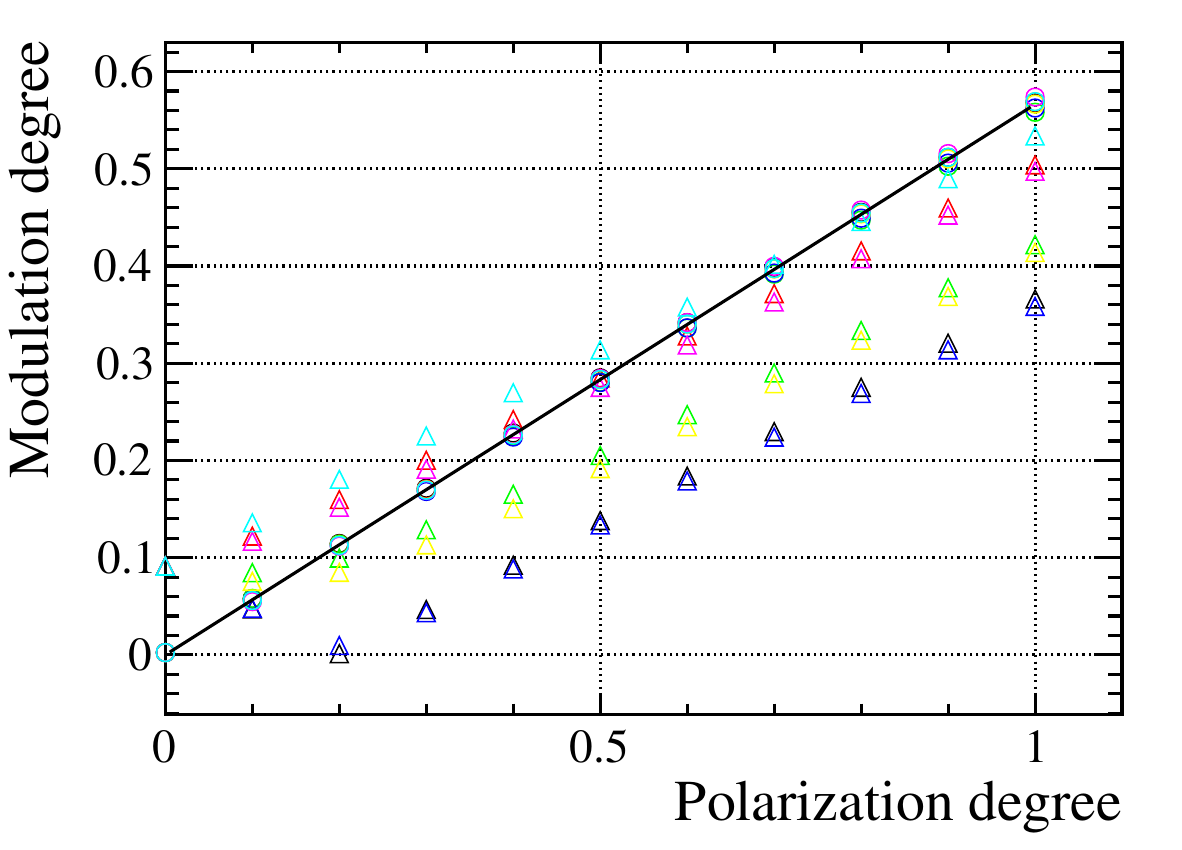}}
  \hfill
  \subfigure[]{\includegraphics[width=0.29\textwidth]{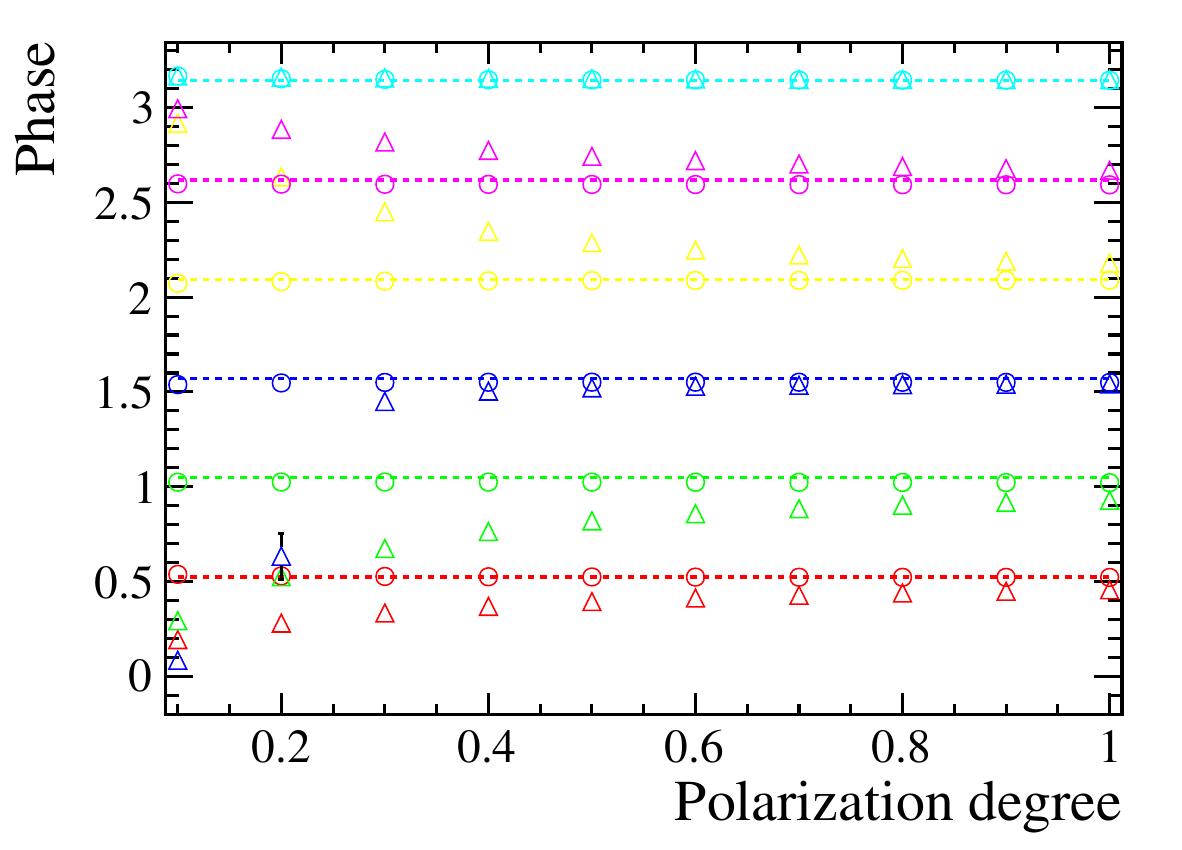}}
  \hfill
  \subfigure[6.40\,keV]{\includegraphics[width=0.29\textwidth]{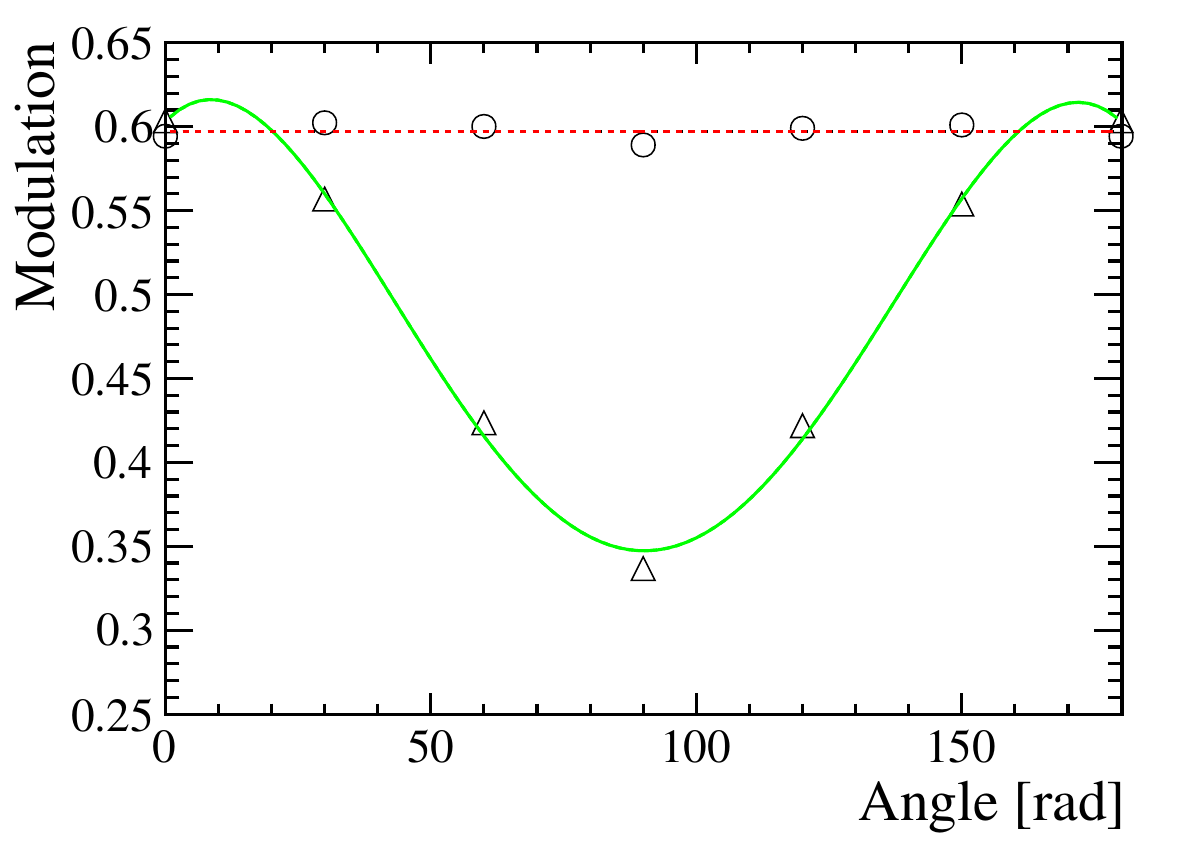}}
  \hfill
  \subfigure[]{\includegraphics[width=0.29\textwidth]{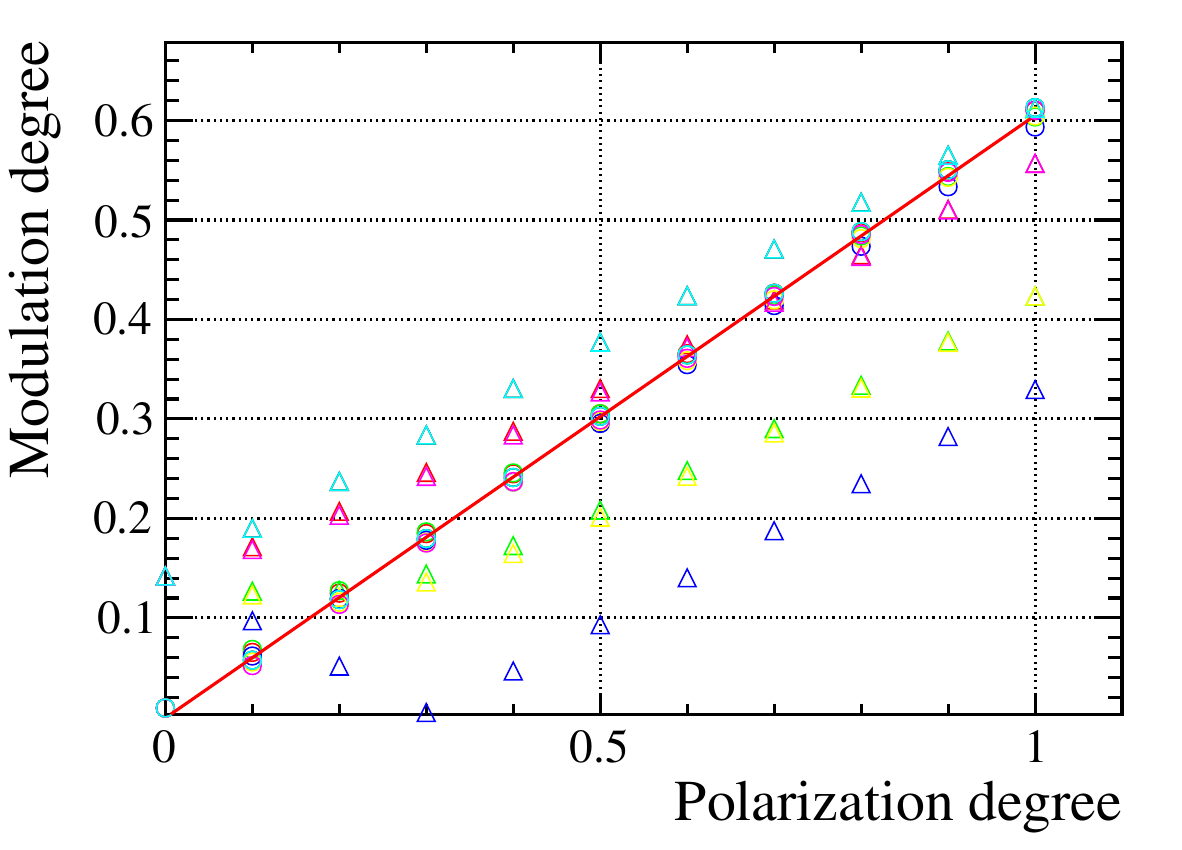}}
  \hfill
  \subfigure[]{\includegraphics[width=0.29\textwidth]{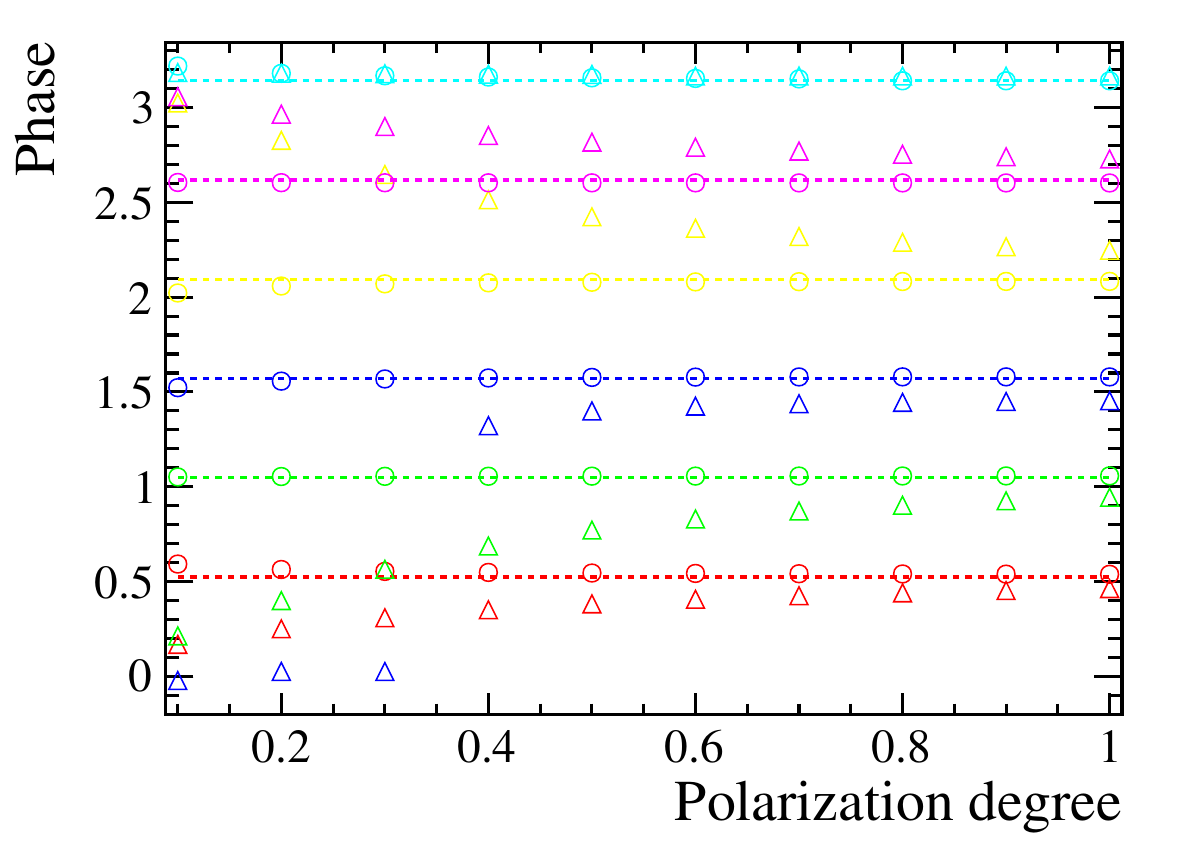}}
  \hfill
  \subfigure[8.05\,keV]{\includegraphics[width=0.29\textwidth]{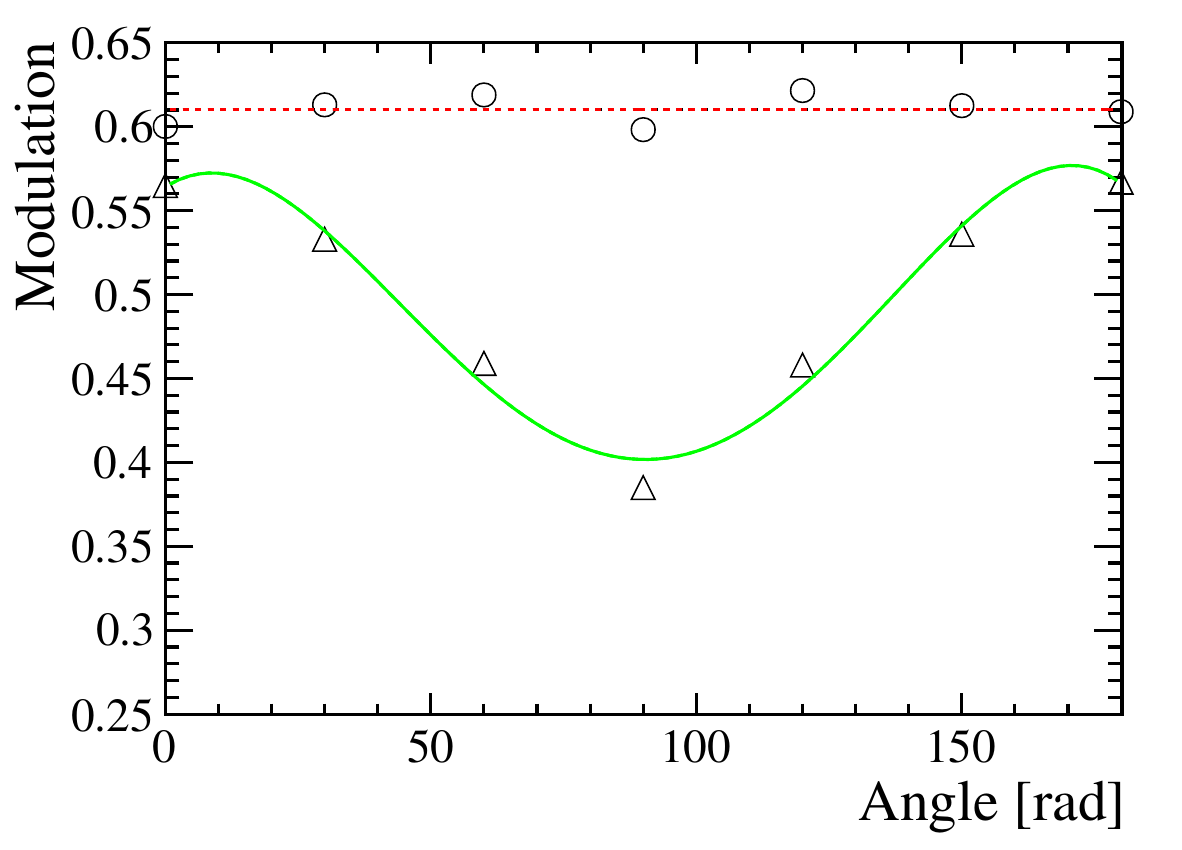}}
  \hfill
  \subfigure[]{\includegraphics[width=0.29\textwidth]{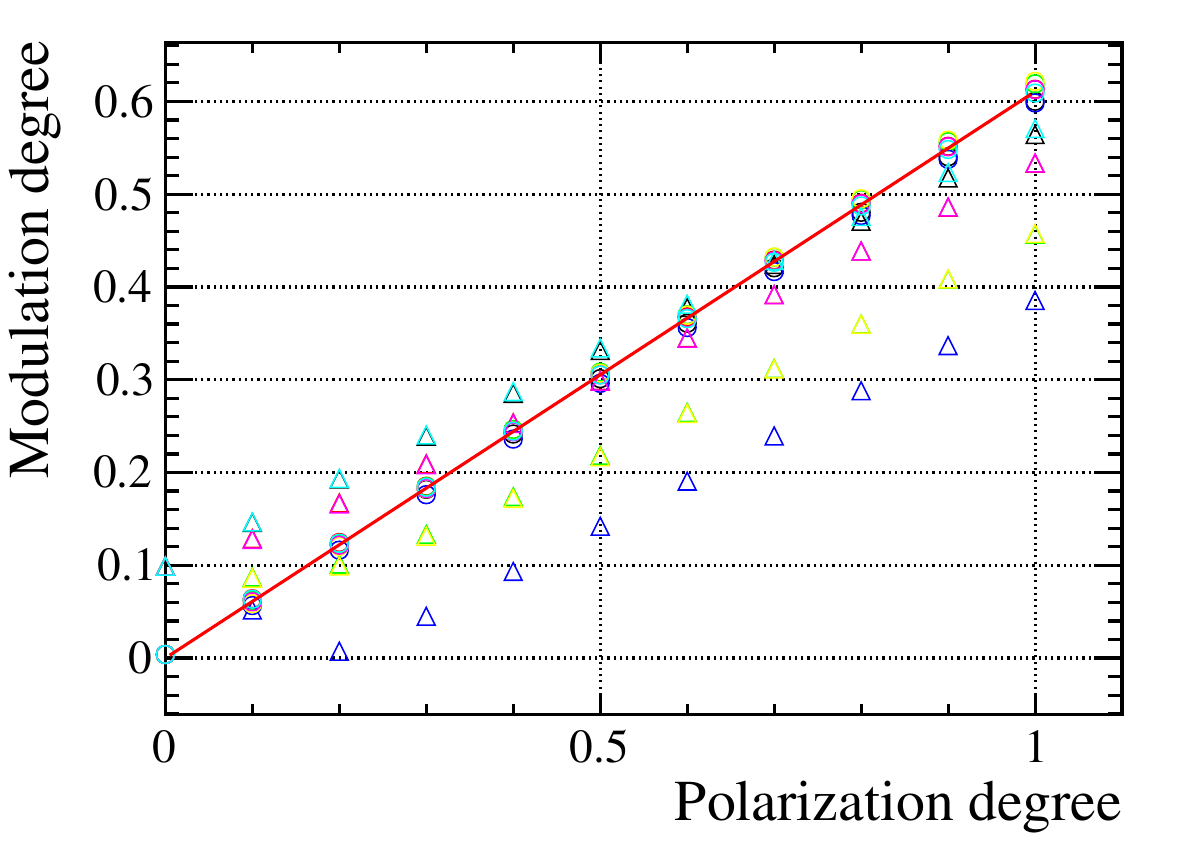}}
  \hfill
  \subfigure[]{\includegraphics[width=0.29\textwidth]{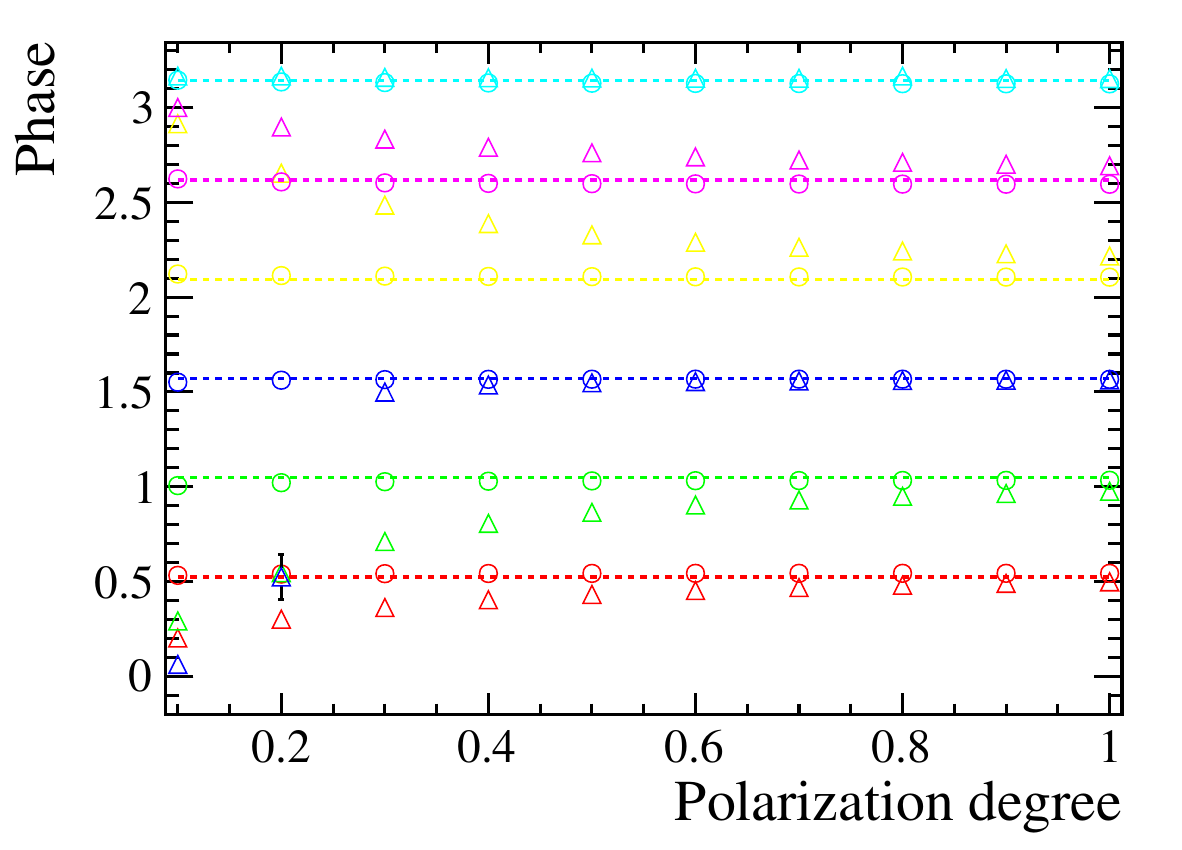}}
  \caption{The results are for 2.98\,keV, 4.51\,keV, 5.40\,keV, 6.40\,keV, and 8.05\,keV, respectively. (a), (d), (g), (j), (m) represent the modulation degree measured at different polarization phases of polarized sources and the corrected modulation degree. (b), (e), (h), (k), (n) represent the comparison before and after correction for data at different polarization phases and degrees. The hollow triangles represent the uncorrected results, the hollow circles represent the corrected results, with different colors representing different polarization phases, and the red line represents the linear fitting of the corrected data. (c), (f), (i), (l), (o) represent the reconstruction of the polarization phase before and after correction, with the dashed line representing the true polarization phase of the polarized source.}
  \label{fig:DifMod}
\end{figure*}

At a specific energy point, using the aforementioned method, we only need to calibrate one corresponding parameter, namely $\Bar{\eta}$, and simulate the response matrix at that energy point. This allows for the application of Bayesian iteration to correct the modulation curves at different polarization degrees and phases. The corrected polarization degree and modulation level exhibit a good linear relationship, and the modulation levels at different phases also show good consistency. Fig.\ref{fig:RecFix_90deg} illustrates the comparison of the modulation distribution before and after correction for experimental data at 5.40\,keV and 90° polarization phase, ranging from unpolarized to 99.9\% polarized. When the polarization degree is low, the residual modulation will dominate the distribution of the modulation curve. At this point, the unrevised experimental data reconstruction results will exhibit significant deviations. In contrast, the corrected data maintains good stability in the reconstruction of the polarization phase, while also exhibiting a strong linear relationship between the modulation level and the polarization degree. Fig.\ref{fig:DifMod} displays the comparison of experimental data before and after correction at several energy points, including 2.98\,keV, 4.51\,keV, 5.40\,keV, 6.40\,keV, and 8.05\,keV, for different polarization phases and degrees. The uncorrected data, due to the residual modulation not being eliminated, exhibit significant differences in the reconstructed modulation degree at different phases. For example, at 4.51\,keV, the modulation degrees at 0° and 90° for the same fully polarized source differ by approximately 18\%. The polarization phase reconstruction results from the uncorrected data also show a significant deviation from the true values, especially in the direction that differs by 90° from the polarization direction of the residual modulation. However, after Bayesian iterative correction, the polarization data show good consistency in modulation degree across different phases. Additionally, the polarization degree and modulation degree exhibit a good proportional relationship, meeting the calibration requirements of the LPD.

\subsubsection{Error and Comparison}
The error in the modulation degree of the corrected data distribution mainly arises from two sources. One part originates from the statistical error of the data, which can be obtained through fitting. The other part of the error arises from the process of using the Bayesian method for correction:
\begin{enumerate}
\item Error propagation in the Bayesian iteration process: This error can be calculated through the error propagation matrix \ref{eq:err_propMatr}.

\item Termination of the Bayesian iteration: Although the chi-square calculation results show good convergence after 10 iterations for all experimental data, the convergence levels of the data at different polarization phases are inconsistent due to the fixed number of iterations. This results in slight differences in the reconstructed modulation degree at different polarization phases after correction.

\item Parameterized response matrix: The error in the estimation of the parameter $\Bar{\eta}$ provided by the simulator will be propagated to the response matrix, and during the Bayesian iteration process using the response matrix, the error will be propagated to the corrected data's modulation degree.
\end{enumerate}

The error propagation in point 1 is calculated by the RooUnfold package. For the statistical error of the data and points 1 and 2, due to the dependence of the Bayesian method iteration process on the original data itself, it is difficult to decouple and analyze the contributions of these two parts. Therefore, a unified error estimation is provided using the modified Bootstrap method, and this part of the error is denoted as $\sigma_{\text{unfold}}$:
Sampling 10,000 times at a certain polarization degree (taking fully polarized data as an example).
\begin{enumerate}
\item Each sampling involves 100,000 with-replacement samplings of the data at 0°, 30°, 60°, 90°, 120°, and 150° phases in the experiment.
\item Reconstruction of the sampled data at the six phases is performed, and the Bayesian method is used to correct the reconstructed angular distribution results. Six sets of corrected data are fitted to obtain six modulation degrees.
\item Random weights are assigned to the six modulation degrees, with the total sum of the six weights equaling 1. The weighted sum yields the modulation degree for this sampling.
\end{enumerate}

\begin{figure*}[htbp]
\includegraphics
  [width=1.\hsize]
  {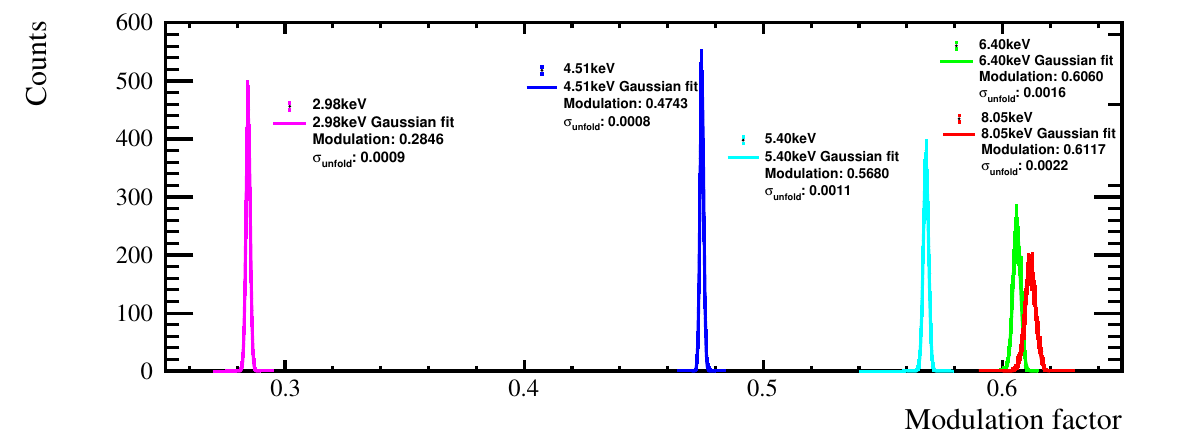}
\caption{The modified Bootstrap method is used to sample datasets at different energy points, which are then iteratively corrected through Bayesian inference to obtain the weighted modulation degree distribution. The modulation degree distribution is estimated by fitting it with a Gaussian function to quantify the error introduced by the Bayesian iteration, denoted as $\sigma_{\text{unfold}}$.}
\label{fig:Bootstrap}
\end{figure*}

After 10,000 samplings, a distribution of the modulation degrees is plotted, and a Gaussian fit is applied. The fitted sigma represents the $\sigma_{\text{unfold}}$. Fig.\ref{fig:Bootstrap} illustrates the modulation distribution of several energy points sampled using the modified Bootstrap method from completely polarized data, and the $\sigma_{\text{unfold}}$ obtained from Gaussian fitting.

For the point 3, the error introduced by parameterization can be propagated to the error of the response matrix by providing the error of the parameter $\Bar{\eta}$. This error is ultimately propagated to the error of the modulation degree. The error introduced by parameterization is denoted as $\sigma_{\text{para}}$. We obtain the error of $\Bar{\eta}$ as follows: by making a slight adjustment to the value of $\Bar{\eta}$ corresponding to a specific energy point, denoted as $\Delta\Bar{\eta}$. After the adjustment, we incorporate the parameter $\Bar{\eta}$ + $\Delta\Bar{\eta}$ into the Star-XP simulation software to simulate 1,500,000 events, and reconstruct their angular distributions. When the $\chi^2/\text{ndf}$ between this adjusted angular distribution and the angular distribution obtained from the simulation with $\Bar{\eta}$ equals 1, the $\Delta\Bar{\eta}$ represents our estimated error of $\Bar{\eta}$. We apply the modified Bootstrap method to resample the data 10,000 times using the response matrix obtained from the parameter value $\Bar{\eta}\pm\Delta\Bar{\eta}$, and then perform Gaussian fitting to obtain the total error $\sigma_{\text{unfold+para}}$ or denoted as $\sigma_{\text{sys}}$. The contributions of the various error terms at different energy points and the modulations corrected are presented in table \ref{tab:compare}.

\begin{figure}[htbp]
\includegraphics
  [width=1.\hsize]
  {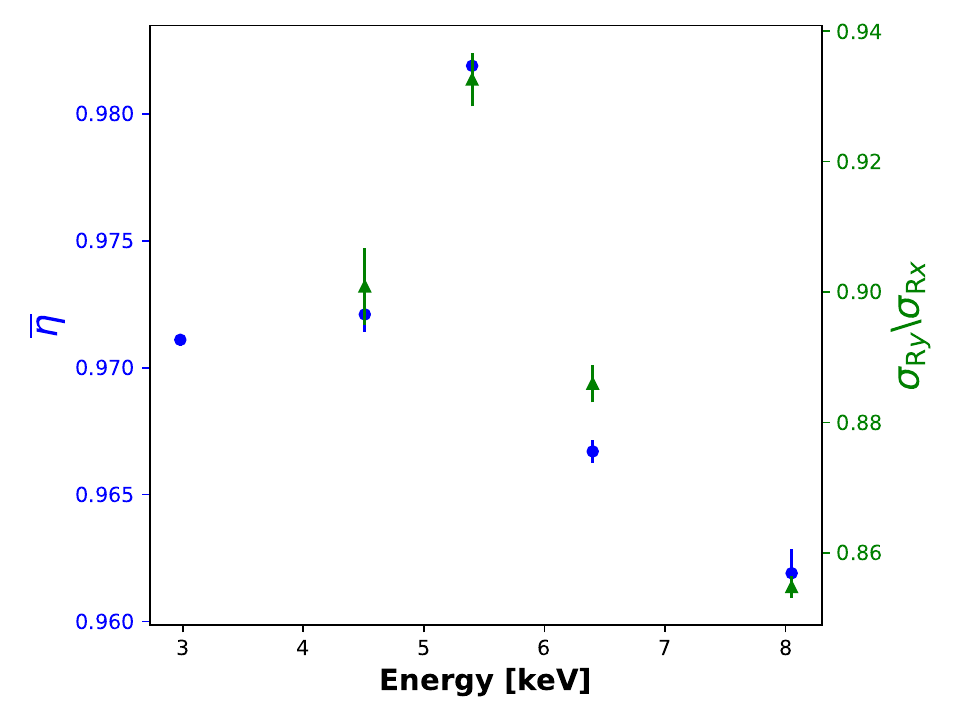}
\caption{The Comparison of the position resolution ratio in the Y($\sigma_{\text{R}y}$) and X($\sigma_{\text{R}x}$) directions with $\Bar{\eta}$ at different energies. Blue circles represent $\Bar{\eta}$, and green triangles represent $\sigma_{\text{R}y} \backslash \sigma_{\text{R}x}$.}
\label{fig:RatioCompare}
\end{figure}

\begin{figure*}[htbp]
  \centering
  \subfigure[]{\includegraphics[width=0.45\textwidth]{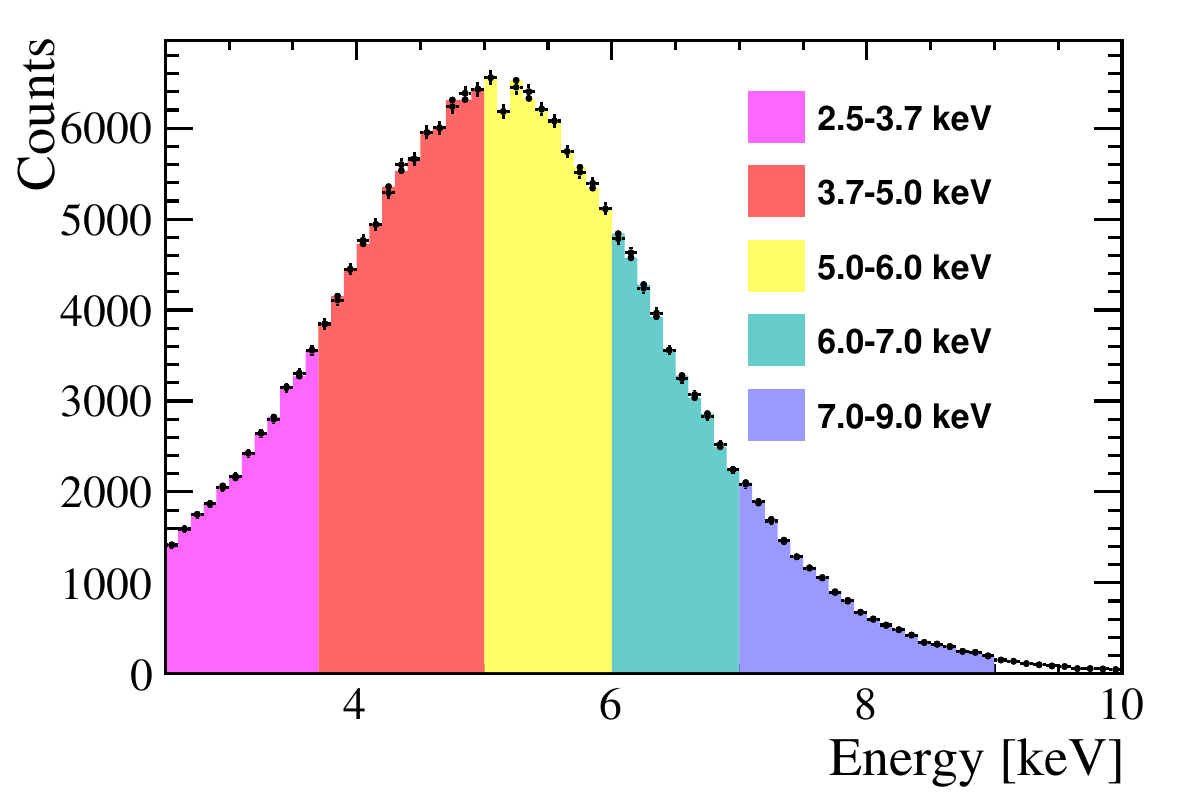}}
  \hfill
  \subfigure[]{\includegraphics[width=0.45\textwidth]{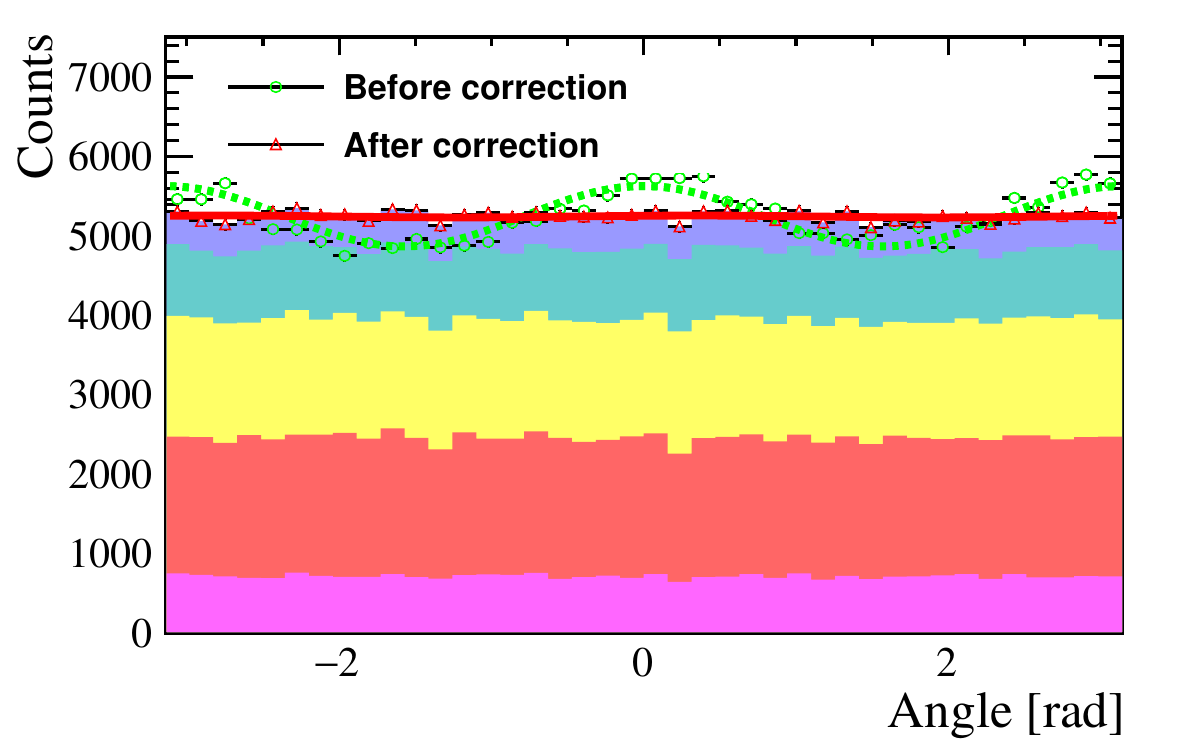}}
  \hfill
  \subfigure[]{\includegraphics[width=0.45\textwidth]{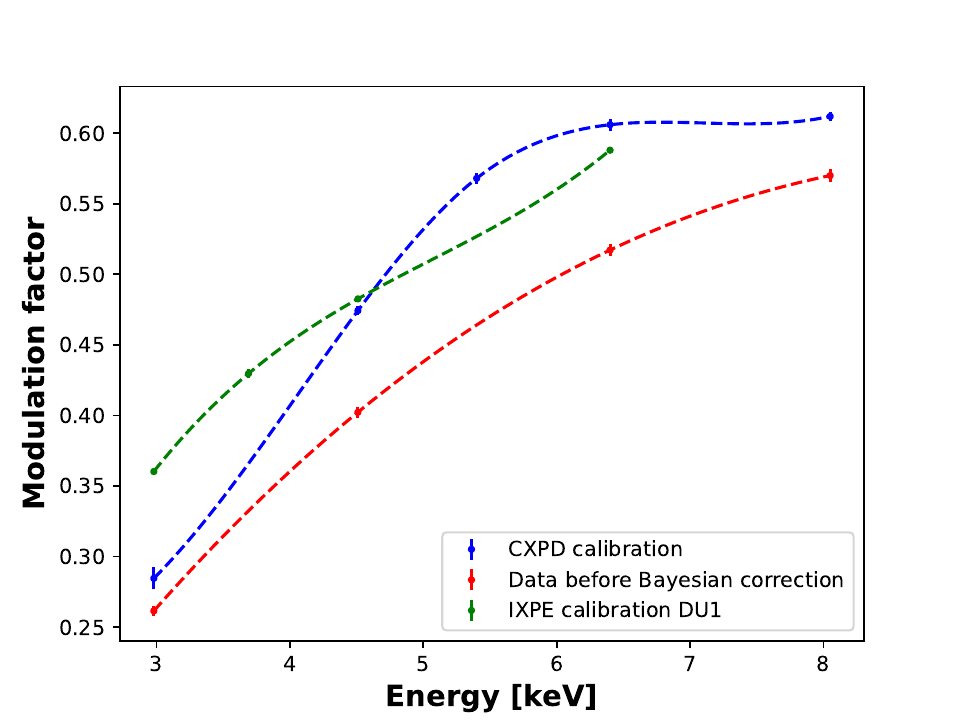}}
  \hfill
  \subfigure[]{\includegraphics[width=0.45\textwidth]{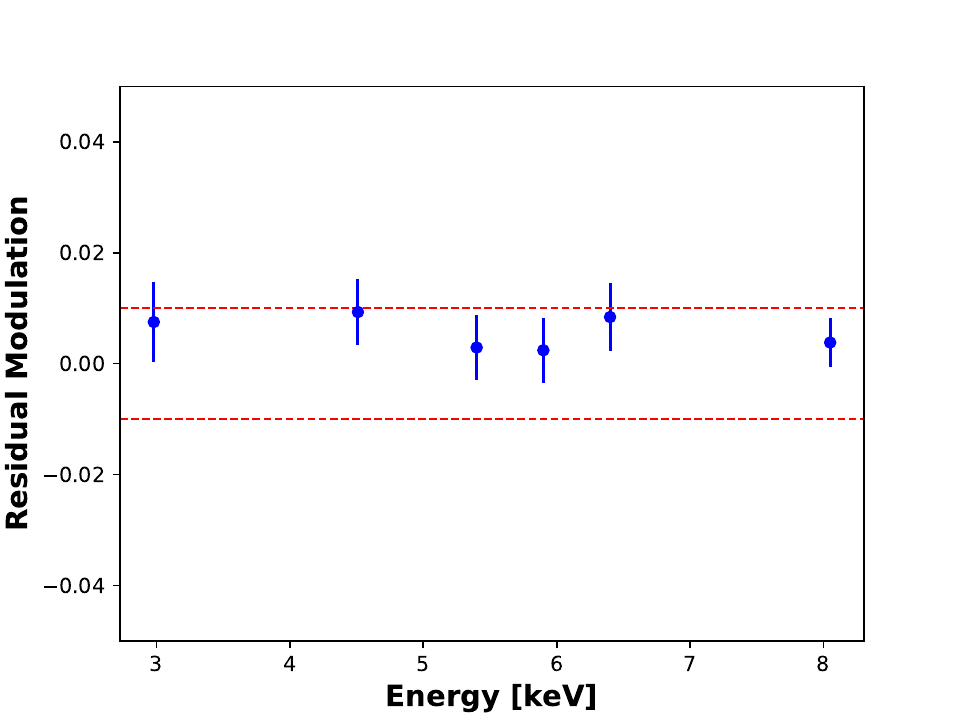}}
  \caption{(a),(b) Residual modulation of the continuous X-ray energy spectrum before and after partition correction. The green dashed line represents the fitting result of the residual modulation before correction, while the red solid line represents the fitting result after correction. (c) The red points represent the modulations at different energy points before correction \cite{Feng_2023}, the blue points represent the results after correction, and the green points represent the calibration results of the IXPE Detection Unit 1 (DU1). (d) Residual modulation after correction of unpolarized data.}
  \label{fig:Mod_Compare}
\end{figure*}

Furthermore, we compared the variation trends of $\Bar{\eta}$ and the ratio of position resolution in the Y and X directions of the detector at different energies, as shown in Fig.\ref{fig:RatioCompare}, the trends are in good agreement. Using existing calibration data, we collected data from unpolarized continuous X-ray sources in the energy range of 2.5-9.0\,keV and performed reconstruction and correction. Based on the energy resolution of the detector and the energy points of calibration, we divided the continuous spectrum into five intervals for correction, as shown in Fig.\ref{fig:Mod_Compare} (a) and (b). The residual modulation of the continuous spectrum decreased from 6.95\% before correction to 0.22\%. Fig.\ref{fig:Mod_Compare}(c) demonstrates the modulations at several energy points before \cite{Feng_2023} and after Bayesian correction, and compares them with the calibration results of the IXPE detector \cite{Di_Marco_2022}. Comparing the modulations before and after correction, the modulations after correction are higher at each energy point than before correction. Comparing the corrected modulations with the results from IXPE, when the energy is below 4.51\,keV, our detector's modulations are lower than the IXPE calibration results. The charge induction chip pixel size of IXPE (60\,$\mu$m) is smaller than the pixel size of our current Topmetal-II (83\,$\mu$m), resulting in less pixel reconstructed tracks for low-energy photons and higher reconstruction accuracy requirements for resolution. Therefore, the modulation of the reconstructed tracks for the low-energy part is lower than the IXPE results. However, when the energy is higher than 4.51\,keV, the corrected modulations are higher than the IXPE results, possibly due to the better signal-to-noise ratio of Topmetal-II. For longer tracks, the pixel resolution no longer plays a decisive role in reconstruction accuracy, and factors such as chip noise, the diffusion coefficient of secondary ionization electrons, detector gain, track length, and others begin to have a greater impact on the reconstruction. More importantly, after correction, the residual modulations of the detector at several energy points have been reduced to levels below 1\%. In addition, the residual modulation result at 5.9\,keV in Fig.\ref{fig:Mod_Compare}(d) is obtained using the $\Bar{\eta}$ at 5.4\,keV. The result of the response matrix correction at 5.9\,keV is 0.24\% ± 0.59\%. The energy resolution at 5.4\,keV, corresponding to the detector, is approximately 19.5\% \cite{Feng_2023}, and 5.9\,keV coincides with the boundary value of the 5.4\,keV energy resolution. This result indicates that the calibration parameter $\Bar{\eta}$ can be extended to the energy resolution range of the detector at several energy points, while still maintaining good correction results.

\begin{table*}
\centering
\caption{The values of $\Bar{\eta}$, $\Delta\Bar{\eta}$, $\sigma_{\text{unfold}}$, $\sigma_{\text{para}}$, and the statistical error $\sigma_{\text{stat}}$ at different energy points, as well as the corrected modulations or residual of polarized and unpolarized source. The detailed parameters of the X-ray polarizing source can be found in \cite{Xie2023}. }
\label{tab:compare}
\resizebox{\linewidth}{!}{
\tiny
\begin{tabular}{*{9}{c}}
\toprule[0.25mm]
Energy  & $\Bar{\eta}$ & $\Delta\Bar{\eta}$ & $\sigma_{\text{unfold}}$ & $\sigma_{\text{sys}}$ & Polarization degree &$\sigma_{\text{stat}}$ & $\sigma_{\text{total}}$ & Modulation/Residual \\ \toprule[0.25mm]
\multirow{2}{*}{2.98\,keV}  & \multirow{2}{*}{0.9711} & \multirow{2}{*}{0.00024} & \multirow{2}{*}{0.0009} & \multirow{2}{*}{0.0010} & 97.4\%              & 0.0079     & 0.0080       & 0.2846              \\
                         &                         &                          &                         &                         & 0.0                   & 0.0072     & 0.0073       & 0.0075              \\
\multirow{2}{*}{4.51\,keV}  & \multirow{2}{*}{0.9721} & \multirow{2}{*}{0.00068} & \multirow{2}{*}{0.0008} & \multirow{2}{*}{0.0009} & 99.8\%              & 0.0033     & 0.0034       & 0.4743              \\
                         &                         &                          &                         &                         & 0.0                   & 0.0058     & 0.0059       & 0.0093              \\
\multirow{2}{*}{5.40\,keV}  & \multirow{2}{*}{0.9819} & \multirow{2}{*}{0.00052} & \multirow{2}{*}{0.0011} & \multirow{2}{*}{0.0013} & 99.9\%              & 0.0037     & 0.0039       & 0.5680               \\
                         &                         &                          &                         &                         & 0.0                   & 0.0057     & 0.0058       & 0.0029              \\
\multirow{2}{*}{6.40\,keV}  & \multirow{2}{*}{0.9667} & \multirow{2}{*}{0.00044} & \multirow{2}{*}{0.0016} & \multirow{2}{*}{0.0022} & 99.8\%              & 0.0038     & 0.0044       & 0.6060               \\
                         &                         &                          &                         &                         & 0.0                   & 0.0057     & 0.0061       & 0.0084              \\
\multirow{2}{*}{8.05keV} & \multirow{2}{*}{0.9619} & \multirow{2}{*}{0.00097} & \multirow{2}{*}{0.0022} & \multirow{2}{*}{0.0023} & 99.8\%              & 0.0026     & 0.0034       & 0.6117              \\
                         &                         &                          &                         &                         & 0.0                   & 0.0039     & 0.0045       & 0.0038              \\ \bottomrule[0.25mm]
\end{tabular}
}
\end{table*}

\section{Summary and outlook}
\label{sec:summary}
    This paper discusses the systematic effects of GMPD and corrects the residual modulation of modulation curves caused by various systematic effects. GMPD is a prototype detector designed for POLAR-2/LPD, and the study of GMPD systematic effects is of great significance for the subsequent design and performance optimization of LPD, reducing systematic effects, and calibrating detector polarization performance. In the second section, we list several main systematic effects that lead to residual modulation, including differences in gain and layout of chip pixels, signal attenuation in electronics, track truncation, and charge accumulation effects. For these known systematic effects, we corrected them through calibration, setting threshold conditions, and time positioning. For the remaining residual modulation caused by a part of the systematic effects, we obtained the response matrix through parameterization combined with Monte Carlo simulation and used the Bayesian method to eliminate the contribution of residual modulation in the modulation curve. The final results show that the residual modulation of the data corrected by our algorithm has been reduced to below 1\% at various calibration energy points. The reconstructed modulation degrees of data at different polarization phases show good consistency, and the polarization degree and modulation degree also exhibit a good linear relationship. At the same time, we discussed the errors of the correction algorithm proposed in this paper and compared the corrected modulation results with the IXPE calibration results. The data results of GMPD after correction by our algorithm show higher polarization detection performance than IXPE above 5\,keV. 

The results of this paper indicate that the correction algorithm proposed by us can be well applied to the correction of systematic effects in the LPD detector. Additionally, our parameterized correction algorithm can naturally be extended to the study and correction of oblique incidence systematic effects. The correction algorithm that introduces Stokes parameters in IXPE is established under the condition of normal incidence. When photons are obliquely incident, the description of photoelectrons using the Stokes parameter system is incomplete \cite{Muleri_2014}, making it difficult to extend to the correction of oblique incidence systematic errors. The large field-of-view design of LPD implies that most of the time we need to analyze obliquely incident data results. Based on the method proposed in this paper, we will also carry out the reconstruction and study of oblique incidence systematic effects in the future.

\appendix
\section{Bayesian iterative method}
\label{sec::Bayesian}
The Bayesian iterative method is a statistical technique used to estimate a probability distribution by iteratively updating prior beliefs with new evidence. It is employed to correct for detector effects and estimate the true distribution of a physical variable from the measured data.
    
In Bayes method, the unfolded distribution, $\hat{n}(C_i)$, is given by applying the unfolding matrix $M_{ij}$ to the measured distribution, $\hat{n}(E_j)$, as shown in the following equation \ref{eq:Bayes_unfold}:

\begin{equation} 
\label{eq:Bayes_unfold} 
\hat{n}(C_i) = \sum\limits_{j=1}^{n_{\text{E}}}M_{ij}n(\text{E}_{j}). 
\end{equation}

The unfolding matrix is given by:
\begin{equation} 
\label{eq:unfold_Matrix} 
M_{ij} = \frac{P(\text{E}_{j}|\text{C}_{i})n_{0}(\text{C}_{i})}{\epsilon\Sigma^{n_{\text{C}}}_{k=1}P(\text{E}_{j}|\text{C}_{k})n_{0}(\text{C}_{k})}. 
\end{equation}
$P(\text{E}_{j}|\text{C}_{k})$ is the element of response matrix $R$. $\epsilon$ is defined as $\epsilon_{i} \equiv\Sigma^{n_{\text{E}}}_{j=1}P(\text{E}_{j}|\text{C}_{i})$. In the first round of Bayesian iteration, $n_{0}(\text{C})$ is set based on prior knowledge, while in subsequent iterations, $n_{0}(\text{C})$ will be replaced by the $\hat{n}(C)$ obtained from equation \ref{eq:Bayes_unfold}. $\hat{n}(C)$ will be updated with each iteration.

The computation of the error propagation matrix in the Bayesian method is given by the following equations \ref{eq:err_prop} \ref{eq:err_propMatr}:

\begin{multline} 
\label{eq:err_prop} 
\frac{\partial\hat{n}(\text{C}_i)}{\partial{n}(\text{E}_{j})} = M_{ij} + \sum\limits_{k=1}^{n_{\text{E}}}M_{ik}n(\text{E}_{k}) \\
\times\left(\frac{1}{n_{0}(\text{C}_{i})}\frac{\partial{n}_{0}(\text{C}_i)}{\partial{n}(\text{E}_{j})} - \sum\limits_{l=1}^{n_{\text{C}}}\frac{\epsilon_{l}}{n_{0}(C_{l})}\frac{\partial{n}_{0}(\text{C}_{l})}{\partial{n}(\text{E}_{j})}M_{lk} \right). 
\end{multline}

\begin{equation} 
\label{eq:err_propMatr} 
V(\hat{n}(\text{C}_{k}),\hat{n}(\text{C}_{l})) = \sum\limits_{i,j=1}^{n_{\text{E}}}\frac{\partial{\hat{n}}(\text{C}_{k})}{\partial{n}(\text{E}_{i})}V(n(\text{E}_{i}),n(\text{E}_{j}))\frac{\partial{\hat{n}}(\text{C}_{l})}{\partial{n}(\text{E}_{j})} . 
\end{equation}
Here, $V(n(\text{E}_{i}),n(\text{E}_{j}))$ is computed from the measurement data.

\bibliographystyle{nst}

\small\bibliography{sample}

\end{document}